\PassOptionsToPackage{pdfpagelabels=false}{hyperref}
\documentclass[fleqn,usenatbib,useAMS]{mnras}

\usepackage[T1]{fontenc}
\usepackage{ae,aecompl}

\usepackage{graphics,graphicx}	
\usepackage{amsmath}	
\usepackage{amssymb}	
\usepackage{pdflscape}

\usepackage{times}
\usepackage{txfonts}
\usepackage{natbib}
\usepackage{array}
\usepackage{multirow}
\usepackage{float}
\usepackage{natbib}
\usepackage{multirow}
\usepackage{afterpage}
\usepackage{mathtools}
\usepackage{listings}
\usepackage{caption}
\usepackage{subcaption}
\usepackage{hyperref}
\usepackage{grffile}

\title[Extreme dust-obscured star formation in quasar host galaxies]{Gravitational lensing reveals extreme dust-obscured star formation in quasar host galaxies}
\author[H. R. Stacey et al.]{H. R. Stacey,$^{1,2}$\thanks{h.r.stacey@astro.rug.nl} J. P. McKean,$^{1,2}$ 
N. C. Robertson,$^{3}$ R. J. Ivison,$^{4,5}$ K. G. Isaak,$^{6}$
\newauthor D. R. G. Schleicher,$^{7}$ P. P. van der Werf,$^{8}$ W. A. Baan,$^{1}$ A. Berciano Alba,$^{1,8}$
\newauthor M. A. Garrett$^{9}$ and A. F. Loenen$^{8}$\\
$^{1}$ASTRON, Netherlands Institute for Radio Astronomy, Oude Hoogeveensedijk 4, 7991 PD Dwingeloo, The Netherlands \\
$^{2}$Kapteyn Astronomical Institute, University of Groningen, P.O. Box 800, 9700 AV Groningen, The Netherlands\\
$^{3}$Department of Physics, University of Oxford, Keble Road, Oxford OX1 3RH, UK \\
$^{4}$Institute for Astronomy, University of Edinburgh, Royal Observatory, Blackford Hill, Edinburgh EH9 3HJ, UK \\
$^{5}$European Southern Observatory, Karl-Schwarzschild-Str. 2, D-85748 Garching bei M\"unchen, Germany \\
$^{6}$Science Support Office, ESTEC/SCI-S, Keplerlaan 1, 2201 AZ Noordwijk, The Netherlands \\
$^{7}$Departamento de Astronom\'ia, Facultad Ciencias F\'isicas y Matem\`aticas, Universidad de Concepci\'on, Av. Esteban Iturra s/n Barrio Universitario, \\ 
Casilla 160-C Concepci\'on, Chile \\
$^{8}$Leiden Observatory, Leiden University, PO Box 9513, NL-2300 RA Leiden, The Netherlands \\
$^{9}$Jodrell Bank Centre for Astrophysics, University of Manchester, Manchester M13 9PL, UK\\
}

\date{Accepted 2018 February 19. Received 2018 February 19; in original form 2017 May 19}

\pubyear{2017}

\begin{document}
\label{firstpage}
\pagerange{\pageref{firstpage}--\pageref{lastpage}}
\maketitle

\begin{abstract}
We have observed 104 gravitationally-lensed quasars at $z\sim1$--4 with {\it Herschel}/SPIRE, the largest such sample ever studied. By targeting gravitational lenses, we probe intrinsic far-infrared (FIR) luminosities and star formation rates (SFRs) more typical of the population than the extremely luminous sources that are otherwise accessible. We detect 72 objects with {\it Herschel}/SPIRE and find 66 percent (69 sources) of the sample have spectral energy distributions (SEDs) characteristic of dust emission. For 53 objects with sufficiently constrained SEDs, we find a median effective dust temperature of $38^{+12}_{-5}$~K. By applying the radio-infrared correlation, we find no evidence for an FIR excess which is consistent with star-formation-heated dust. We derive a median magnification-corrected FIR luminosity of $3.6^{+4.8}_{-2.4} \times 10^{11}\ {\rm L_{\odot}}$ and median SFR of $120^{+160}_{-80}\ {\rm M_{\odot}\ yr^{-1}}$ for 94 quasars with redshifts. We find $\sim10$~percent of our sample have FIR properties similar to typical dusty star-forming galaxies at $z\sim2$--3 and a range of SFRs $<$20--10000 ${\rm M_{\odot}\ yr^{-1}}$ for our sample as a whole. These results are in line with current models of quasar evolution and suggests a coexistence of dust-obscured star formation and AGN activity is typical of most quasars. We do not find a statistically-significant difference in the FIR luminosities of quasars in our sample with a radio excess relative to the radio-infrared correlation. Synchrotron emission is found to dominate at FIR wavelengths for $<15$~percent of those sources classified as powerful radio galaxies.
\end{abstract}

\section{Introduction}
\label{intro}

Key to the study of galaxy formation and evolution is understanding the physical processes that drive star formation and the growth of active galactic nuclei (AGN). The concurrence of these phenomena is thought to relate a coevolution driven by feedback from the AGN, which may quench or induce star formation in the host galaxy through interactions with the interstellar medium. The mechanism of feedback may involve mechanical energy injection via AGN-driven jets, called `jet-mode' or `radio-mode' \mbox{\citep{Bicknell:2000,Klamer:2004}}, or radiative energy injection via winds, called `quasar-mode', although these processes are not well understood \cite[see][for review]{Alexander:2012}.

Hydrodynamical simulations of galaxy formation \citep{Matteo:2005,Hopkins:2005,Bower:2006} and various observational studies \protect\citep[for example][]{Page:2004,Stevens:2005,Coppin:2008} support an evolutionary model, initially proposed by \cite{Sanders:1988} and developed more recently by \protect\cite{Hopkins:2008}, in which quasars are formed as a result of gas-rich major mergers. According to this scenario, luminous dusty star-forming galaxies (DSFGs) are merger-driven starbursts that represent a transition phase into dust-obscured quasars. Over time, feedback effects strip the quasar host galaxies of gas and dust, and the quasars become unobscured and ultraviolet (UV) luminous. These leave passive spheroidal galaxies when the quasar exhausts its supply of cold gas.

Quasars that are luminous in the far-infrared (FIR) to mm regime are therefore predicted to be in a transition phase of their evolution with high rates of dust-obscured star formation. Studying the properties of these sources can provide important information about the evolutionary process, particularly when compared to the large population of extreme starburst galaxies that were discovered through blind surveys with the Submillimetre Common-User Bolometer Array (SCUBA), {\it Herschel Space Observatory} and now the Atacama Large Millimetre/sub-millimetre Array (ALMA).

Studies of FIR-luminous quasars, such as those in the SCUBA Bright Quasar Survey \citep{Isaak:2002,Priddey:2003} and MAMBO/IRAM-30~m Survey \citep{Omont:2001,Omont:2003}, and more recent studies of quasars detected with {\it Herschel}/SPIRE \citep[for example]{Pitchford:2016} have found that these quasars are embedded within gas- and dust-rich starbursting galaxies, with star formation rates of $\sim$1000 M$_{\odot}$ yr$^{-1}$, comparable to FIR-detected DSFGs. The low spatial density of FIR-luminous quasars, relative to DSFGs and UV-luminous quasars, has led some to argue for a quick transition from starbursting DSFG to an AGN-dominated quasar, with the FIR-luminous quasar phase being less than 100~Myr, and perhaps as short as $\sim$1~Myr \citep[for example]{Simpson:2012}. However, studies of  individually-detected quasars have mostly focused on significantly bright sources due to limitations in sensitivity or source confusion. While some recent progress has been made with the improved sensitivity and resolution of ALMA \citep{Harrison:2016,Banerji:2016,Scholtz:2018}, resolutions of 100-pc are required to spatially resolve regions of star formation and AGN-heating, which are still difficult to attain for the high-redshift Universe. 

Other studies have instead used stacking to investigate the mean star formation properties of quasar host galaxies. These studies, which account for redshift and stellar mass, find no significant correlation between star formation and AGN activity, and find SFRs comparable to normal star-forming galaxies that lie on the galaxy main sequence \citep{Rosario:2013,Azadi:2015,Stanley:2017}.

The next logical step in understanding the properties of quasar host galaxies at all luminosities requires an investigation of lower surface-brightness sources. Many of the limitations of confusion and sensitivity can be mitigated by observing quasars that have been magnified by a gravitational lens.

The advantages of observing strong gravitationally-lensed quasars are three-fold. The first is that magnification effects increase the apparent flux-density such that a magnification factor of $\sim$10 the reduces integration time by a factor of $\sim$100. Sources with intrinsic flux densities below the confusion limit of field quasars can therefore be observed, probing the fainter end of the luminosity function \citep[for example]{Impellizzeri:2008}. The second advantage is the increase in apparent surface area, which combined with source reconstruction methods, allow source structure to be probed on much smaller physical scales \citep[for example]{Rybak:2015a,Rybak:2015b}. A third advantage is that gravitational lensing has different systematic biases compared to field sources; while field observations tend to bias high luminosity or low-redshift sources, gravitationally-lensed sources are more biased towards compact higher redshift sources (typically $z > 1$) and less biased towards high intrinsic luminosities\footnote{Although these biases are dependent on whether the gravitational lens systems are selected via the lens or source populations.} \citep[for example]{Swinbank:2010}. In combination, these methodologies allow for a more complete view of the quasar population to be constructed.

In this paper, we have targeted a sample of strong gravitationally-lensed quasars with the {\it Herschel Space Observatory} \citep{Pilbratt:2010} and derive their dust temperatures, intrinsic FIR luminosities and dust-obscured SFRs. Previous work in this area has been undertaken by \cite{Barvainis:2002}, who detected 23 of 40 gravitationally-lensed quasars and radio galaxies in their sample at 850~$\umu$m with SCUBA. They found dust emission broadly comparable to radio galaxies, in line with the AGN unification model, and no statistically-significant difference AGN classified as powerful radio galaxies, as would be expected if they have the same host galaxy properties. We have observed 104 lensed quasars, including 37 of the \citeauthor{Barvainis:2002} sample, detecting 72 sources in at least one band with the {\it Herschel}/SPIRE. As our data cover shorter wavelengths, we are also able to determine the dust temperatures for the first time and infer whether the heated dust is due to star formation or AGN activity.

In Section \ref{sample}, we present our sample selection, the relevant properties of the quasars in our sample, the parameters of the observations, and our data reduction methods. In Section \ref{results}, we report the results of the photometric measurements and the analysis of the radio-to-FIR spectral energy distributions (SEDs) of the sources. In Section \ref{discussion}, we show that the SEDs are consistent with dust heating due to star formation in the quasar host galaxies, and we compare our results with a sample of DSFGs at similar redshifts. Here, we also consider the contribution to the total radio emission from star formation processes for these quasars by considering the infrared--radio correlation. Finally, in Section~\ref{conc} we present a summary of our results and discuss the future work that we will carry out with this sample.

Throughout, we assume the \cite{Planck:2016} instance of a flat $\Lambda$CDM cosmology with $H_{0}= 67.8~$km\,s$^{-1}$ Mpc$^{-1}$, $\Omega_{M}=0.31$ and $\Omega_{\Lambda}=0.69$.

\section{Sample \& Observations}
\label{sample}

In this section, we describe our sample of gravitationally-lensed quasars and present the observations that were carried out using the {\it Herschel Space Observatory}.

\subsection{Sample selection}

Our sample consists of all of the gravitationally-lensed quasars that were observed with the {\it Herschel Space Observatory} \citep{Pilbratt:2010} using the Spectral and Photometric Imaging Receiver (SPIRE) instrument \citep{Griffin:2010}. The vast majority of the observations came from our own open time project (Proposal ID: OT1\_abercian\_1). At the time of the proposal, these included all  known quasars lensed by foreground galaxies. The majority of the sample are identified spectroscopically to be quasars, although some are identified as powerful radio galaxies without detections of prominent emission lines\footnote{We refer to all these objects as quasars in this paper for simplicity.}. These sources are listed in the Sloan Digital Sky Survey Quasar Lens Search (SQLS) catalogue and CASTLES database \citep{Inada:2012,Kochanek:1999} and come from a variety of surveys at optical and radio wavelengths. Our sample is quite heterogeneous given the nature of the different surveys from which the targets were selected, but its size will allow us to draw representative conclusions on the relative FIR properties of jet-dominated and SF-dominated quasars, and provides a large parent sample from which further higher resolution observations of interesting individual objects can be made.

\begin{figure} 
\includegraphics[width=0.5\textwidth]{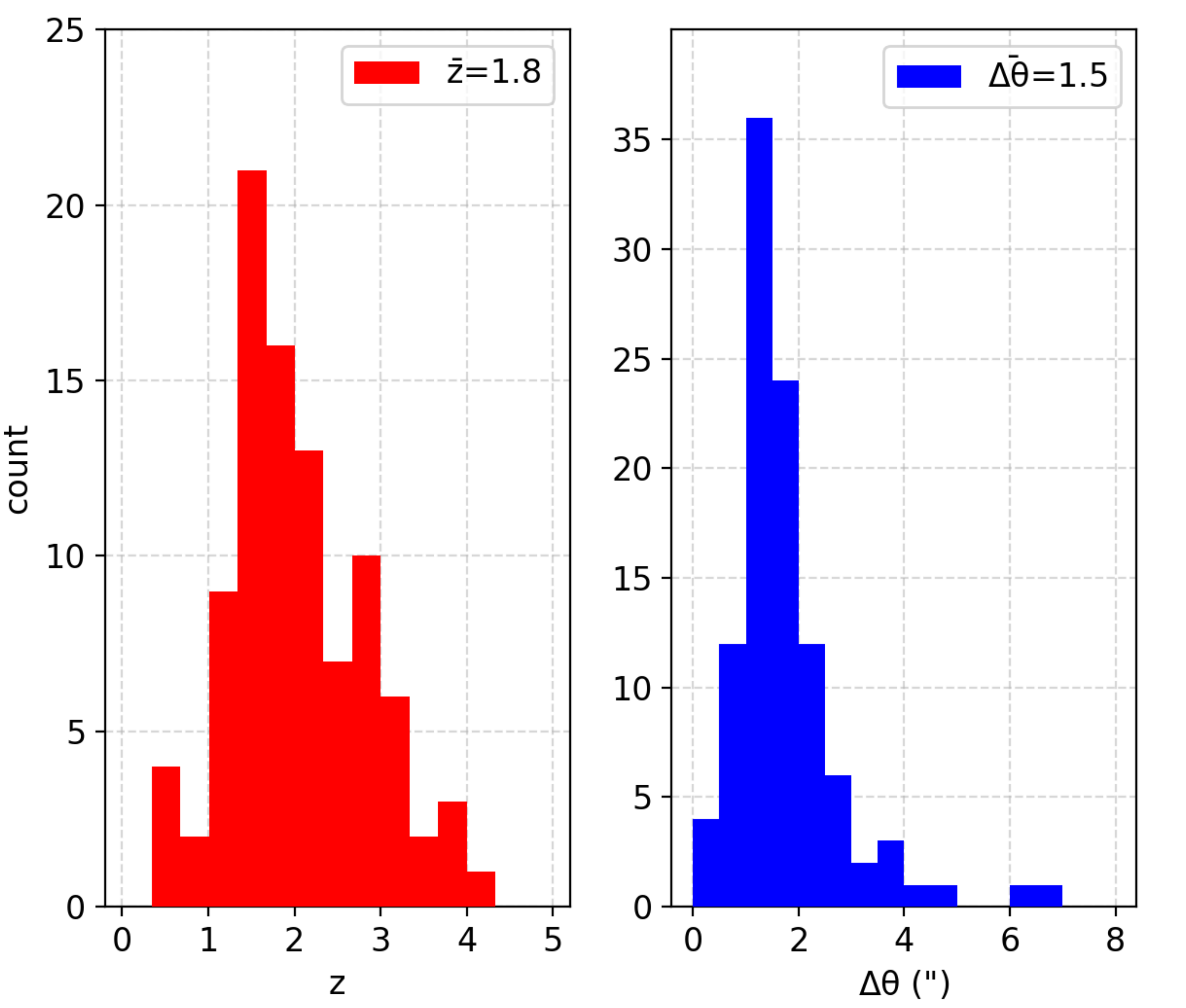}
\caption{\small (left) The redshift distribution of 94 objects in our sample with known redshift, which has a median redshift of 1.8. (right) The lensed image separations, in arcsec, which have a median of 1.5 (excluding SDSS~J1029+2623 which has a maximum image separation of 22.5~arcsec).}
\label{fig:sep_z}
\end{figure}

In total, there are 104 lensed quasars in our sample, the relevant properties of which are presented in Table \ref{table:spireflux} of the Appendix. The redshift distribution and maximum image separations of the lensed quasars in our sample are presented in Fig.~\ref{fig:sep_z}. The full width at half maximum (FWHM) of the point spread function (PSF) in each band is 18, 24 and 35 arcsec for the 250, 350 and 500 $\umu$m bands, respectively. Therefore, all but 3 of our sample (Q0957+561, RX~J0921+4529 and SDSS~J1029+2623) have separations between the lensed images that are $<$1/3 of the smallest {\it Herschel}/SPIRE beam size, and can therefore be considered point sources for our study. The sample was observed in small map mode with one scan repetition per source, with a total integration time of 2--3 min per target, such that a source of 50~mJy will be detected at the 5$\sigma$ level in the 500~$\umu$m band.

Of our quasar sample, 21 have 850~$\umu$m detections and 11 have 450~$\umu$m detections with SCUBA by \cite{Barvainis:2002}. Assuming magnifications from the literature and spectral energy distributions (SEDs) described by \cite{Yun:2002} ($T_{\rm d}=58$ K, $\beta=1.35$), nearly all of these detected sources (15 at 250 and 350~$\umu$m, 18 at 500~$\umu$m) would be below the confusion limits of {\it Herschel}/SPIRE were they not gravitationally-lensed. It is therefore likely that the quasar population with intrinsic fluxes below those of previously detected field sources will be revealed in this study. Moreover, while SCUBA measurements lie on the Rayleigh-Jeans side of the thermal SED, the {\it Herschel}/SPIRE bands allow for better constraints on the peak of the SED, and thus, more accurate estimates of the characteristic FIR-luminosities and dust temperatures of the sample. We note that the previous study by \citeauthor{Barvainis:2002} assumed a dust temperature of 30~K for their sample, which may have biased their estimates of the FIR luminosities and inferred star formation rates.

\subsection{Radio properties}

\begin{figure*}
\centering
\includegraphics[width=0.85\textwidth]{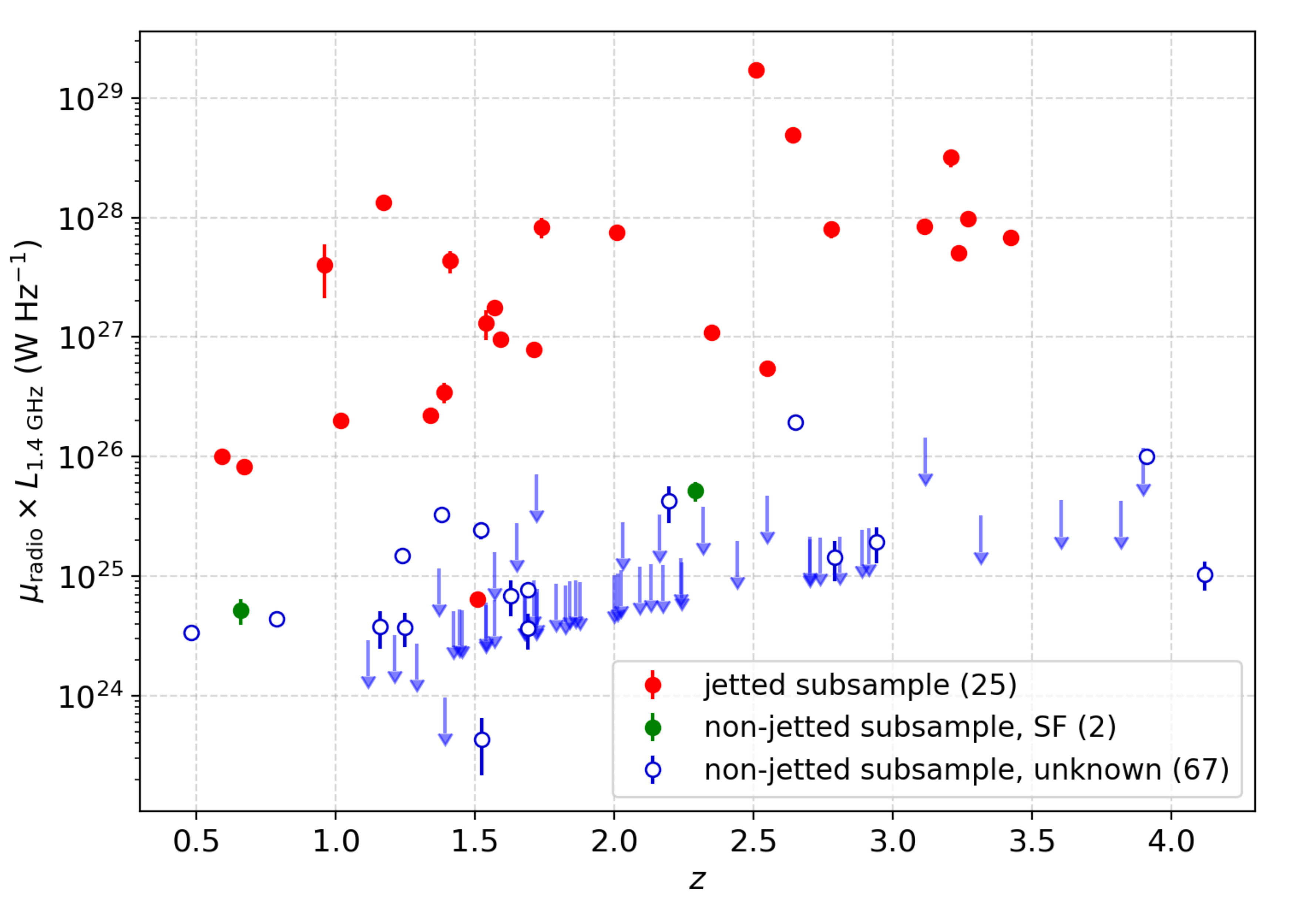}
\caption{The rest-frame 1.4~GHz radio luminosity-density (interpolated or extrapolated from existing data) as a function of redshift for 92 objects in our quasar sample with radio measurements and a known redshift. Most of the upper limits are taken from FIRST or NVSS. The jetted subsample, with known jet-dominated radio emission, are shown in red. The non-jetted subsample includes two quasars with star-formation-dominated radio emission (shown in green) and 67 with unknown radio emission mechanism (shown in blue).}
\label{fig:radio-z}
\end{figure*}

Radio emission from quasars may be associated with AGN (synchrotron) or star formation (synchrotron, free-free) processes. Quasars with radio jets are associated with jet-mode feedback, whereas quasars without these features are primarily radiative, so it is convenient to classify our sample based on their radio properties. There is a range of terminology and methods of classification employed in the literature to distinguish these groups, typically radio-`loud' and radio-`quiet' based on radio luminosity or radio-optical ratio. However, we find it more appropriate to group these by considering the operative feedback mechanisms. We have divided the sample into jetted (quasars with known jet-dominated radio emission) and non-jetted (quasars with star-formation-dominated radio emission and those where the dominant radio emission mechanism is unknown) dependent on whether there has been confirmation of the existence of a radio jet component with high-resolution radio data. For this, we have used the data from targeted observations for individual objects in the literature. Of the 34 quasars within the sample that we classify as jet-dominated radio sources, 31 are from the MIT-Green Bank Survey \citep[MG]{Langston:1990}, the Jodrell Bank-VLA Astrometric Survey \citep[JVAS]{Patnaik:1992}, the Cosmic Lens All-Sky Survey \citep[CLASS]{Myers:2003,Browne:2003}, the Parkes-NRAO-MIT survey \citep[PMN]{Griffith:1993} and other radio surveys, all of which are dominated by radio-luminous AGN due to their respective flux-density limits. The remaining three sources are Q0957+561 \citep{Garrett:1994}, H1413+117 (Stacey et al. in prep) and HS~0810+2554 (Hartley et al. in prep).

At low radio luminosities, composite AGN and star formation emission are likely, and differentiating between these possibilities is difficult. We define only two sources in our sample with established star-formation-dominated radio emission, RX~J1131$-$1231 and IRAS~F10214+5255. VLBI experiments to detect the radio core of these quasars suggest the radio emission is primarily due to star formation \citep{Wucknitz:2008,Deane:2013}. In all other cases, the emission mechanism is undetermined, either because they are not detected at radio wavelengths or the detections are at too low an angular resolution to discriminate between compact (AGN) or extended (star formation) emission. We obtain the majority of these measurements from the National Radio Astronomy Observatory (NRAO) Very Large Array (VLA) Sky Survey \citep[NVSS]{Condon:1998} and the Faint Images of the Radio Sky at Twenty-Centimeters \citep[FIRST]{Becker:1995}, both at 1.4~GHz and with beam sizes of 45 and 5~arcsec, respectively.

We show the rest-frame 1.4~GHz radio luminosities for the sample in Fig.~\ref{fig:radio-z}. We include all quasars without evidence of jet-dominated radio emission in the non-jetted subsample for the time being, but refine these classifications using the radio--infrared correlation in Section~\ref{section:radioinfrared}.

\begin{table}
\caption{Number of detections in each {\it Herschel}/SPIRE band for the jetted and non-jetted subsamples.}
\begin{tabular}{l c c c c}  \hline
 & N & 250 $\umu$m & 350 $\umu$m & 500 $\umu$m \\ \hline \noalign {\smallskip}
Jets & 34 & 24 (71~percent) & 23 (68~percent) & 16 (47~percent) \\
No jets & 70 & 47 (67~percent) & 41 (59~percent) & 23 (33~percent) \\ 
Total & 104 & 71 (68~percent) & 64 (62~percent) & 39 (38~percent) \\  \noalign {\smallskip} \hline
\end{tabular}
\label{table:spiredetections}
\end{table}

\subsection{Photometry}
\label{section:photometry}

The sources have been observed with the {\it Herschel}/SPIRE instrument in three bands centred on 250, 350 and 500~$\umu$m, which effectively cover the rest-frame spectrum from  40 to 394~$\umu$m for the redshift range of our sample. The calibrated data were obtained from the {\it Herschel} Science Archive using the {\it Herschel} Interactive Processing Environment (HIPE) \citep{Ott:2010} version 14.0.0. 

The photometry was performed using the SUSSEXtractor and Timeline Fitter algorithms within HIPE \citep{Savage:2007,Bendo:2013} using recommendations in the SPIRE Data Reduction Guide\footnote{\url{http://herschel.esac.esa.int/hcss-doc-14.0/print/spire_drg/spire_drg.pdf}}. The Timeline Fitter performs point source photometry by fitting Gaussians to the baseline subtracted timeline samples, given source locations on the sky. The SUSSEXtractor method extracts point sources from the beam-smoothed, calibrated maps. We set a threshold of 3$\sigma$, where $\sigma$ is the RMS noise of the background around the source. While the Timeline Fitter gives more precise measurements and was the preferred method, SUSSEXtractor was occasionally more successful at extracting lower flux-density sources ($S_{\nu}\lesssim30$ mJy). We place a detection limit of $3\sigma$ on the photometric measurements, where $\sigma$ is the RMS noise of the map, including confusion, given that we know the positions of the gravitational lens systems.

We explored the possibility of fixing the positions of source extraction with SUSSEXtractor to the `true' sky positions in order to avoid an upward bias due to fitting to random noise spikes. While the effect of this is reduced as SUSSEXtractor fits to beam-smoothed maps, it has been noted to cause a bias in submillimetre measurements where the signal-to-noise ratio is low \cite[for example]{Ivison:2002,Coppin:2005}. As we would expect, there is a systematic upward shift in the flux densities measured when the position is left free. However, the change is generally not more than 10~percent and within the photometric errors. We choose not to employ this method as we find there are often significant uncertainties on both the {\it Herschel} astrometry and the `true' source position from the literature. We compared the extracted source positions of the five FIR-bright objects from our sample detected in ALMA to their ALMA positions (which have accurate and precise astrometry due to phase-referencing) and find offsets up to several arcsec. This is consistent with other findings in the literature \cite[for example]{Melbourne:2012}. We also find differences as much as several arcsec in the positions from optical or X-ray positions in the literature relative to the ALMA positions. There is an additional positional uncertainty as the targets are gravitationally-lensed with a range of image separations (Fig.~\ref{fig:sep_z}). Thus, the result of fixing the position for source extraction would be a systematic down-shift of the extracted source flux densities. This bias can be more significant than the bias due to noise spikes and, as many sources are close to the detection limit, this would have a negative effect on the analysis. In any case, these uncertainties in the photometry are far lower than the uncertainties in the FIR luminosity and SFR due to SED fitting and the unknown magnification factor of the lensed systems (see Sections~\ref{section:sedfitting} and \ref{section:magnifications}).

\begin{figure*}
\centering
\includegraphics[width=0.95\textwidth]{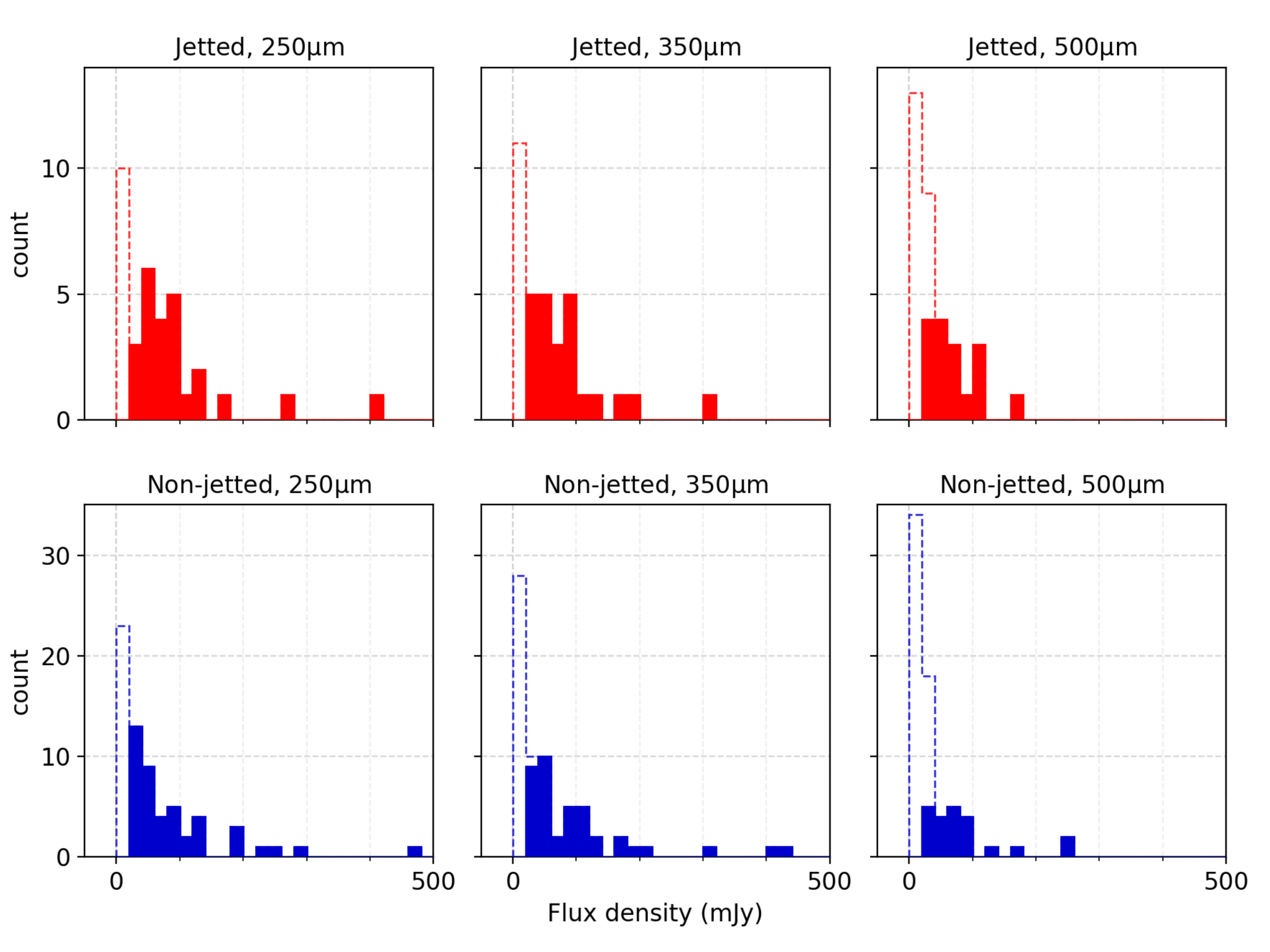}
\caption{Number of sources binned by measured flux-density for the three {\it Herschel}/SPIRE bands, divided into jetted (blue) and non-jetted (red) sub-samples. 1.5~$\sigma$ limits of non-detections are stacked on top of the measured values and outlined with a dashed line. Note that PKS~1830$-$211 is excluded, for clarity, due to its high flux-density (S$_{\rm 250\ \umu m}=537$ mJy, S$_{\rm 350\ \umu m}=670$ mJy, S$_{\rm 500\ \umu m}=806$ mJy).}
\label{fig:spirefluxes}
\end{figure*}

\subsection{Source matching and confusion}
\label{section:confusion}

Due to the sizes of the {\it Herschel}/SPIRE beams, we must also consider the contribution to the measured flux densities from field galaxies, including sources not associated with the target quasars or their lensing galaxies. For example, this could be due to dust-obscured star formation or AGN activity within the lensing galaxy at mm-wavelengths as has been seen in three gravitational lenses observed at high angular resolution with ALMA \citep[{McKean et al. in prep}]{ALMA:2015,Paraficz:2017}. We also note that at radio wavelengths, about 10 percent of lensing galaxies have detected synchrotron emission from an AGN \citep{McKean:2005,McKean:2007}.

We take a series of steps to match the photometric data with our target quasars. We compare the extracted source position from SUSSEXtractor or the Timeline Fitter algorithm with the `true' position of the lensing galaxy (where there is good astrometry, else the brightest lensed image) taken from the NASA Extragalactic Database (NED). We rejected extracted sources whose positional offsets are larger than half the FWHM of the SPIRE beam. We allow for some freedom in source fitting to allow for the combined uncertainties on the {\it Herschel} pointing, the `true' source position, and source fitting. The extracted source positions are then cross-checked with nearby sources listed on NED to minimise the possibility of mis-matching. In addition to this, we use detection in the 250~$\umu$m band (which has the highest resolution and lowest confusion noise) as a prior to confirm a match at 350~$\umu$m and 500~$\umu$m\footnote{We make an exception for PMN~J1632$-$0033 because of the completeness of the SED (see Fig.~\ref{fig:syncseds}).}. This strategy reduces the likelihood of contamination from field sources, but does not exclude the possibility of emission from the lensing galaxy or a nearby unknown FIR-bright field source being included in our photometric measurements.

While most of the targets appear uncontaminated, in some cases, blending is evident in the level 2 (fully calibrated) maps by visual inspection. For example, individual sources may be resolved in the higher resolution 250~$\umu$m maps, but become blended in the 500~$\umu$m band where the beam is largest. The blending results in over-fitting by the source extractor and returns incorrect flux densities. We attempt to overcome this by simultaneously fitting to both the known target position and the blended source using the Timeline Fitter, where possible, then applying the same source matching criteria. Where this fails, we use SUSSEXtractor and fix the extracted position to the `true' target position and blended source position (if known).

We can further identify confusion or mismatching by comparison of the {\it Herschel}/SPIRE data with the source SED. It is likely, based on inspection of their spectra, that we are unable to remove blended emission completely for 8 sources: CLASS~B0712+472, CLASS~B0850+054, SDSS~J0903+5028, CLASS~B1152+200, Q~1208+101, CLASS~B1359+154, SBS~1520+530 and Q~2237+030 (SEDs for all but CLASS~B0712+472 are given in Fig.~\ref{figure:SEDs} of the Appendix). In the cases of CLASS~B0712+472 and Q~2237+030, there is too much blended emission to confidently measure the quasars, so we assume upper limits for all three bands by measuring the off-source RMS noise of the maps. For CLASS~B1152+200 this is the case at 350~$\umu$m and 500~$\umu$m. The remaining sources appear to have an additional contribution to their 500~$\umu$m measurement that is inconsistent with thermal dust emission or with synchrotron emission, based on their radio measurements. This may be due to errors in fitting to the blended source or further blending with nearby field sources. We identify known sources within a few arcsec of SDSS~J0903+5028, Q~1208+101, Q~2237+030 that could be responsible and do not find evidence of confusion from the lensing galaxy for these objects. While SBS~1520+530 does have a star-forming lensing galaxy, the measured 500~$\umu$m flux-density implies a flat spectrum that is inconsistent with the upper limits in the sub-mm/mm. In these cases, we assume upper limits for the 500~$\umu$m measurements that include confusion.

Almost all of the ancillary data that is used to derive the source SEDs is taken from literature, which typically consists of high resolution, targeted observations at mm-to-radio wavelengths, and lower resolution surveys at radio wavelengths. Where sources are detected and unresolved, we cannot be certain that they relate to a single source (the target, as opposed to a nearby companion or the lensing galaxy) without higher resolution observations on about arcsec-scales. The detections at 250~$\umu$m are matched to unique radio detections using the same matching criteria described previously, and these are assumed to relate solely to the quasar based on the assumptions that i) the spatial density of quasars is lower than DSFGs, and so we are likely observing a single source rather than multiple sources, and ii) as these quasars are intrinsically bright and gravitationally-magnified, any companion would have to be similarly bright to contaminate our measurements, which is unlikely. Of course, further observations at higher angular resolution with mm-wavelengths interferometers will better match the FIR emission detected here with the optical-to-radio counterparts of the quasars. However, throughout this paper we assume that the quasar is the sole source of the position-matched FIR emission detected with {\it Herschel}/SPIRE.

\section{Results \& analysis}
\label{results}

In this section, we present the photometric results and describe the SED fitting analysis used to determine the physical properties for each gravitationally-lensed quasars within our sample.

\subsection{\textit{\textbf{Herschel}}/SPIRE measurements}
\label{section:spiremeasurements}

The {\it Herschel}/SPIRE photometry for all of the sources observed in our sample is detailed in Table~\ref{table:spireflux} of the Appendix, and their SEDs, using all available data points, are shown in Fig.~\ref{figure:SEDs} in the Appendix. Of the 104 sources observed, 72 are detected in at least one band down to a detection threshold of $3\sigma$. Upper limits are given for those sources not detected at this confidence level. Of the sample, 10 targets suffer from contamination from the lensing galaxy or nearby field sources, which is apparent from their spectral properties and known properties of the lensing galaxies or nearby sources. This mostly affects the 500~$\umu$m band, due to the larger FWHM of the point spread function and their rising synchrotron spectra at longer wavelengths (see Section~\ref{section:confusion}).

The measured flux-density distribution for each of the bands, separated by their radio properties, is shown in Fig. \ref{fig:spirefluxes} and the number of detections is given in Table~\ref{table:spiredetections}. We use the two-sample Kolmogorov--Smirnov (K--S) test to compare whether the measured flux densities of the subsamples are consistent with the same underlying distribution. For all K--S tests in this work we employ a Peto--Prentice Generalized Wilcoxon method\footnote{We employ this method as it is usually the most reliable and least affected by differences in the censoring patterns.} for censored data using the {\it twosampt} task in the STSDAS statistics package within {\sc iraf}. The test returns a probability ($p$) for the null hypothesis, for which $p<0.05$ we take as statistically-significant. For our subsamples, the test returns probabilities of 0.44, 0.57 and 0.75 for the distributions of measured flux densities at 250~$\umu$m, 350~$\umu$m and 500~$\umu$m, respectively. While the detection rates are slightly higher for the jetted quasars, the differences between the subsamples are not statistically-significant.

\subsection{Spectral slopes}
\label{section:spectralslopes}

\begin{figure}
\includegraphics[width=0.5\textwidth]{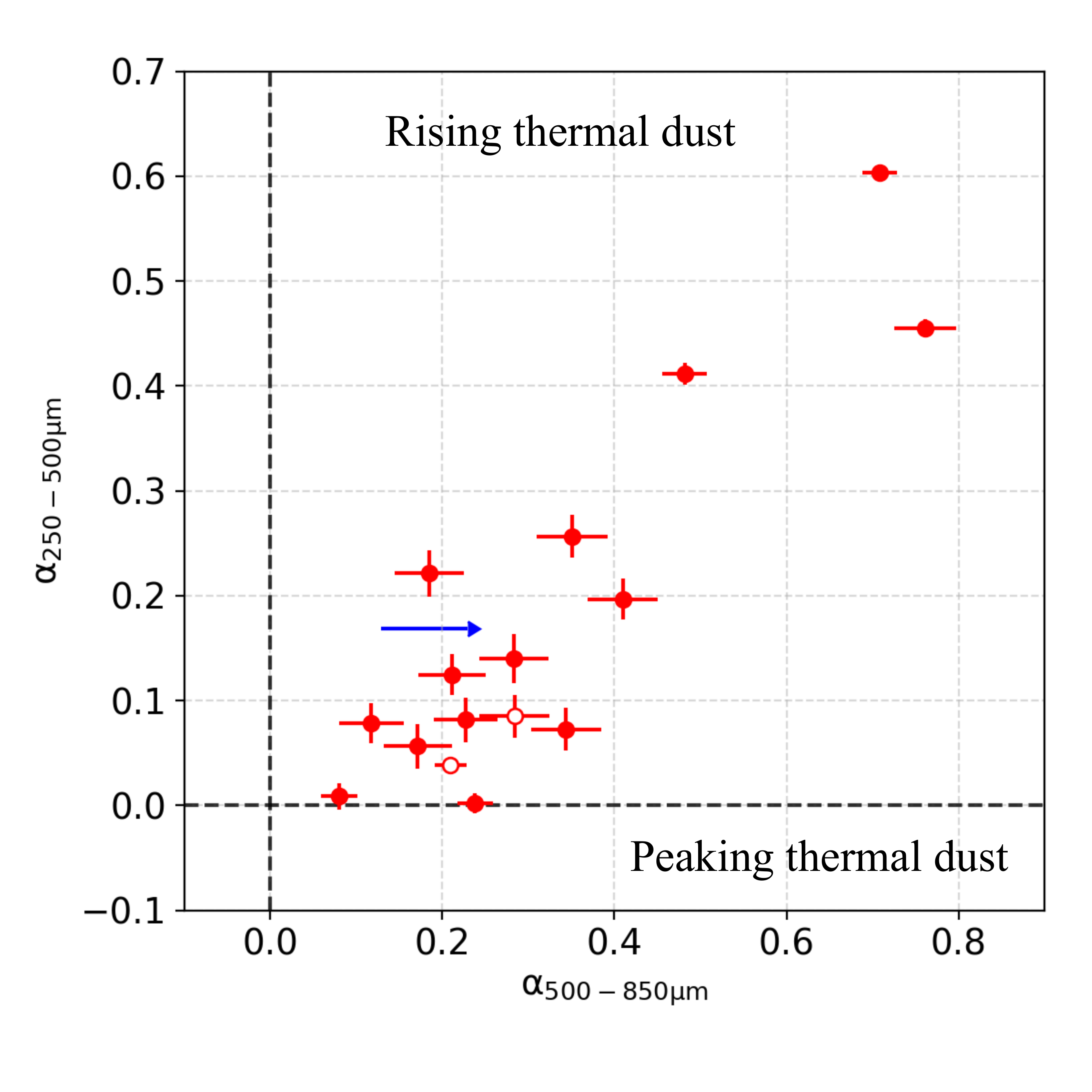}
\caption{Spectral index with frequency of the high and low {\it Herschel}/SPIRE bands relative to 850~$\umu$m, for the 17 sources in the sample with previous sub-mm detections and three SPIRE detections. Open circles are measurements at the same wavelength, but not from SCUBA. Limits due to non-detections at 850~$\umu$m are shown in blue. The plot excludes PKS~1830$-$211, for clarity, due to its large negative spectral index ($\alpha=-0.5$).}
\label{fig:spectral-index-scuba}
\includegraphics[width=0.5\textwidth]{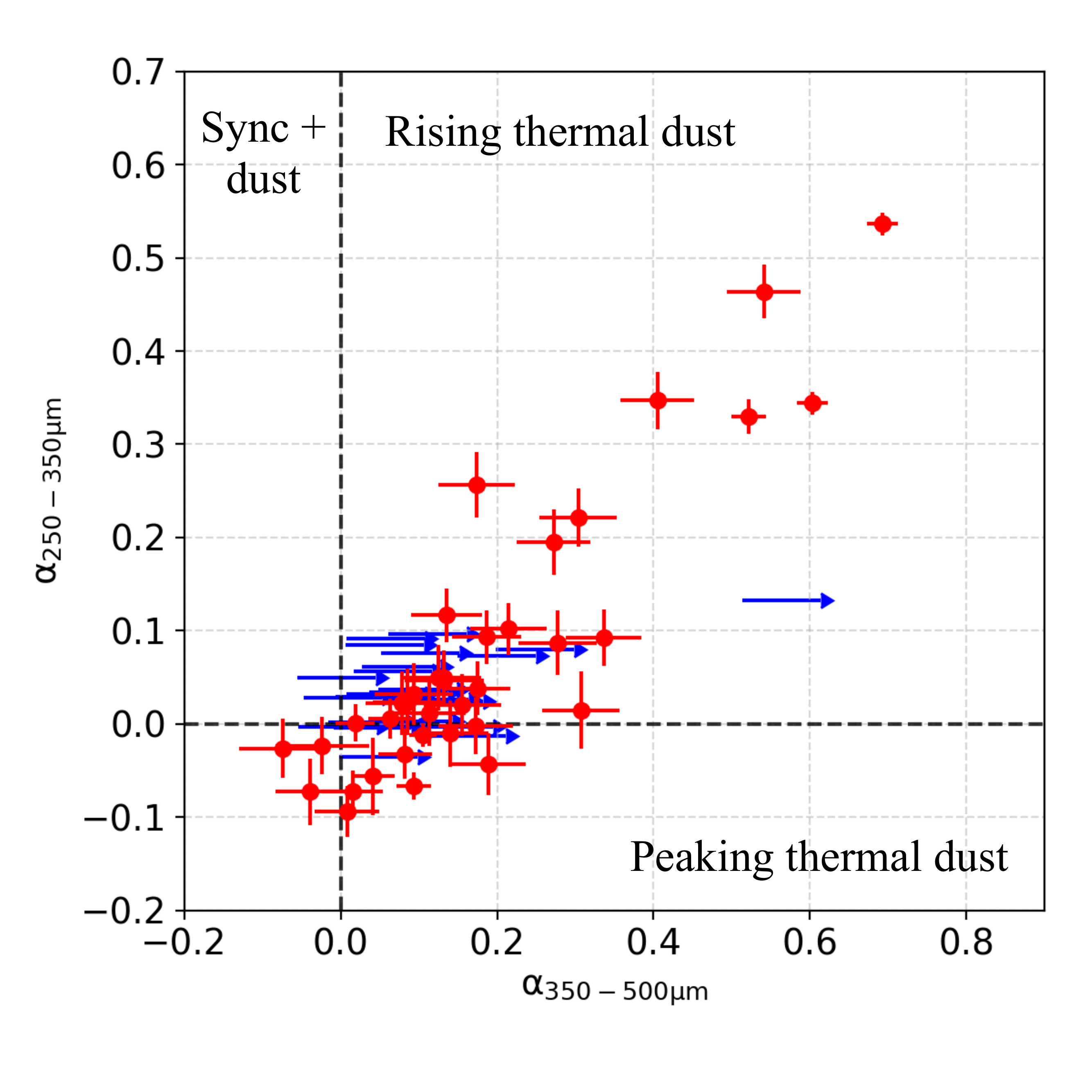}
\caption{Spectral index of high-to-mid against mid-to-low {\it Herschel}/SPIRE for 63 sources with 250~$\umu$m and 350~$\umu$m detections. The positive quadrants contain rising spectra associated with dust. 3 sources with falling spectra from 350 to 500~$\umu$m may have contamination from synchrotron emission, as discussed in Section~\ref{section:spectralslopes}. Lower limits due to non-detections at 500~$\umu$m are shown in blue. The plot excludes PKS~1830$-$211 due to its steep negative spectral index ($\alpha=-0.5$).}
\label{fig:spectral-index-spire}
\end{figure}

In Figs.~\ref{fig:spectral-index-scuba} and \ref{fig:spectral-index-spire}, we show the spectral index between 850 and 500~$\umu$m ($\alpha_{\rm 500~\umu m}^{\rm 850~\umu m}$) and 500 and 250~$\umu$m ($\alpha_{\rm 250~\umu m}^{\rm 500~\umu m}$)\footnote{The spectral index is defined as a power-law, $S_{\nu}\propto\nu^{\alpha}$, where $S_{\nu}$ is flux-density and $\nu$ is frequency.}. 74 objects in our sample have detections in the FIR to sub-mm, including the 72 {\it Herschel}/SPIRE detected sources and a further 2 which have only sub-mm detections. In most cases, we find evidence for heated dust emission: of these 74 objects, we ascribe the emission in 69 cases (66~percent of the sample) as being due to thermal dust emission from their rising or peaking spectra in FIR with frequency, relative to their sub-mm/mm/radio emission.

Of the five remaining sources that are detected in at least one band, there is no clear evidence for heated dust emission in the current data (see Fig.~\ref{fig:syncseds} for their SEDs). These sources are CLASS~B1030+074, JVAS~B0218+357, PKS~1830$-$211, PMN~J1838$-$3427 and PMN~J1632$-$0033. These sources do not have rising spectra in the FIR and have strong flat-spectrum synchrotron emission in the radio. Unfortunately, these sources were not observed by the SCUBA and MAMBO surveys or were discovered too late to be part of the \cite{Barvainis:2002} sample. Without measurements in the sub-mm regime, it is not clear how the synchrotron component falls off towards the FIR. In the cases of CLASS~B1030+074, JVAS~B0218+357 and PKS~1830$-$211, the flat-spectrum component continues into the mm-regime, so it is likely there will be a significant contribution from optically-thin synchrotron emission in the {\it Herschel}/SPIRE measurements (SED fitting of PKS~1830$-$211 is discussed further in \ref{section:1830} of the Appendix). PMN~J1838$-$3427 and PMN~J1632$-$0033 do not have enough high-frequency data to extrapolate their spectra into the FIR. It is possible these sources have spectra comparable to CLASS~B1127+385 or CLASS~B1152+200, where sub-mm measurements or upper-limits dictate that synchrotron emission does not have a significant contribution in the FIR (see Fig.~\ref{fig:syncseds} of the Appendix for their SEDs). JVAS~B0218+357 and PMN~J1838$-$3427 have measurements that appear characteristic of peaking dust emission, but this could also be explained by variability or a self-absorbed synchrotron component. We fit thermal SEDs to the {\it Herschel}/SPIRE measurements for these five quasars to place upper limits on a possible contribution of heated dust to the FIR emission.

\begin{table*}
\bgroup
\def\arraystretch{1.1}
\setlength{\tabcolsep}{1.2em}
\caption{Magnification values from the literature, with errors where given, and the data with which the lens modelling is performed (line or continuum). CO line emission is assumed to have a similar location and extent as star-formation-heated dust emission, thus a similar magnification.}
\begin{tabular*}{0.7\textwidth}{l l l l}
\hline \noalign {\smallskip}
Source & $\mu_{\rm SF}$ & Method & Reference \\ 
\hline \noalign {\smallskip}
APM~08279+5255 & 4.2 & CO(1-0) & \cite{Riechers:2009} \\ 
RX~J0911+0551 & $18.7\pm1.3$ & 360~GHz continuum & P. Tuan-Anh, private communication \\ 
Q~0957+561 & $7\pm1$ & CO(2-1) & \cite{Krips:2005} \\
IRAS~F10214+4724 & $6\pm1.5$ & CO(1-0) & \cite{Deane:2013} \\
RX~J1131-1231 & 7.3 & CO(2-1) & \cite{Paraficz:2017} \\ 
H1413+117 & 11.0 & 690~GHz continuum & M. Rybak, private communication\footnotemark
\\
PSS~J2322+1944 & 2.5 & CO(2-1) & \cite{Carilli:2003} \\ \hline
\end{tabular*}
\label{table:magnifications}
\egroup
\end{table*}
\footnotetext{\protect\cite{Venturini:2003} also find a factor of 11 based on CO(7-6) line observations.}

\subsection{Magnifications}
\label{section:magnifications}
To derive the {\it intrinsic} properties of the quasars in the sample, the measurements must be corrected for their lensing magnification. Generally, these are obtained from the literature and are typically derived from an analysis of optical or radio gravitational lensing data. However, optical and radio components of quasars tend to be compact (size scales of $\leq$ pc to a few 10s of pc), and can result in very high magnification factors if the source is close to a lensing caustic. For example, JVAS~B1938+666 has a radio magnification factor of 173 \citep{Barvainis:2002}, whereas the 2.2~$\umu$m infrared emission from the AGN host galaxy has a magnification of about 13 \citep{Lagattuta:2012}. This presents a problem for accurately estimating the properties of this sample of gravitationally-lensed quasars at FIR to sub-mm wavelengths, as the size scales of AGN emitting regions may be anywhere from $\sim$pc (in the case of the  AGN core) to $\sim$kpc (radio jets or a star-forming disk). Where the radio or optical magnifications are high, it is likely that the dust emission (assuming it is coincident with the quasar) will be differentially magnified as only a small region will be close to the caustic and the overall magnification will be lower. Magnifications derived from optical or radio data are therefore unlikely to be accurate indicators of the actual dust magnification. 

Only a few quasars in our sample have high-resolution observations in the FIR to sub-mm regime, thus we list the source properties given in Tables \ref{table:spireflux} and \ref{table:luminosities} uncorrected for lensing magnification. Known magnifications ($\mu_{\rm SF}$) in the FIR to sub-mm, based on dust or molecular gas tracers relating to star-forming regions, are given in Table~\ref{table:magnifications}. We assume that cold molecular gas has a similar extent, thus similar average magnification, as the star formation heated dust emission. Only two of the sources in the sample have resolved dust emission related to star formation. These are the Cloverleaf quasar and RX~J0911+0551, which have magnifications of 11 and 19, respectively\footnote{M. Rybak and P. Tuan-Anh, private communication.}. For the intrinsic properties discussed below, we conservatively assume a magnification of $\mu_{\rm est.} = 10_{-5}^{+10}$ for the sources without known magnifications. This is consistent with lens modelling of dust emission in {\it Herschel}-selected strongly-lensed star-forming galaxies: \cite{Bussmann:2013} find total magnification factors 2--15 ($\bar{\mu}=8$) for a sample of 20 observed with the Submillimeter Array (SMA), and \cite{Dye:2017} find magnifications factors 4--24 ($\bar{\mu}=12.5$) in a sample of six observed with ALMA. 

Magnifications of more than 20 are unlikely if the sources are extended more than $\sim200$~pc, as discussed in the \citeauthor{Barvainis:2002} study. The two sources in our sample with reconstructed dust emission, RX~J0911+0551 and the Cloverleaf quasar, both have dust emitting regions of $\sim1$~kpc in size (\citealt{Tuan-Anh:2017}, Stacey et al. in prep). Assuming these sizes are characteristic, our assumption of $\mu_{\rm est.} = 10_{-5}^{+10}$ is likely representative of the magnifications of the sample, including a conservative uncertainty to account for outliers, and will provide an indication of the unlensed properties of the sample as a whole. The median values of the intrinsic properties we derive in the following analyses do not account for the factor of 2 error in the magnification because the assumption is taken for all but seven objects.

\subsection{SED modelling}
\label{section:sedfitting}
To constrain the physical properties of the FIR emission in each quasar host galaxy, we fit a combined non-thermal and thermal SED model to the {\it Herschel}/SPIRE data, along with any available data in the literature, excluding our measurements that are affected by confusion, as noted in Table~\ref{table:spireflux}. This model will account for any synchrotron component, in the case of the jetted targets, and any heated dust component of the SED. We use a power-law with spectral index $\alpha$,
\begin{equation}
S_{\nu} \propto \nu^{\alpha},
\label{eq:spectrum}
\end{equation}
to describe the flux-density ($S_\nu$) as a function of frequency ($\nu$) in the case of synchrotron emission. We do this only to estimate the contribution of synchrotron to the FIR spectrum, so do not attempt more complex fitting describing spectral turn-overs (for example CLASS~B1422+231 shown in Fig.~\ref{figure:SEDs} of the Appendix). The SEDs of flat spectrum radio sources will likely turn down at higher frequencies and have a negligible synchrotron contribution in the FIR (for example, CLASS~B1127+385 in Fig.~\ref{figure:SEDs}). In some cases, there is a suggestion that the synchrotron emission turns down towards the sub-mm (for example, Q0957+561, suggested by an upper limit at 230~GHz) so we assume this does not contribute substantially to the FIR emission. We choose not to fit a synchrotron component where there is a single radio detection as we have no knowledge of the spectral behaviour, and, as these single measurements typically correspond to lower luminosities, the FIR contribution will be small.

We use a characteristic modified black-body,
\begin{equation}
S_{\nu}\propto \frac{\nu^{3 + \beta}}{e^{h\nu /kT_{\rm d}}-1},
\end{equation} 
to describe the heated dust component, where $h$ is the Planck constant, $k$ is the Boltzmann constant, $T_{\rm d}$ is effective dust temperature, and $\beta$ is the emissivity index, which determines the steepness of the Rayleigh-Jeans slope of the spectrum.

\begin{figure}
\includegraphics[width=0.48\textwidth]{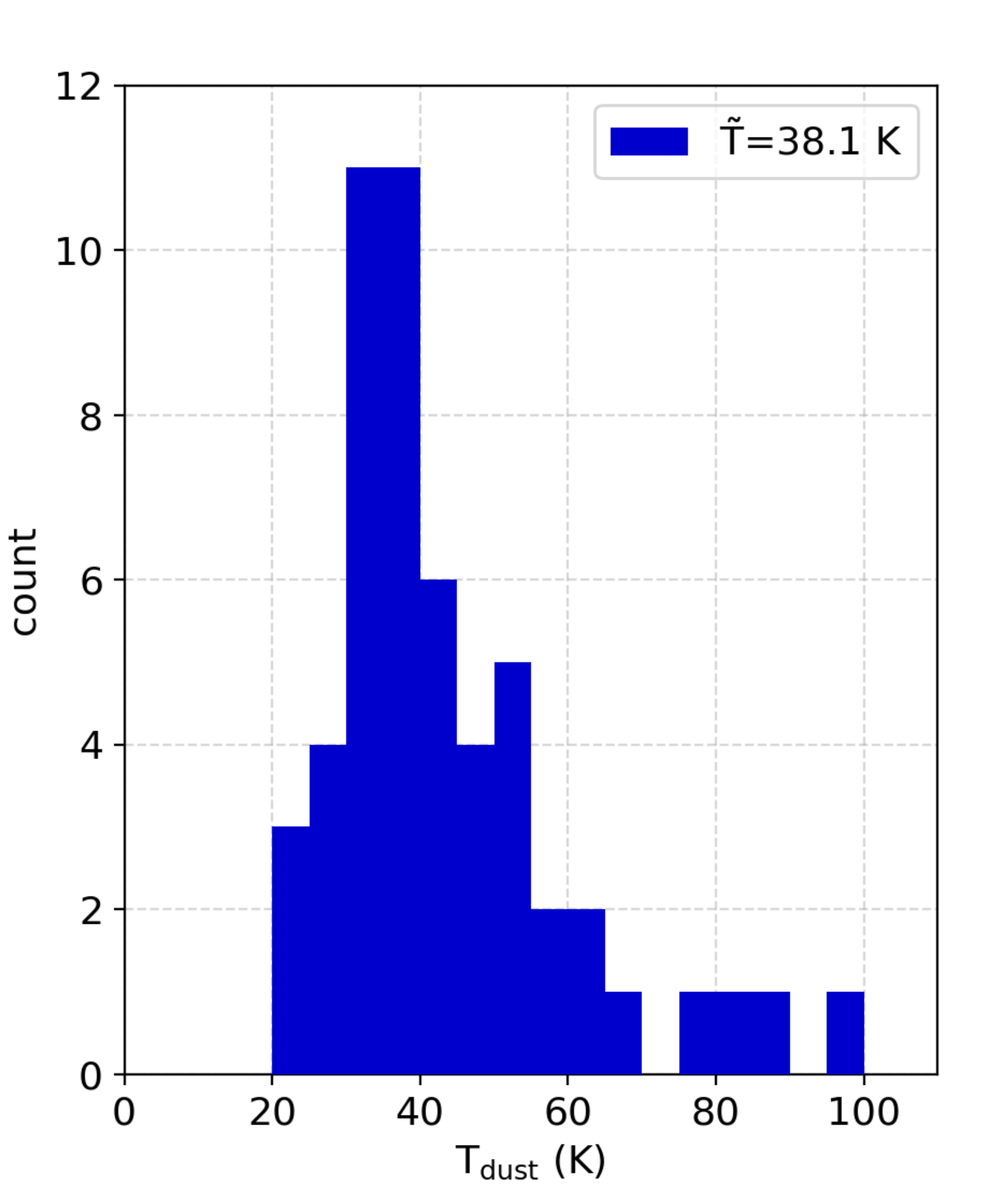}
\caption{Histogram of effective dust temperatures for 53 quasars in the sample with temperature fitting, excluding those without a known redshift and synchrotron-dominated sources. The median dust temperature is $38^{+12}_{-5}$~K. Where two dust temperatures are fit, only the colder component is included here.}
\label{fig:dusttemps}
\end{figure}

The Python implementation {\sc{emcee}} \citep{Foreman-Mackey:2013} was used to build a Markov Chain Monte Carlo (MCMC) analysis of the fitted SED for each data set, allowing the dust temperature and normalization as free parameters to sample the posterior probability distribution of the model. Where possible, we also leave $\beta$ as a free parameter in the model, allowing for a test of the range of dust emissivities that are consistent with the data. However, fitting for $\beta$ requires at least four data points to constrain the peak and Rayleigh-Jeans slope of the modified black-body function. For many sources, our {\it Herschel}/SPIRE data are the only measurement in the FIR--sub-mm regime. Thus, we assume a value of $\beta=1.5$ for these sources, as is frequently applied in the literature \citep[for example]{Magnelli:2012}. Various combinations of $T_{\rm d}$ and $\beta$ have been found for samples of high-redshift quasars. For example, \cite{Priddey:2001} find an average of $T_{\rm d}=41\pm5$~K and $\beta=1.95\pm 0.3$, whereas \cite{Beelen:2006} find an average of $T_{\rm d}=47\pm3$~K and $\beta=1.6\pm0.1$ for their sample. For the sources that were not detected, or had only one detection to constrain the fit, we assume the median fitted dust temperature of the sample (38~K, see Fig.~\ref{fig:dusttemps}) and $\beta=1.5$, and fit only for the normalization. For these sources with only one detection, we fit the 16th and 84th percentile values of the median fitted temperatures to estimate our errors on the FIR luminosity. As $\beta$ is highly correlated with $T_{\rm d}$, errors on the derived properties of sources without $\beta$-fitting may be underestimated. However, the FIR luminosity is not strongly affected by our assumptions due to the joint dependency of $T_{\rm d}$, $\beta$ and normalization, so this will not have a significant effect on the inferred values of $L_{\rm FIR}$ or SFR. This is unsurprising, as the luminosity is derived by the integral of the fit defined by the data points.

For the purpose of spectral fitting, the ten sources without a known redshift are assumed to have $z=1.8$, equivalent to the median redshift of the sample. The choice of redshift significantly affects the luminosity-distance, thus these objects are not included in the overall statistics.

For three sources (APM~08279+5255, H1413+117 and IRAS~F10214+4724) there is sufficient data in the mid-IR (MIR) to motivate fitting a two-temperature dust model. Table~\ref{table:sedfits} shows the number of sources fitted with each combination of spectral parameters.

\begin{figure*}
\includegraphics[width=\textwidth]{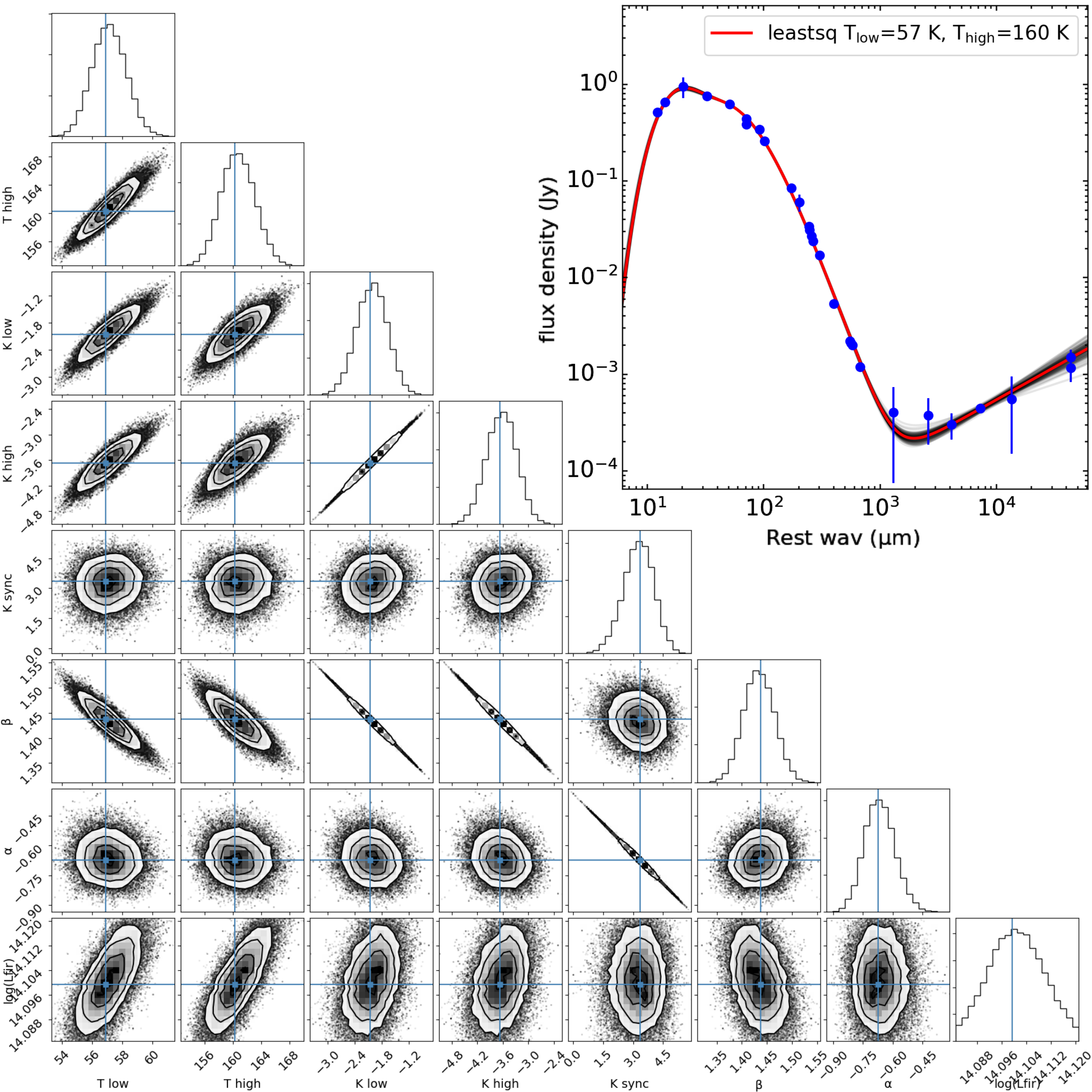}
\caption{Two-dimensional probability densities of the MCMC output for the SED fitting of APM~08279+5255, fit with all parameters: T1, low dust temperature (observed); T2, high dust temperature (observed); K1, K2, K3, log normalizations of the dust and synchrotron fits; $\beta$, the emissivity index; $\alpha$, the synchrotron power-law index. Also shown is $L_{\rm FIR}$. The blue points on the corner plot show the least-squares parameters. The SED is shown above, with 100 random samples of the MCMC in black and the least-squares model in red.}
\label{fig:apmcornerplot}
\end{figure*}

\begin{figure*}
\includegraphics[width=\textwidth]{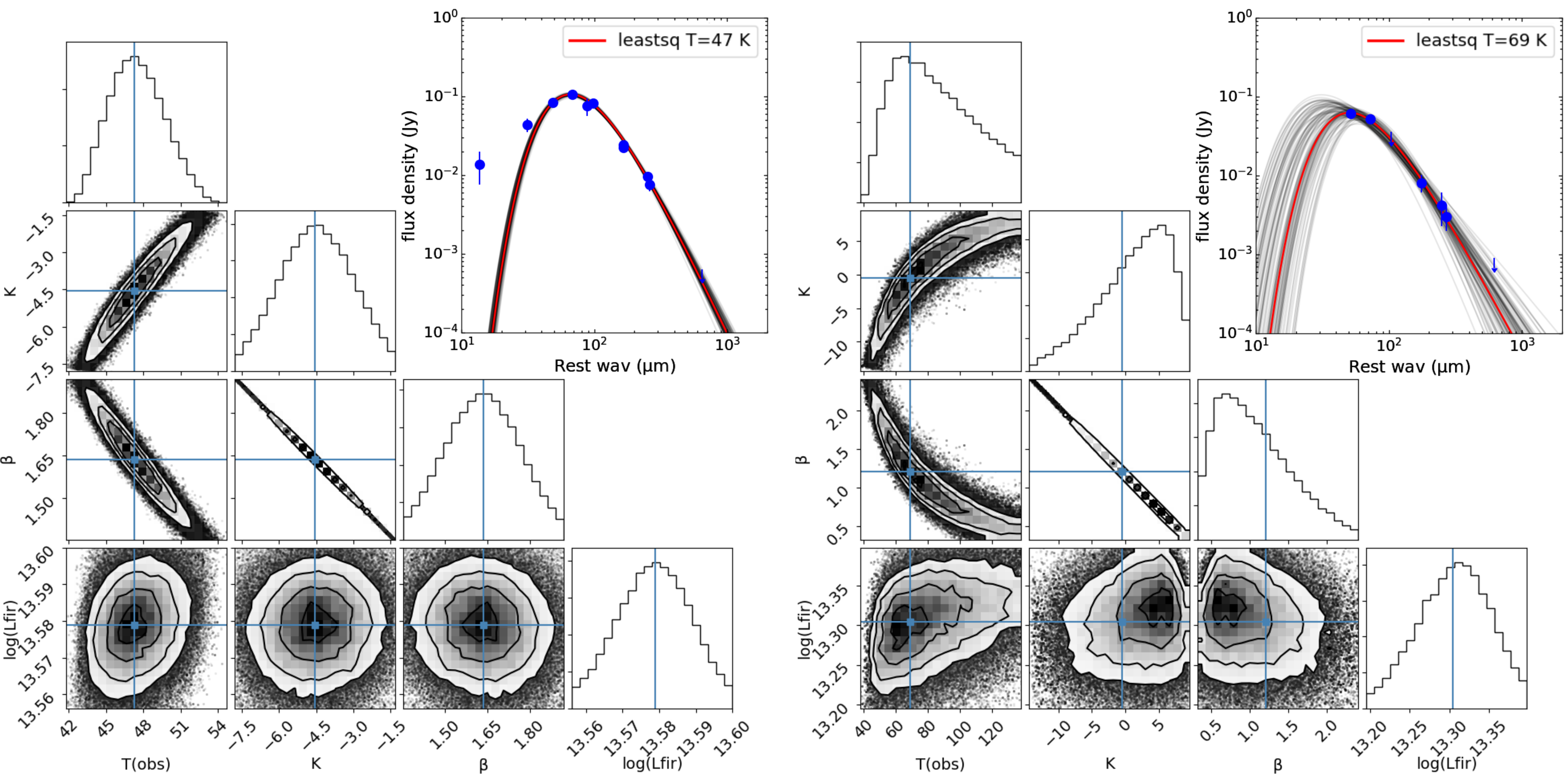}
\caption{Two-dimensional probability densities of the MCMC results of SED fitting for PSS~J2322+1944 (left) and Q~1208+101 (right), showing the correlations between the spectral parameters: T, observed dust temperature; K, normalization; $\beta$, the emissivity index. Also shown is $L_{\rm FIR}$. The blue points on the corner plots show the least-squares parameters. The SEDs are shown above the corner plots, with 100 random samples of the MCMC in black and the least-squares model in red.}
\label{fig:betacornerplots}
\end{figure*}

We have included posterior probability distributions of the MCMC output of the SED fit for three sources, to show the correlation between the various fitting parameters and highlight the effect of sparse sampling of the SED. Fig.~\ref{fig:apmcornerplot} shows the result for APM~08279+5255, where there is sufficient data to fit seven spectral parameters. In Fig.~\ref{fig:betacornerplots}, we compare the results for two sources: PSS~J2322+1944, where the peak of the dust emission and the Rayleigh-Jeans slope are both well constrained, and Q~1208+101, where the peak is poorly constrained.

\begin{table}
\caption{Number of sources fitted with each set of spectral parameters. The number of upper limits are given in brackets. Five synchrotron-dominated sources (JVAS~B0218+357, CLASS~B1030+074, PMN~J1632$-$0033, PKS~1830$-$211 and PMN~J1838$-$3427) are fitted with single temperature modified black bodies to compute upper limits on contributions from any dust emission to their FIR spectra (for PKS~1830$-$211 this includes a synchrotron component). Of the sources with no temperature fitting, 30 sources have no detections and 10 have only one detection.}
\begin{center}
\begin{tabular}{p{3.9cm} c} \hline
Spectral fit & Number of sources \\ \hline
Two $T_{\rm dust}$ + $\beta$ + synchrotron & 3 \\
Single $T_{\rm dust}$ + $\beta$ + synchrotron & 8 \\
Single $T_{\rm dust}$ + $\beta$ & 10 \\
Single $T_{\rm dust}$ + synchrotron & 5 \\
Single $T_{\rm dust}$ & 33(5) \\
Fixed $T_{\rm dust}$ & 10(30) \\ \hline
Total & 69(35) \\ \hline
\end{tabular}
\label{table:sedfits}
\end{center}
\end{table}

\subsection{Physical properties}
\label{section:properties}

The dust temperature, FIR luminosity and SFR of 69 gravitationally-lensed AGN in our sample are listed in Table~\ref{table:luminosities}. We give both the results from the MCMC analysis and those from least-squares fitting. The values from the MCMC analysis are the median, 16th and 84th percentile of the posterior probability distributions. We give upper limits for the remaining 30 sources of the sample with insufficient detections, and the further 5 that appear to be synchrotron dominated. For clarification, a summary of median values of various properties we derive and an explanation of the objects included in these statistics are given in Table~\ref{table:properties}.

A histogram of the dust temperatures derived directly from the modified black-body fits is shown in Fig.~\ref{fig:dusttemps}. Where a model with two dust components are fit, we include only the colder component here. We find a median of $T_{\rm d} = 38^{+12}_{-5}$~K for the 53 sources with fitted dust temperatures and known redshifts, 51 of which have dust temperatures $<60$~K that, as we discuss in Section \ref{discussion}, can be reasonably attributed to be due to heating by star formation.

The FIR luminosity ($L_{\rm FIR}$) is derived for all sources with fitted modified black-body spectra by integrating the fitted modified black-body spectra between the rest-frame wavelengths 40 and 120 $\umu$m, using the definition of the FIR regime given by \cite{Helou:1988}, that is,
\begin{equation}
L_{\rm FIR} = \frac{4 \pi D_{L}^2}{(1+z)} \int_{40~{\rm \umu m}}^{120~{\rm \umu m}} S_{\nu,\rm rest}\,d\nu,
\end{equation} where $z$ is the redshift and $D_{L}$ is the luminosity distance. We then extrapolate to the total infrared luminosity (8 to 1000~$\umu$m; rest-frame) using the colour correction factor of 1.91 given by \cite{Dale:2001} (i.e. $L_{\rm IR}=1.91\,L_{\rm FIR}$) to correct for the contribution from MIR spectral features. The methodology used to calculate the star formation rate (SFR) is that given by \cite{Kennicutt:1998}, assuming a Salpeter initial mass function,
\begin{equation}
{\rm SFR\,(M_{\odot}\ yr^{-1})} = \frac {L_{\rm IR}}{5.8 \times 10^{9}},
\end{equation} where $L_{\rm IR}$ is in units of L$_{\odot}$. 

\begin{figure*}
\centering
\includegraphics[width=0.75\textwidth]{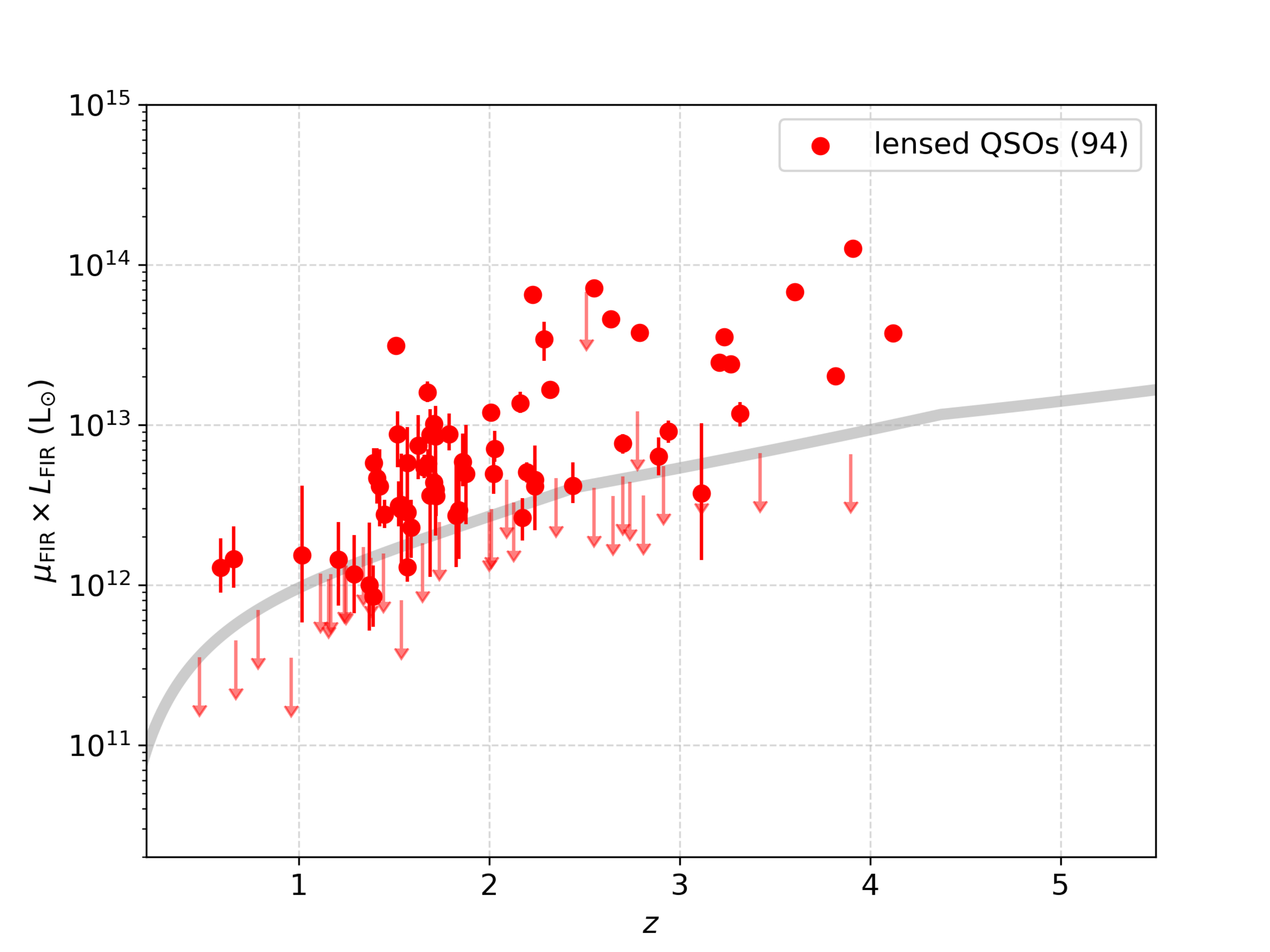}
\caption{\small FIR luminosity (40--120~$\umu$m) against redshift for the 94 objects in our sample with known redshift. This includes 53 with fitted dust temperatures, 10 with fixed dust temperatures, and 31 upper limits. The measured luminosities are shown in red, with no magnification correction. The grey line shows the estimated luminosity detection limit for a source with $T_{\rm d}=38$~K and $\beta=1.5$, assuming a 3$\sigma$ detection limit based on the mean RMS noise in each {\it Herschel}/SPIRE band. This is an overestimate at low redshift, as sources with lower dust temperatures will be preferentially detected, likewise, this is an overestimate at high-redshift where there will be bias towards higher temperature sources.}
\label{fig:lfir-z}
\centering
\includegraphics[width=0.75\textwidth]{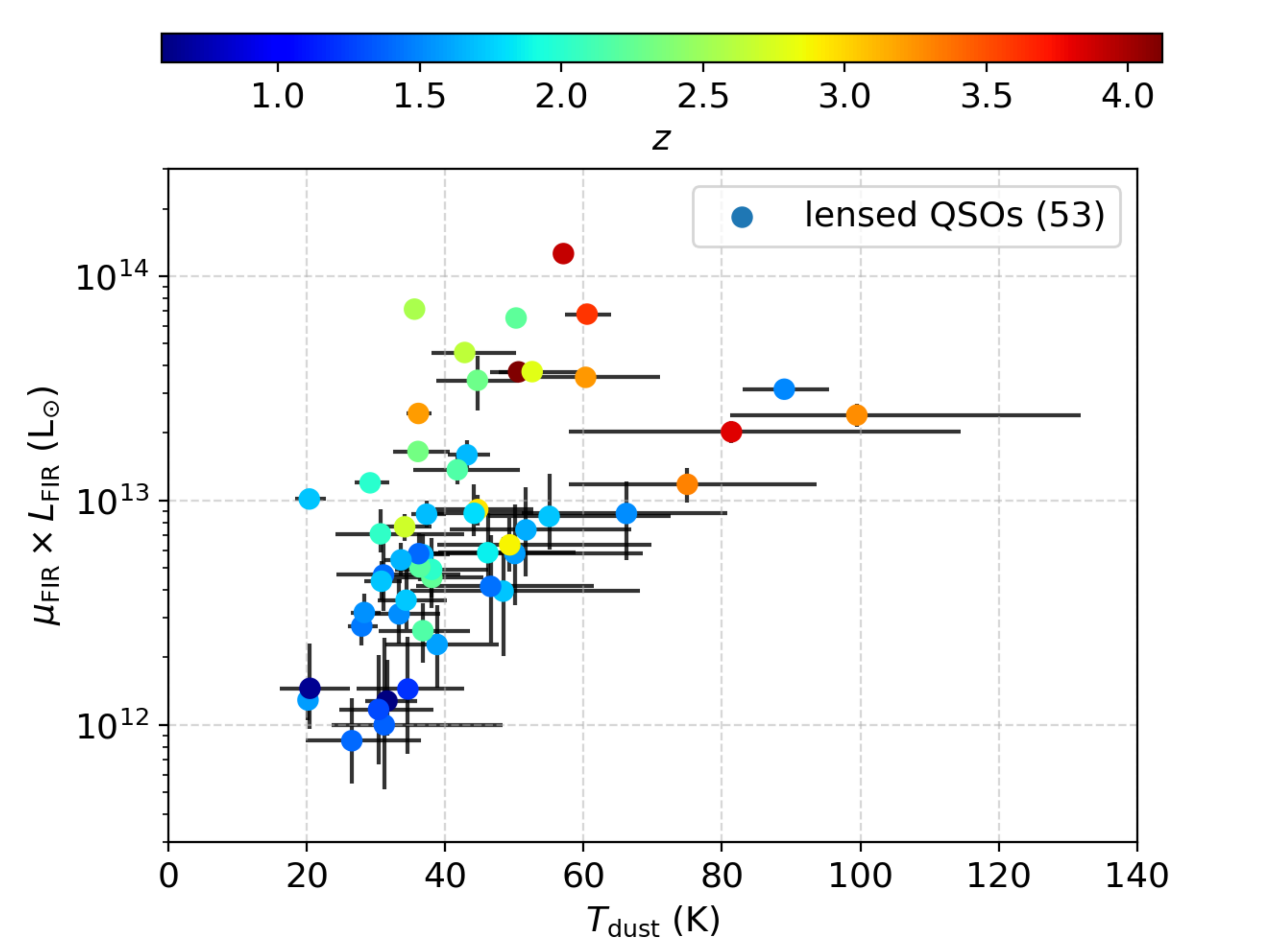}
\caption{\small FIR luminosity against fitted dust temperature for the 53 objects in our sample with fitted dust temperature and known redshift. The colour scale indicates source redshift. The luminosities are not corrected for lensing magnification.}
\label{fig:lfir-t}
\end{figure*}

In Figs.~\ref{fig:lfir-z} and \ref{fig:lfir-t}, we show the FIR luminosity uncorrected for lensing magnification based on the fitted SED models as a function of redshift and dust temperature, respectively. Note that the dust temperature is invariant to the lensing magnification in the absence of strong differential magnification.  The uncorrected luminosity as a function of redshift (Fig.~\ref{fig:lfir-z}) shows a clear trend in the data, from $\sim10^{12}$ L$_{\odot}$ at redshift 0.5--1 to $\sim10^{13}$--$10^{14}$ L$_{\odot}$ at redshift 3--4.

We use the {\it bhkmethod} task in the STSDAS statistics package to compute the Kendall correlation test, taking into account the luminosity upper limits. The Kendall statistic $\tau$ quantifies the degree of correlation (from $-1$ for a strong anti-correlation, 0 for no correlation, to 1 for a strong positive correlation) and the significance of this is given by the probability ($p$), for which $<0.05$ we take as statistically-significant. Our data shows a correlation in temperature with redshift ($\tau=0.64$, $p=4\times10^{-4}$) and in temperature with $L_{\rm FIR}$ ($\tau=0.77$, $p<1\times10^{-4}$).

We find a large spread of $L_{\rm FIR}$, as is clear from Fig~\ref{fig:lfir-t}: the low luminosities are associated with low temperatures and low redshifts, and the high luminosities with high redshifts and generally higher dust temperatures. The 6 sources with measured luminosities $<1.5\times10^{12}$~L$_{\odot}$ (corresponding to magnification-corrected SFRs $<50$  M$_{\odot}$ yr$^{-1}$) are associated with dust temperatures $<25$~K and/or redshifts $z<1.5$. These trends can be explained by observational bias given the wavelength limits of the {\it Herschel}/SPIRE bands and the flux-limits of our observations, which approximately correspond to the luminosity detection limit shown in Fig.~\ref{fig:lfir-z}.

\begin{figure} 
\includegraphics[width=0.5\textwidth]{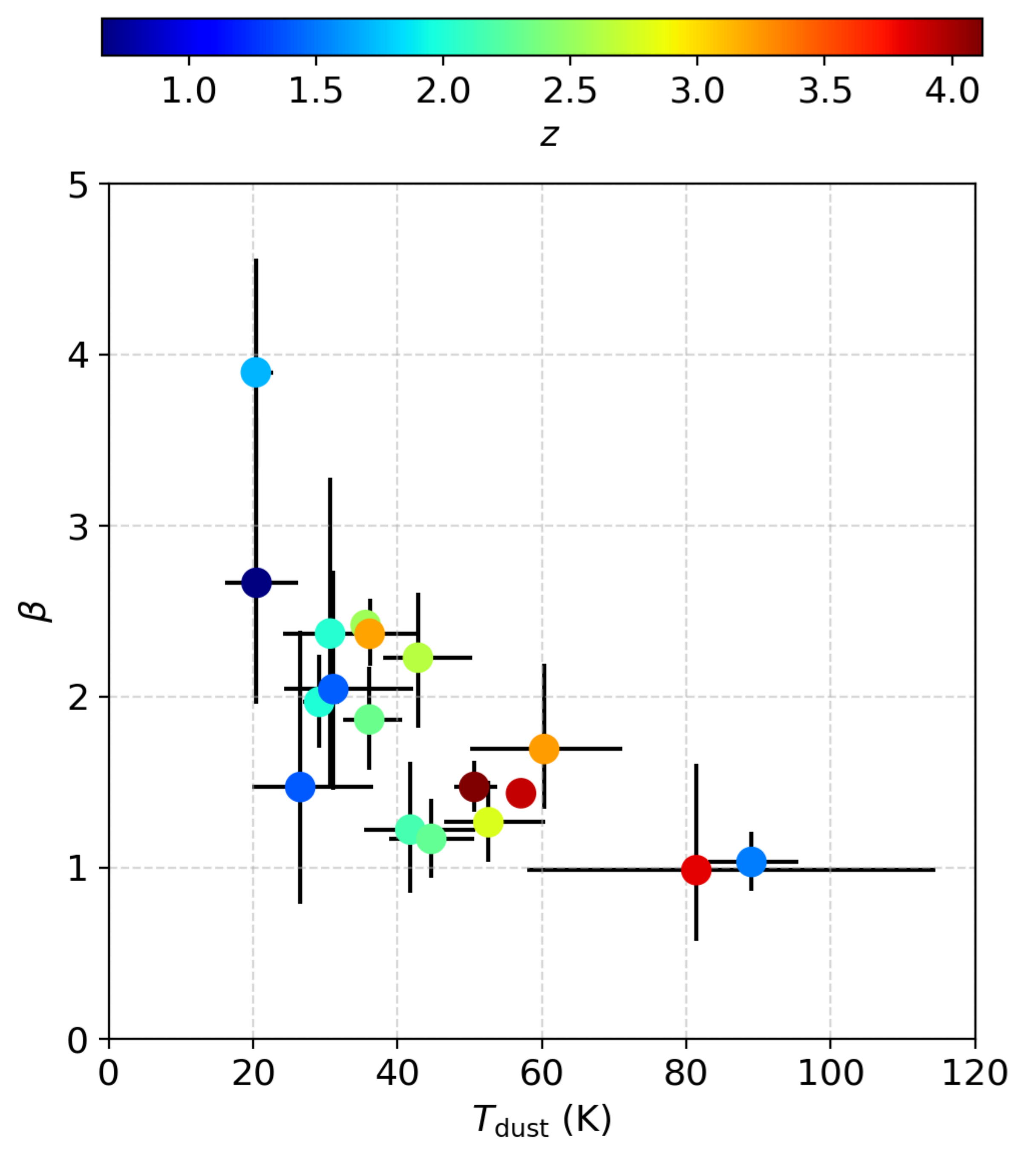}
\caption{\small $\beta$ against effective dust temperature for 18 objects in our sample with fitted $\beta$ and known redshift. The colour scale indicates source redshift.}
\label{fig:beta-t}
\end{figure}

In Fig.~\ref{fig:beta-t}, we present the dust emissivity as a function of dust temperature. We find a strong anti-correlation ($\tau=-1.30$, $p<1\times10^{-4}$) between the parameters. This effect is expected for internally-heated dust clouds \citep{Juvela:2012b}. However, it is not clear to what extent this correlation is a reflection of the `true' $\beta$--$T_{\rm d}$ relation, as the effect of source blending or observational noise may cause an artificial steepening of the anti-correlation, as noted by \cite{Juvela:2012a}. The shallow $\beta$ values may be a result of fitting a composite of dust emission from star formation and AGN-heating with a single greybody, when multiple components are required (see HS~0810+2554, Section~\ref{section:0810} in the Appendix). These problems further highlight the need for more multi-frequency data to better sample the SED and to reduce fitting errors due to observational noise.

\section{Discussion}
\label{discussion}

In this section, we investigate the global properties of the sample of gravitationally-lensed quasars and compare them with samples of unlensed FIR bright quasars and star-forming galaxies.

\subsection{Comparison to DSFGs}

We select the unlensed DSFGs observed with {\it Herschel}/SPIRE by \cite{Magnelli:2012} (hereafter, M12) for comparison with our lensed quasar sample, after correcting for the magnifications (see Section~\ref{section:magnifications}). The M12 objects are canonical DSFGs selected in ground-based submillimetre surveys with no evidence of a strong AGN component. We select the 46 unlensed objects with known redshifts from the M12 sample, which have redshifts $2.2^{+0.5}_{-0.7}$, FIR luminosities of $5.4^{+3.1}_{-3.7} \times 10^{12}~{\rm L_{\odot}}$ and star formation rates of $1800^{+1000}_{-1200}~{\rm M_{\odot}~yr^{-1}}$ (median, 25th and 75th percentiles). If the dust emission we detect is related to dust-obscured star formation, we expect dust temperatures comparable to DSFGs. Further to this, if DSFGs are antecedent to quasars, we would expect some fraction of our quasar sample to be FIR-luminous with star formation rates that are comparable to DSFGs.

\begin{figure*}
\centering
\includegraphics[width=0.75\textwidth]{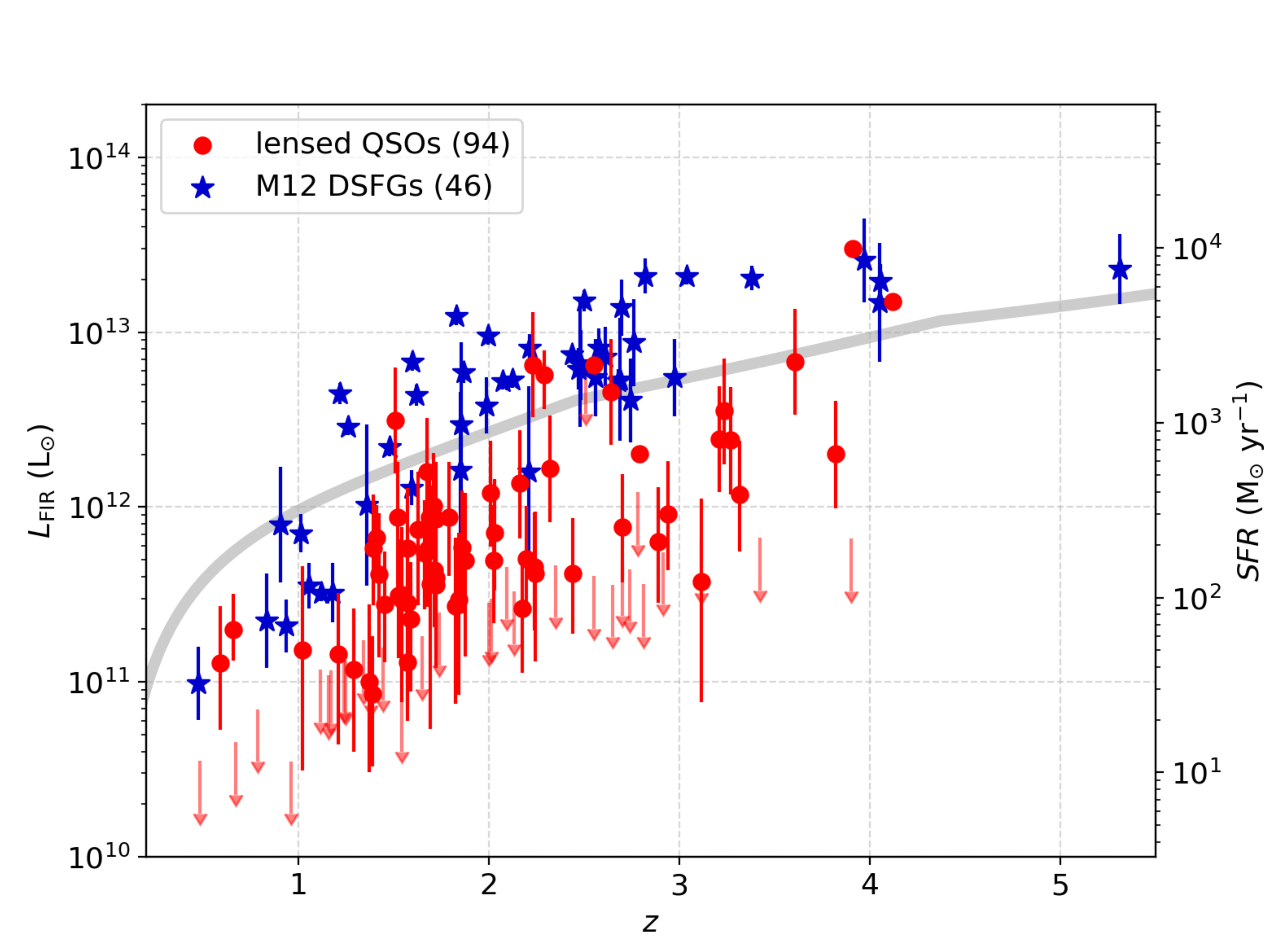}
\caption{\small FIR luminosity (40--120~$\umu$m) and equivalent SFR against redshift for lensed quasars in this sample (94 objects, excluding those without known redshifts) and for the M12 DSFGs (46 objects). The quasar luminosities are magnification-corrected (see Section~\ref{section:magnifications}). Magnification factors for seven sources are given in Table~\ref{table:magnifications}. Where the magnification factor is unknown, a we assume a value of $10^{+10}_{-5}$. The grey line shows the luminosity detection limit for a source with $T_{\rm d}=38$~K and $\beta=1.5$, assuming a 3$\sigma$ detection limit.}
\label{fig:lfir-z-smgs}
\centering
\includegraphics[width=0.75\textwidth]{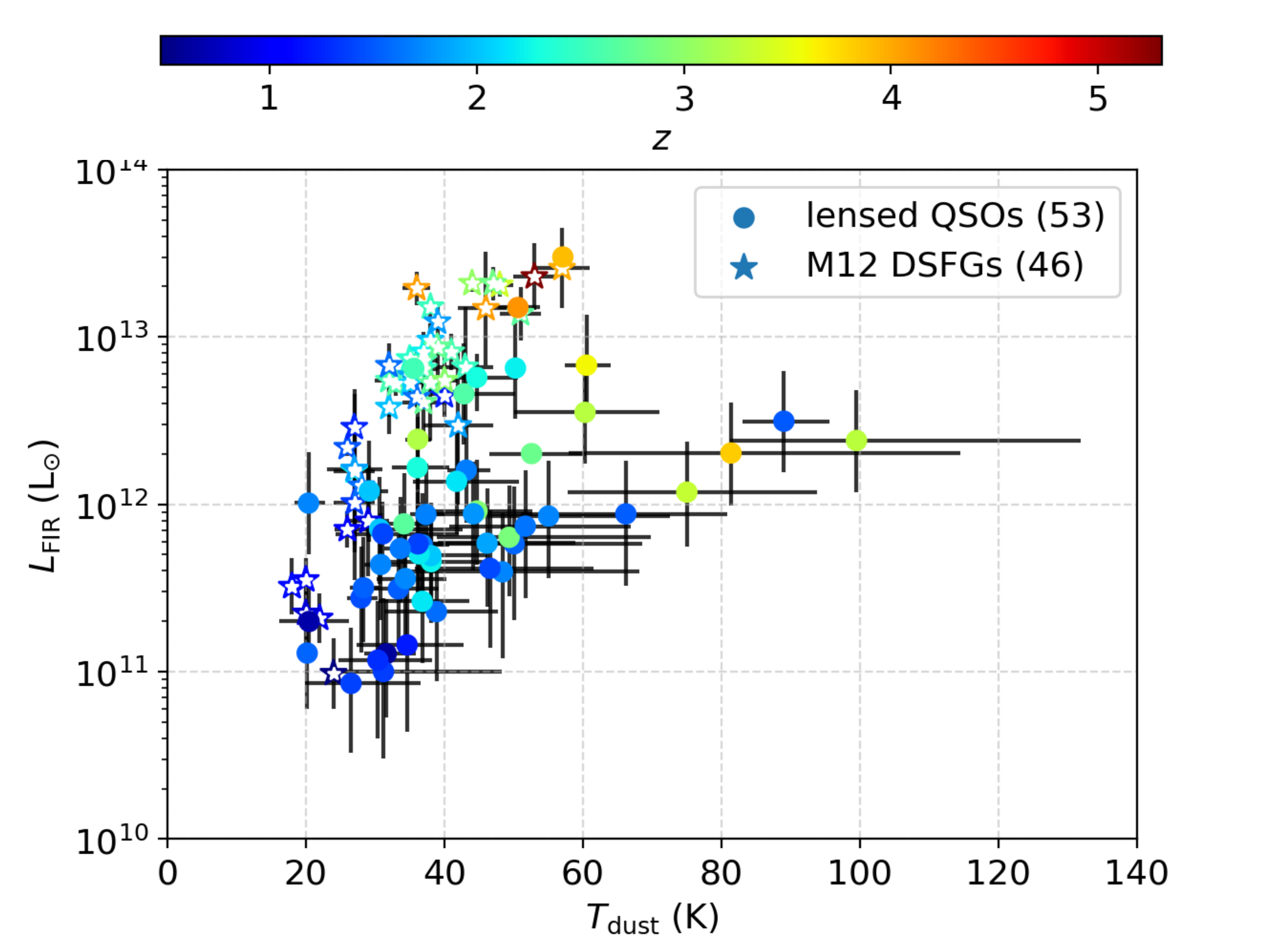}
\caption{\small FIR luminosity against dust temperature for our lensed quasar sample (53 objects, with fitted dust temperature and known redshift) and for the M12 DSFGs (46 objects). Quasar luminosities are magnification-corrected (see Section~\ref{section:magnifications}). The colour scale indicates source redshift.}
\label{fig:lfir-t-smgs}
\end{figure*}

We show in Figs. \ref{fig:lfir-z-smgs} and \ref{fig:lfir-t-smgs} the FIR luminosity against redshift and against dust temperature, respectively, for our sample and the M12 DSFGs. The median fitted dust temperature of our sample is $38^{+12}_{-5}$~K (ranges are the 25th and 75th percentiles) for 53 objects with sufficiently constrained SEDs and redshifts. This is consistent with the M12 DSFGs, which have a median temperature of $36^{+4}_{-9}$ K, typical of star-forming galaxies at $z\sim2$.

We apply the Kaplan--Meier (K--M) method to estimate the underlying distribution of FIR luminosities, taking into account the upper limits, using the task {\it kmestimate} in the STSDAS statistics package. This method assumes a randomly censored distribution: while this seems counter-intuitive as we have a fixed flux-density limit, our redshift range spans several orders of magnitude in luminosity distance so the sample is effectively randomly censored. We find a K--M estimated median, 25th and 75th percentiles of $3.6^{+4.8}_{-2.4} \times 10^{11}$ L$_{\odot}$ for the intrinsic luminosities for 94 objects with redshifts, including 63 detections and 31 upper limits, compared to $5.8^{+7.1}_{-2.7} \times 10^{11}$ L$_{\odot}$ for just the 63 objects with detected dust emission and known redshifts. The K--M estimated median of SFRs in our sample is $120^{+160}_{-80}\ {\rm M_{\odot}~yr^{-1}}$, with $190^{+230}_{-90}\ {\rm M_{\odot}~yr^{-1}}$ for just the objects with detected dust emission.

The SEDs determined here clearly demonstrate that 69 objects (66~percent) of our sample show evidence for heated dust emission at FIR to sub-mm wavelengths (these SEDs are shown in Fig.~\ref{figure:SEDs} of the Appendix). Also, given the similar dust temperatures of our lensed quasar sample and the DSFGs studied by M12, there is at least circumstantial evidence that this dust heating is due to star formation activity. Approximately 10~percent of the sample have extreme star-formation rates $>1000~{\rm M_{\odot}~yr^{-1}}$ comparable to typical, unlensed DSFGs at $z\sim 2$--4 detected in {\it Herschel}/SPIRE. The SFRs of these lensed quasars are consistent with sources that are transitioning from DSFGs to UV-bright quasars according to the \citeauthor{Sanders:1988} evolutionary model. The rest of the detected sample have still extreme SFRs similar to the lower luminosity DSFGs selected at $z<1.5$, but there is no clear cut-off at low SFR. The range of SFRs we find ($<$ 20--10000~${\rm M_{\odot}~yr^{-1}}$) is consistent with sources at different stages of evolution, and is not too surprising given the heterogeneous nature of our sample. Nevertheless, the high detection rate in the {\it Herschel}/SPIRE bands implies that most quasars are FIR-luminous sources with a strong coexistence of extreme dust-obscured star formation and AGN activity. This result implies a transition time from quasar-starburst to unobscured, gas-poor system of the order of the lifetime of the quasar (i.e. $\lesssim 100$~Myr), rather than much shorter timescales of $\sim1$~Myr, as has been suggested \citep{Simpson:2012}. Further studies, including spectral line data of the molecular gas in these systems, are required to understand anything of the gas reservoirs and depletion times, or make any conclusions regarding possible implications for AGN and stellar feedback.

\subsection{Comparison by radio properties}

Using the two-sample K--S test (described in Section~\ref{section:spiremeasurements}) we compare the derived $L_{\rm FIR}$ distributions of the jetted subsample with the remaining quasars in the sample. We do not correct for the lensing magnification here to prevent any bias due to our assumptions about the magnification factor of the heated dust. Including all measured luminosities and upper limits (excluding those without redshifts), the test returns a probability of 0.23 that these subsamples are drawn from the same underlying distribution. The K--M estimated median, 25th and 75th percentiles of the FIR luminosity is $1.6^{+10}_{-1.5} \times10^{12}$~L$_{\odot}$ for the jetted subsample and $3.7^{+3.5}_{-2.4} \times10^{12}$~L$_{\odot}$ for the remaining quasars. While the jetted subsample has a larger distribution of luminosities, the test suggests that the difference in the luminosity distributions is not statistically-significant.

A combination of systematic biases and the smaller size of the jetted subsample may affect our findings. We note that quasars with radio jets tend to be hosted in more massive galaxies \mbox{\citep{Mandelbaum:2009}}, thus our jetted subsample may be biased towards larger FIR luminosities and hence larger SFRs. At present, we do not have data to account for the stellar mass of the galaxies in our sample.

The median, 25th and 75th percentiles of the redshifts are $1.7^{+0.9}_{-0.3}$ for the jetted subsample and $1.8^{+0.6}_{-0.3}$ for the non-jetted. Despite the fact that the sample selection between these groups is different, with the jetted quasars generally selected in the radio by source properties and the non-jetted objects typically selected by lens population, the redshift distributions of the two groups are similar and therefore not a substantial source of systematic bias.

\begin{table}
\begin{center}
\bgroup
\def\arraystretch{1.5}
\caption{Kaplan--Meier (K--M) estimated 50th, 25th and 75th percentile ranges of the FIR luminosity distributions of the jetted and non-jetted subsamples. This includes the 92 objects in our sample with radio measurements and known redshifts. We give the number of values, $N$, and in brackets the number of upper limits. We also give the results when the subsamples are selected by $q_{\rm IR}$ value, and the K--S test probability that these samples are drawn from the same underlying distribution.}
\begin{tabular*}{0.45\textwidth}{p{1.3cm}p{1.35cm}m{1.2cm}m{1.2cm}c}
\hline
 & & N(lims) & K--M~$L_{\rm FIR}$ ($10^{12}$~L$_{\odot}$) & K--S test \\ 
\hline 
\multicolumn{1}{ c| }{\multirow{2}{*}{uncorrected}} & jetted & 15(10) & $1.6^{+10}_{-1.5}$ & \multirow{2}{*}{0.23} \\
\multicolumn{1}{ c| }{} & non-jetted & 48(19) & $3.7^{+3.5}_{-2.4}$ & \\ 
\hline 
\multicolumn{1}{ c| }{\multirow{2}{*}{corrected}} & jetted & 19(17) & $1.3^{+7.8}_{-1.3}$ & \multirow{2}{*}{0.06}\\
\multicolumn{1}{ c| }{}  & non-jetted & 43(13) & $4.1^{+3.3}_{-1.4}$  & \\ 
\hline 
\end{tabular*}
\label{table:medians}
\egroup
\end{center}
\end{table}

Our result is consistent with the conclusions of \protect\cite{Barvainis:2002} who found no statistically-significant difference in 850~$\umu$m luminosity between their samples of quasars and radio galaxies. Other studies have found no significant differences in the star-forming properties of quasars by radio mode. \cite{Harris:2016} analysed a sample of optically-luminous quasars at redshifts between 2 and 3 through stacking, of which 95~percent are undetected individually with {\it Herschel}/SPIRE. They find a mean SFR of $300\pm100~{\rm M_{\odot}~yr^{-1}}$, consistent with our overall result, but find no correlation with black hole accretion. A recent study by \cite{Pitchford:2016} of higher luminosity quasars with {\it Herschel}/SPIRE also find no relation between AGN accretion/outflows and the FIR properties of their host galaxies. Alternatively, \cite{Kalfountzou:2014} studied a stacked sample of quasars and do find a positive correlation between jet activity and FIR luminosity for jetted quasars, defined by a 5~GHz/4000~\AA\ ratio $>$10. However, they find average SFRs to be comparable for jetted and non-jetted quasars, except at low optical luminosities.

Overall, our results do not point towards there being an enhancement in the FIR luminosity of jetted quasars and radio galaxies, relative to the non-jetted quasars in our sample. However, a more complete understanding of this result, particularly given the contradictory studies discussed above, will require detailed observations of individual objects. In this respect, our investigation of lensed quasars from within our sample will again be important since it will allow the radio-jets, (stellar) host galaxy and the heated dust to be mapped on small angular-scales.

\subsection{Radio--infrared correlation}
\label{section:radioinfrared}
The radio--infrared luminosity correlation for star-forming galaxies has been well established for several decades. The relation is described by the parameter $q_{\rm IR}$, the ratio between the total infrared luminosity (8 to 1000~$\umu$m; rest-frame) and the 1.4 GHz rest-frame luminosity, defined by \cite{Condon:1991} as,
\begin{equation}
q_{\rm IR} = \log_{10}\left (\frac{L_{\rm IR}}{3.75\times10^{12}\ L_{\rm 1.4\ GHz}}  \right ).
\end{equation} 

\begin{figure*}
\centering
\includegraphics[width=0.8\textwidth]{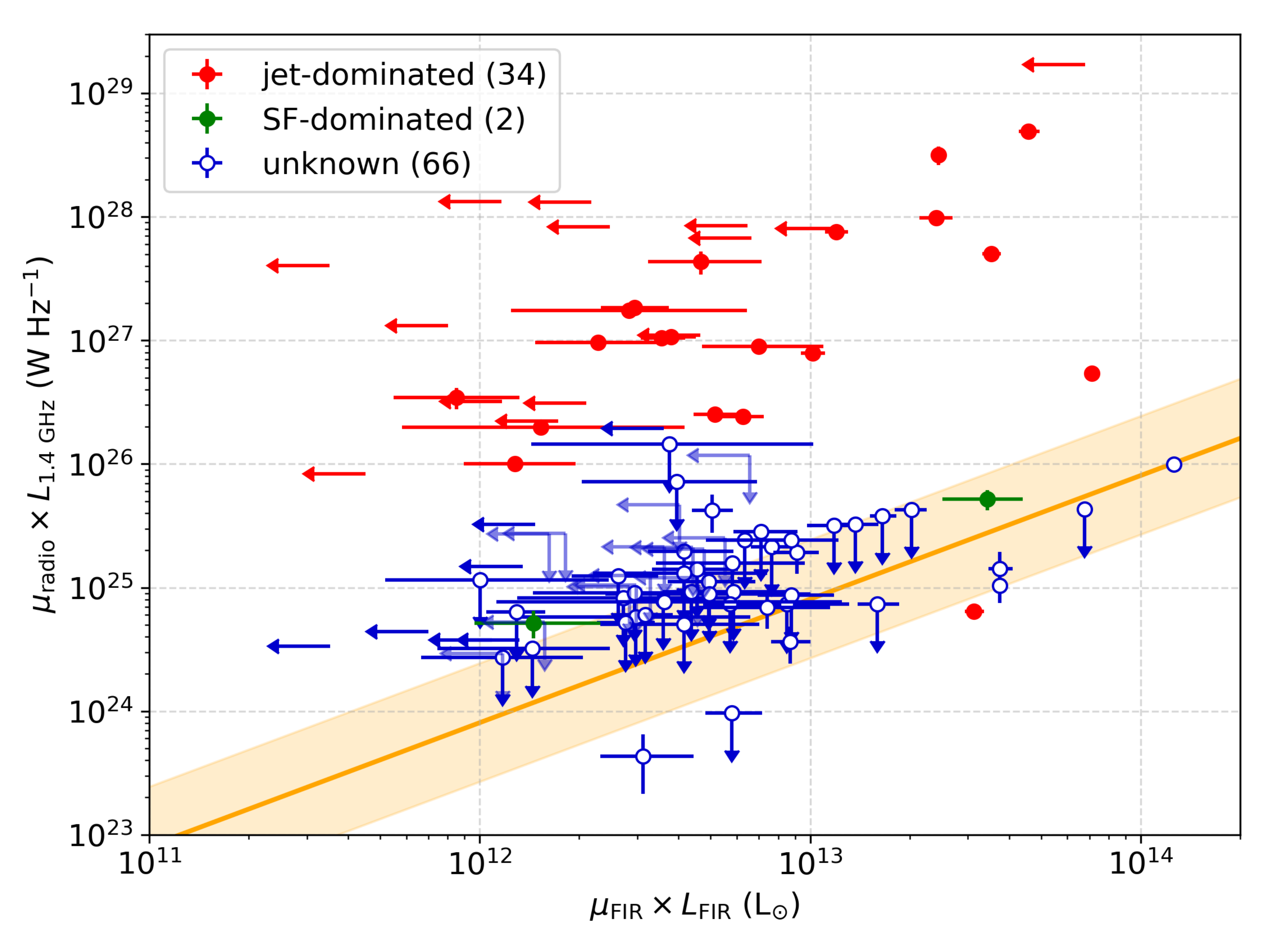}
\caption{\small The radio-infrared correlation for 102 quasars in our sample, excluding WFI~J2026$-$4536 and WFI~J2033$-$4723 for which there are no radio data available. The median $q_{\rm IR}$ for star-forming galaxies from \protect\cite{Ivison:2010} is shown in yellow; the shaded region is $2\sigma_{\rm qIR}$.}
\label{fig:irrc}
\centering
\includegraphics[width=0.8\textwidth]{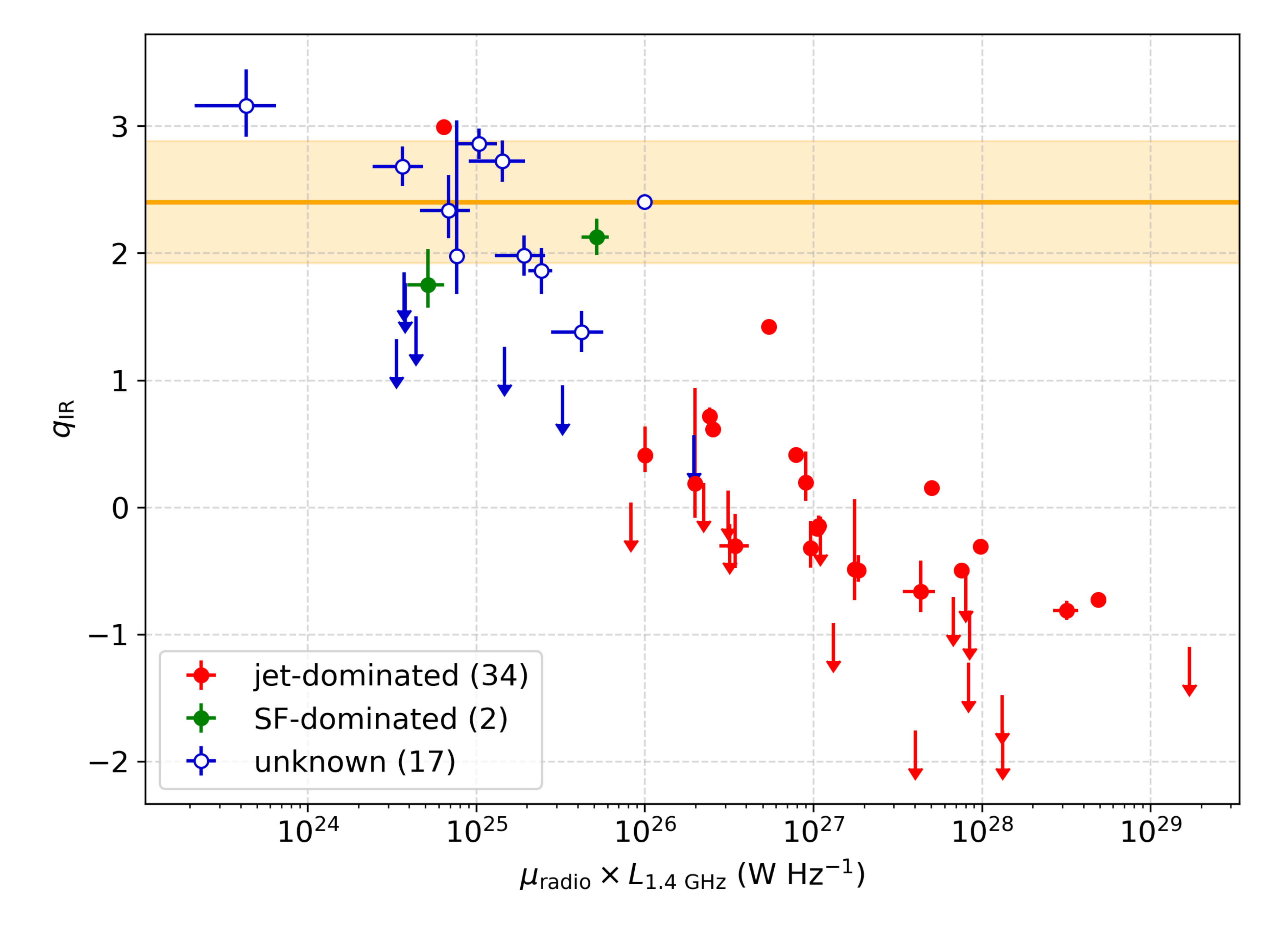}
\caption{\small The radio-infrared factor, q$_{\rm IR}$, for 53 quasars in the sample with radio detections. The median q$_{\rm IR}$ for star-forming galaxies from \protect\cite{Ivison:2010} is shown in yellow; the shaded region is $2\sigma_{\rm qIR}$.}
\label{fig:qir}
\end{figure*}

We explore the radio--infrared correlation for our sample to evaluate the contributions from star formation to the radio and FIR emission; for example, those sources above the correlation would have an excess of non-thermal synchrotron emission and those below the correlation would have an excess of thermal dust emission in the FIR, both of which could be related to the presence of a significant AGN contribution to the respective wavelength regimes \citep{Sopp:1991}. \cite{Ivison:2010} find $q_{\rm IR}=2.40 \pm 0.24$ for a flux-limited sample of sources selected from {\it Herschel}/SPIRE at 250~$\umu$m with VLA flux-densities at 1.4~GHz. We plot the rest-frame radio and IR-luminosity for the sample in Fig.~\ref{fig:irrc} and the median $q_{\rm IR}$ from \citeauthor{Ivison:2010} for reference. We interpolate or extrapolate to the rest-frame 1.4 GHz (depending on the low-frequency data available) by fitting the radio SEDs with a power-law, as given by equation (\ref{eq:spectrum}). A spectral index of $\alpha=-0.70\pm0.14$ is assumed for those objects with a single radio measurement, which is typically at 1.4~GHz from NVSS or FIRST. As above, we do not account for magnification to prevent bias due to our assumptions about the magnification factor of the dust emission. The $q_{\rm IR}$ values for the radio-detected quasars are shown in Fig.~\ref{fig:qir}.

We find that almost all of the jetted quasars lie significantly above the radio--infrared correlation for star-forming galaxies, by up to 3 to 4 orders of magnitude in rest-frame 1.4 GHz luminosity, and thus have $q_{\rm IR}$ values below that obtained by \cite{Ivison:2010}. This is to be expected as these are all powerful radio sources that are known to be dominated by synchrotron emission associated with AGN activity; the core and jet components of many sources are well studied as part of the CLASS, MG and PMN gravitational lens surveys. We note that the average magnification factors of jetted radio sources will likely be higher than that of the dust, as the radio emission comes from a more compact region, although how much higher will be dependent on where the radio source lies relative to the lens and the lensing caustics. In such cases, the inferred $q_{\rm IR}$ values may be lower if the radio component is boosted relative to the dust heated by star formation. However, we do not expect this to alter our conclusions as the effect will only be significant for sources with extremely compact radio emission associated with jets and, in almost all cases, this will not produce the several orders of magnitude difference needed to account for the offset from the correlation seen in Fig.~\ref{fig:qir}.

Quasars whose radio emission is associated with star formation are not expected to have significantly different magnifications between the radio and FIR, and so should remain close to the radio--infrared correlation for star-forming galaxies. Only 19 quasars in our sample classified as non-jetted have radio detections, of which only 2 are confirmed to have radio emission that is dominated by star formation (IRAS~F10214+5255 and RX~J1131$-$1231), the rest are currently undetermined. We observe a scatter around the radio--infrared relation for the non-jetted quasars. The scatter above the \citeauthor{Ivison:2010} relation may be due to contributions to their radio emission from low-power radio jets, or possibly additional radio emission from the foreground lensing galaxy. As the radio components of these sources have not yet been observed at a high enough angular resolution, it is not clear whether the apparent radio power dichotomy represents a true bi-modality in emission mechanism.

Recent studies point towards synchrotron emission from star formation as the dominant source of radio emission in non-jetted quasars \citep[for review]{Padovani:2016}. However, evidence of a milliarcsecond-scale jet in the classically radio-`quiet' lensed quasar HS~0810+2554 (part of our sample) suggests that it is not correct to assume that star formation is the primary radio emission mechanism in all cases (Hartley et al. in prep). There may be a composite of emission processes, or a further sub-population of quasars with low-power radio jets. Most of the detected radio-`quiet' quasars in our sample have $q_{\rm IR}$ values around the \citeauthor{Ivison:2010} relation, within the expected scatter. Deeper, higher spatial resolution observations of these sources are required to determine whether they hold to the relation, indicating whether the radio and FIR emission are indeed dominated by star formation. The radio upper limits in Fig.~\ref{fig:irrc}, due to non-detections in FIRST and NVSS, indicate that this population of quasars would be found in the $\sim\umu$Jy regime as proposed by \cite{White:2007}, if star formation dominates.

We add eleven of the non-jetted quasars with radio detections more than 2$\sigma$ above the radio--infrared correlation to our jetted subsample to refine our subsamples\footnote{We exclude WFI~2026$-$4536 and WFI~2033$-$4723 in our radio--infrared correlation analysis as there are no radio data for these sources.}, under the assumption that the radio excess is due to AGN activity within the background object and not from the foreground lensing galaxy.

We again perform the K--S test on the FIR luminosity distribution of the samples with measured $L_{\rm FIR}$ and radio emission, as before, and find the probability that they are drawn from the same sample decreases from 0.23 to 0.06, but is still not statistically-significant. The K--M-estimated median is higher for the objects which do not have a radio excess: $1.3^{+7.8}_{-1.2} \times 10^{12}$ L$_{\odot}$ for the new jetted subsample and $4.1^{+3.3}_{-1.4} \times 10^{12}$ L$_{\odot}$ for the remaining sources (as above, the uncertainties are at the 25th and 75th percentiles). It is possible that different FIR properties are simply caused by the small number of detections in the jetted subsample.

Notably, HS~0810+2554 lies below the radio--infrared relation with a $q_{\rm IR}$ value of 2.90 despite having radio jets that dominate its radio emission. In Section~\ref{section:0810} of the Appendix we explore the possibility that this is caused by fitting a single temperature dust model to a composite of AGN and star-formation-heated dust. It is possible this is also the case for another 7 sources with larger fitted dust temperatures ($T_{\rm d}>60$~K) that appear to be outliers from the bulk of the sample (see Fig.~\ref{fig:lfir-t-smgs}) and would explain some of the observed anti-correlation between $\beta$ and $T_{\rm d}$ (as shown in Fig.~\ref{fig:beta-t}). Interestingly, when we apply a two-temperature model to HS~0810+2554 (described in Section~\ref{section:0810}), the corrected $q_{\rm IR}$ lies within the 2$\sigma$ range of the radio--infrared correlation amidst the non-jetted sources. This suggests that the correlation may not be a reliable method of isolating jet-dominated quasars in the case of radio-weak AGN. If indeed quasars with radio jets are misidentified as non-jetted, this may cause or mask differences in the FIR properties between the subsamples.

The fact that no sources lie below the expected $q_{\rm IR}$ range in Fig.~\ref{fig:irrc} implies that we do not significantly overestimate the FIR luminosity due to an additional un-modelled AGN component in the dust emission, with the exception of HS~0810+2554. We note that APM~08279+5255, which has the most well defined SED and is fitted here with a two-temperature dust model, falls exactly on the radio--infrared correlation with a $q_{\rm IR}$ value of 2.40. Therefore, with sufficient data in the MIR and sub-mm it will be possible to better isolate the star-forming contribution to the SEDs. However, the radio and FIR luminosities derived for our current {\it Herschel}/SPIRE dataset, like in the case of the fitted dust temperatures, appear to be consistent with star formation being the dominant mechanism for heating the dust.

\begin{table*}
\caption{A summary of the objects used for our statistics. We give the median, 25th and 75th percentiles of certain properties, the number of measurements (N) this includes (number of limits, in brackets, where included), relevant figures within the paper, and an explanation of which objects are included in the selection. We give  magnification-corrected luminosities here, but the selection is the same for the uncorrected luminosities and SFRs.}
\begin{tabular*}{0.99\textwidth}{c c c c l} \hline
Property & Median & N(lims) & Fig. & Comment \\ \hline \noalign {\smallskip}
$z$ & $1.8^{+0.7}_{-0.3}$ & 94 & \ref{fig:sep_z} & Objects with known redshift. \\ \noalign {\smallskip}
$T_{\rm dust}$ & $38^{+12}_{-5}$~K & 53 & \ref{fig:dusttemps},\ref{fig:lfir-t} & Objects with known redshift, and dust temperature fitted as a free parameter in the SED. \\ \noalign {\smallskip}
$\beta$ & $2.0^{+0.4}_{-0.5}$ & 21 & \ref{fig:beta-t} & Objects with $\beta$ fitted as a free parameter in the SED. 18 of these have known redshift. \\ \noalign {\smallskip}
$L_{\rm FIR}$ & $3.6^{+4.8}_{-2.4} \times 10^{11}$ L$_{\odot}$ & 63(31) & \ref{fig:lfir-z},\ref{fig:lfir-z-smgs} & Objects with known redshifts, of which 63 have measured dust emission and 31 have upper limits. \\
\noalign {\smallskip} \hline
\end{tabular*}
\label{table:properties}
\end{table*}

\section{Conclusions}
\label{conc}

According to the current paradigm of galaxy evolution, some fraction of the quasar population is expected to be transitional sources from DSFGs to UV-luminous quasars. These transition sources will be FIR-luminous, with clear evidence of ongoing star formation. However, only a handful of extremely FIR-luminous sources have been studied thus far due to the limitations in observational sensitivity, and then at wavelengths relatively insensitive to $T_{\rm d}$ and $L_{\rm FIR}$. In order to study the link between DSFGs and quasars, we observed 104 gravitationally-lensed quasars with the {\it Herschel}/SPIRE instrument at 250, 350 and 500~$\umu$m to determine the fraction that are FIR-luminous. Due to the magnification effects of gravitational lensing, we are able to measure the star-forming properties of individual quasars at lower intrinsic luminosities than those previously studied and thus more representative of the quasar population. We find most sources in our sample have magnification-corrected FIR luminosities below the estimated detection limit, a factor of 10 lower on average.

From our study, we detected 72 (69~percent) of the gravitationally-lensed quasars in at least one band with {\it Herschel}/SPIRE, and find evidence for heated dust emission in 69 (66~percent) of the objects given the shape of their SEDs. By fitting modified black bodies to our new measurements and to data in the literature, we derive a median $\mu_{\rm FIR}\times L_{\rm FIR}$ of $3.6^{+4.7}_{-2.4} \times 10^{12}$ L$_{\odot}$ and an implied magnification-corrected $L_{\rm FIR}$ of $3.6^{+4.8}_{-2.4} \times 10^{11}$ L$_{\odot}$ and SFR of $120^{+160}_{-80}\ {\rm M_{\odot}~yr^{-1}}$ for 94 objects with redshifts, under the assumption that for the vast majority of the targets the FIR magnification is $\mu_{\rm FIR} = 10^{+10}_{-5}$. The range of fitted dust temperatures of the sample is $38^{+12}_{-5}$~K for 53 objects with redshifts and sufficiently constrained SEDs to fit for the dust temperature, which is characteristic of ongoing dust obscured star formation. We compare our sample of gravitationally-lensed quasars to a sample of DSFGs observed with {\it Herschel}/SPIRE at similar redshifts and find similar dust temperatures, which gives circumstantial evidence of star formation being the dominant mechanism for heating the dust. We find $\sim10$~percent have similar SFRs to DSFGs, suggestive of sources transitioning from DSFGs to UV-luminous quasar, and a large range of SFRs $<20$--10000~${\rm M_{\odot}~yr^{-1}}$.

Finally, using the radio--infrared correlation for star-forming galaxies, we find that the jetted quasars, selected by measured jet emission based on high resolution radio data, show an excess of radio luminosity by of up to 4 orders of magnitude. Non-jetted quasars lie close to the correlation expected for star-forming galaxies, however, the scatter above the correlation suggests at least some of these sources have contributions to their radio emission from AGN activity. We do not find evidence for an excess of FIR emission in the sample, given their radio luminosities, which again supports the view that the dust heating is dominated by star formation.

We find no significant difference in the $L_{\rm FIR}$ distribution of our jetted and non-jetted subsamples. Non-jetted quasars are three times as luminous on average when we select jetted sources by their radio excess relative to the radio--infrared correlation, however the difference is not statistically-significant. We caution that further data are required to eliminate the possibility of a systematic bias due to stellar mass or small sample sizes. Radio investigations are also required to identify the radio emission mechanism (and hence AGN feedback mechanisms) associated with non-jetted quasars, as our results suggest that the radio--infrared correlation may not be sufficient to identify sources with jet-dominated radio emission.

Our results reveal a strong co-existence of AGN activity and host galaxy star formation in quasars, as proposed in the \citeauthor{Sanders:1988} model. We find that this is true for the majority of quasars, suggesting that the FIR-luminous quasar phase is not distinct from the unobscured quasar phase and there is no sharp transition between them. However, our analysis is limited by the available observational data, which prevents us from making definitive statements about the implications for AGN and stellar feedback models. More photometric data points in the mm and sub-mm are needed to reduce the errors and assumptions in SED fitting, and thus the derived host galaxy properties. With additional MIR data, we can fit the AGN contribution to the dust heating and derive better constraints on the SFR for these sources. Much progress can also be made with the advent of ALMA, where the extent of the heated dust emission can be mapped on 50--500~mas-scales, which given the magnifications provided by the gravitational lenses, will also allow structures $<50$~pc in size to be resolved. In addition, through high resolution imaging of the mm-emission from these sources, it will also be possible to determine robust wavelength-dependent magnifications from lens modelling, which will further reduce the uncertainties in our analysis for individual objects.

\section*{Acknowledgements}

We thank the anonymous referee who helped improve the clarity of our paper. HRS would like to thank Kristen Coppin, Alison Kirkpatrick, Leon Koopmans and Simona Vegetti for helpful discussions. NCR acknowledges support from the ASTRON/JIVE Summer Student Programme. RJI acknowledges support from ERC in the form of the Advanced Investigator Programme, 321302, COSMICISM. DRGS thanks for funding through Fondecyt regular (project code 1161247), through the `Concurso Proyectos Internacionales de Investigaci\'on, Convocatoria 2015' (project code PII20150171), through ALMA-Conicyt (project code 31160001) and the BASAL Centro de Astrof\'isica y Tecnolog\'ias Afines (CATA) PFB-06/2007. 

{\it Herschel} is an ESA space observatory with science instruments provided by European-led Principal Investigator consortia and with important participation from NASA.{\it Herschel}/SPIRE has been developed by a consortium of institutes led by Cardiff University (UK) and including Univ. Lethbridge (Canada); NAOC (China); CEA, LAM (France); IFSI, Univ. Padua (Italy); IAC (Spain); Stockholm Observatory (Sweden); Imperial College London, RAL, UCL-MSSL, UKATC, Univ. Sussex (UK); and Caltech, JPL, NHSC, Univ. Colorado (USA). This development has been supported by national funding agencies: CSA (Canada); NAOC (China); CEA, CNES, CNRS (France); ASI (Italy); MCINN (Spain); SNSB (Sweden); STFC, UKSA (UK); and NASA (USA). HIPE is a joint development (are joint developments) by the Herschel Science Ground Segment Consortium, consisting of ESA, the NASA Herschel Science Center, and the HIFI, PACS and SPIRE consortia.

This research has made use of the NASA/ IPAC Infrared Science Archive, which is operated by the Jet Propulsion Laboratory, California Institute of Technology, under contract with the National Aeronautics and Space Administration.

{\sc iraf} is distributed by the National Optical Astronomy Observatory,
which is operated by the Association of Universities for Research
in Astronomy (AURA) under co-operative agreement with
the National Science Foundation. STSDAS is a product of the
Space Telescope Science Institute, which is operated by AURA for
NASA.


\bibliographystyle{mnras}
\bibliography{references}

\clearpage

\begin{appendix}
\section{S\lowercase{ource tables}}
\begin{landscape}
\begin{table}
\setlength{\tabcolsep}{1.5em}
\caption{{\it Herschel}/SPIRE measurements of the lensed quasar sample. We give the lens name, whether the object is jetted (J) or non-jetted (N), the maximum separation of the lensed images ($\Delta\theta$), the source redshift ($z_s$), the measured flux-densities at 250, 350, and 500~$\umu$m, and in brackets, the uncertainty from the source extraction. We comment on notable features of the lens system or with a $\dagger$ denote any issues with the flux-density measurements not discussed in Section~\ref{section:confusion}. Redshifts and image separations are from the CASTLES catalogue \citep{Kochanek:1999}, unless another reference is given. \newline}
\label{table:spireflux}
\begin{tabular*}{1.3\textwidth}{lccccccll}
\hline
Lens Name	&	Type & $\Delta\theta$ (arcsec)	&	$z_s$	& $S_{\rm 250~\umu m}$ (mJy)	&	$S_{\rm 350~\umu m}$ (mJy)	&	$S_{\rm 500~\umu m}$ (mJy)	& Comments & References \\	\hline \noalign {\smallskip}
HE~0047$-$1756	&N	&1.44	&1.676	&197(9)	&130(8)	&60(9) & &  \\%
CLASS~B0128+437	&J	&0.55	&3.114	&$<33$	&$<33$	&$<36$ & &  \\%
Q~J0158$-$4325	&N	&1.22	&1.29	&39(9)	&38(8)	&$<27$ & &	\\%
JVAS~B0218+357	&J	&0.34	&0.96	&89(7)	&122(7)	&120(8) & &\\%
HE~0230$-$2130	&N	&2.05	&2.162	&126(9)	&109(8)	&77(9)	\\%
SDSS~J0246$-$0825	&N	&1.19	&1.68	&88(7)	&75(7)	&30(8)	\\%
CFRS~03.1077	&N	&2.1	&2.941	&43(7)	&44(7)	&$<37$	\\%
MG~J0414+0534	&J	&2.4	&2.64	&266(7)	&190(8)	&112(10)	\\%
HE~0435$-$1223	&N	&2.42	&1.689	&133(7)	&101(7)	&53(9)	\\%
CLASS~B0445+123	&J	&1.35	&--		&$<38$	&$<50$	&$<42$	\\%
HE~0512$-$3329	&N	&0.65	&1.57	&60(7)	&39(9)	&$<32$	\\%
CLASS~B0631+519	&J	&1.16	&--		&63(12)	&82(7)	&71(12)	\\%
CLASS~B0712+472	&J	&1.46	&1.34	&$<33$	&$<28$	&$<36$ & & \\%
CLASS~B0739+366	&J	&0.53	&--		&69(6)	&61(10) &$<40$ & &	\\%
SDSS~J0746+4403	&N	&1.11	&2.0	&$<30$	&$<27$	&$<27$ & & \cite{Inada:2007}\\%
MG~J0751+2716	&J	&0.7	&3.21	&102(4)	&105(3)	&78(4)	\\%
SDSS~J0806+2006	&N	&1.4	&1.54	&40(9)	&$<21$	&$<26$ & & \cite{Inada:2006} \\%
HS~0810+2554	&N	&0.96	&1.5	&186(9)	&98(8)	&53(10)$^\dagger$ & $\dagger$Possible blending & \\%
HS~0818+1227	&N	&2.83	&3.115	&$<30$	&$<28$	&$<30$	\\%
SDSS~J0819+5356	&N	&4.04	&2.239	&40(9)	&40(8)	&$<40$ & & \cite{Inada:2009} \\%
SDSS~J0820+0812	&N	&2.2	&2.024	&49(9)	&54(8)	&$<32$ & & \cite{Jackson:2009} \\%
APM~08279+5255	&N	&0.38	&3.91	&621(3)	&437(3)	&259(4) & & \cite{Downes:1999} \\%
SDSS~J0832+0404	&N	&1.98	&1.116	&$<30$	&$<30$	&$<32$ & & \cite{Oguri:2008} \\%
CLASS~B0850+054	&J	&0.68	&--		&57(9)	&45(7)	&$<32$ & & \\%
SDSS~J0903+5028	&N	&2.99	&3.605	&227(11)&182(8)	&$<50$ & & \cite{Johnston:2003}\\%
SDSS~J0904+1512	&N	&1.13	&1.826	&27(7)	&$<30$	&$<47$ & & \cite{Kayo:2010} \\
SBS~J0909+523	&J	&1.17	&1.38	&$<27$	&$<30$	&$<47$	\\%
RX~J0911+0551	&N	&2.47	&2.79	&181(11)&176(9)	&97(9) & & \\%
RX~J0921+4529	&N	&6.97	&1.65	&$<26$	&$<30$	&$<30$	\\%
SDSS~J0924+0219	&N	&1.75	&1.524	&67(8)	&56(8)	&32(10) & & \cite{Inada:2003} \\%
FBQS~J0951+2635	&J	&1.11	&1.24	&$<30$	&$<30$	&$<41$	\\%
Q~0957+561		&J	&6.26	&1.41	&108(10)& 81(7)	&$<30$ & &	\\%
SDSS~J1001+5027	&N	&2.82	&1.84	&30(9)	&$<27$	&$<30$	\\%
SDSS~J1004+1229	&J	&1.54	&2.65	&$<26$	& $<26$	&$<41$	\\%
LBQS~J1009$-$0252	&N	&1.54	&2.74	&$<25$	&$<27$	&$<27$	\\%
SDSS~J1011+0143	&N	&3.67	&2.701	&$<28$	&$<30$	&$<40$	\\%
Q~1017$-$207	&N	&0.85	&2.55	&$<27$	&$<30$	&$<40$	\\%
\hline
\end{tabular*}
\end{table}
\end{landscape}

\begin{landscape}
\begin{table}
\ContinuedFloat
\setlength{\tabcolsep}{1.5em}
\contcaption{}
\begin{tabular*}{1.3\textwidth}{lccccccll}
\hline
Lens Name	&	Type & $\Delta\theta$ (")	&	$z_s$	& $S_{\rm 250~\umu m}$ (mJy)	&	$S_{\rm 350~\umu m}$ (mJy)	&	$S_{\rm 500~\umu m}$ (mJy) & Comments	 & References\\	\hline \noalign {\smallskip}
SDSS~J1021+4913	&N	&1.14	&1.72	&38(9)	&28(8)	&$<30$	\\%
IRAS~F10214+4724	&J	&1.59	&2.29	&416(5)	&303(4)	&169(4)	\\%
SDSS~J1029+2623	&N	&22.5	&2.197	&49(5)	&47(6)	&30(5) & &	\\
JVAS~B1030+074	&J	&1.65	&1.54	&35(6)	&44(9)	&63(11)	& & \\%
SDSS~J1054+2733	&N	&1.27	&1.452	&88(9)	&84(8)	&55(9) & & \cite{Kayo:2010} \\%
SDSS~J1055+4628	&N	&1.15	&1.249	&$<28$	&$<30$	&$<33$ & & \cite{Kayo:2010} \\%
HE~1104$-$1805	&N	&3.19	&2.32	&139(7)	&122(7)	&88(10)	\\%
PG~1115+080		&N	&2.32	&1.72	&69(17)	&40(7)	&$<39$	\\%
CLASS~B1127+385	&J	&0.74	&--		&90(9)	&105(7)	&56(10) &  &	\\%
RX~J1131$-$1231	&N	&3.8	&0.658	&285(7)	&166(8)	&62(9) & &	\\%
MG~J1131+0456	&J	&2.1	&--		&$<25$	&$<32$	&$<32$	\\%
SDSS~J1131+1915	&N	&1.46	&2.915	&$<28$	&$<30$	&$<30$ & & \cite{Kayo:2010} \\%
SDSS~J1138+0314	&N	&1.34	&2.44	&33(7)	&$<27$	&$<38$	\\%
SDSS~J1155+6346	&N	&1.95	&2.89	&29(7)	&30(6)	&$<30$	\\%
CLASS~B1152+200	&J	&1.59	&1.019	&37(10)	&$<30$	&$<32$ & \\%
SDSS~J1206+4332	&N	&2.9	&1.79	&95(7)	&70(8)	&$<31$ & &	\\%
Q~1208+101		&N	&0.48	&3.82	&61(9)	&52(7)	&$<36$	& \\ 
SDSS~J1216+3529	&N	&1.49	&2.013	&$<32$	&$<32$	&$<30$ & & \cite{Oguri:2008} \\%
HST~12531$-$2914	&N	&1.23	&--	&$<28$	&$<26$	&$<32$ & & \\%
SDSS~J1258+1657	&N	&1.28	&2.702	&46(7)	&54(8)	&60(13) & & \cite{Inada:2009}\\%
SDSS~J1304+2001	&N	&1.87	&2.175	&27(6)	&26(7)	&$<30$ & & \cite{Kayo:2010} \\%
SDSS~J1313+5151	&N	&1.24	&1.877	&49(6) & $<30$	& $<36$ & & \cite{Ofek:2007} \\%
SDSS~J1322+1052	&N	&2.0	&1.711	&87(10)	&79(7)	&57(9) & & \cite{Oguri:2008} \\%
SDSS~J1330+1810	&N	&1.76	&1.393	&126(7)	&91(7)	&36(11) & & \cite{Oguri:2008b}	\\%
SDSS~J1332+0347	&N	&1.14	&1.445	&$<27$	&$<30$	&$<32$	\\%
LBQS~J1333+0113	&N	&1.63	&1.57	&90(9)	&80(7)	&55(10) & & \\%
SDSS~J1339+1310	&N	&1.69	&2.241	&27(8)	&$<30$	&$<32$ & & \cite{Inada:2009} \\%
SDSS~J1349+1227	&N	&3.0	&1.722	&59(8)	&52(8)	&32(9) & & \cite{Kayo:2010}	\\%
SDSS J1353+1138	&N	&1.41	&1.629	&69(9)	&43(7)	&$<30$	\\%
Q~1355$-$2257	&N	&1.23	&1.37	&28(7)	&30(8)	&$<32$	\\%
CLASS~B1359+154	&J	&1.71	&3.235	&139(7)	&99(7)	&64(9)$^{\dagger}$ & ${\dagger}$Possible blending & \\%
SDSS~J1400+3134	&N	&1.74	&3.317	&45(7)	&28(9)	&$<42$ & & \cite{Inada:2009}\\%
SDSS~J1402+6321	&J	&1.35	&0.48	&$<27$	&$<25$	&$<30$	\\%
SDSS~J1406+6126	&N	&1.98	&2.13	&$<35$	&$<36$	&$<41$	\\%
HST~14113+5211	&N	&1.8	&2.81	&$<28$	&$<30$	&$<34$	\\%
H~1413+117		&J	&1.35	&2.55	&521(3)	&403(3)	&247(4)	\\%
JVAS~B1422+231	&J	&1.68	&3.62	&36(11)	&$<30$	&$<45$	\\%
SDSS~J1455+1447	&N	&1.73	&1.424	&54(7)	&34(7)	&$<30$ & & \cite{Kayo:2010}\\%
SBS~J1520+530	&N	&1.59	&1.86	&54(9)	&43(8)	&$<33$ & & \\%
SDSS~J1524+4409	&N	&1.67	&1.21	&43(10)	&32(7)	&$<30$ & & \cite{Oguri:2008} \\%
HST~15433+5352	&N	&1.18	&2.092	&$<30$	&$<30$	&$<47$	\\%
MG~J1549+3047	&J	&1.7	&1.17	&$<28$	&$<32$	&$<36$ & Lensed radio lobe \\%
\hline
\end{tabular*}
\end{table}
\end{landscape}

\begin{landscape}
\begin{table}
\ContinuedFloat
\setlength{\tabcolsep}{1.5em}
\contcaption{}
\begin{tabular*}{1.3\textwidth}{l c c c c c c l l}
\hline
Lens Name	&	Type & $\Delta\theta$ (")	&	$z_s$	& $S_{\rm 250~\umu m}$ (mJy)	&	$S_{\rm 350~\umu m}$ (mJy)	&	$S_{\rm 500~\umu m}$ (mJy) & Comments & References	\\	\hline \noalign {\smallskip}
CLASS~B1555+375	&J	&0.42	&--	&$<29$	&$<30$	&$<30$ & \\%
CLASS~B1600+434	&J	&1.4	&1.59	&36(6)	&27(6)	&$<39$ & Star-forming lens galaxy &\\%
CLASS~B1608+656	&J	&2.27	&1.39	&33(6)	&33(4)	&28(5)	\\%
SDSS~J1620+1203	&N	&2.77	&1.158	&$<26$	&$<28$	&$<40$ & & \cite{Kayo:2010} \\%
PMN~J1632$-$0033&J	&1.47	&3.424	&$<24$	&35(6)	&56(9) &  & \\%
FBQS~J1633+3134	&J	&0.75	&1.52	&62(9)	&30(8)	&$<30$	\\%
SDSS~J1650+4251	&N	&1.23	&1.54	&86(9)	&89(8)	&53(9)	\\%
MG~J1654+1346	&J	&2.1	&1.74	&$<34$	&$<30$	&$<43$	& Lensed radio lobe & \\
PKS~J1830$-$211	&J	&0.99	&2.51	&537(9)	&670(9)	&806(11) & & \\%
PMN~J1838$-$3427&J	&0.99	&2.78	&65(5)	&90(6)	&86(8) & &	\\%
CLASS~B1933+503	&J	&1.0	&1.71	&243(7)	&212(8)	&125(10)	\\%
JVAS~B1938+666	&J	&1.0	&2.01	&163(7)	&164(7)	&120(10)	\\%
PMN~J2004$-$1349&J	&1.18	&--	&38(7)$^\dagger$	&49(6)$^\dagger$	&28(7)$^\dagger$ & $\dagger$Diffuse galactic emission & \\%
MG~J2016+112	&J	&3.52	&3.27	&81(7)	&48(6)	&$<32$ & & \\%
WFI~J2026$-$4536&N	&1.34	&2.23	&460(7)	&301(7)	&162(10)	\\%
WFI~J2033$-$4723&N	&2.33	&1.66	&100(7)	&84(7)	&50(11)	\\%
CLASS~B2045+265	&J	&2.74	&2.35$^{\star}$	&$<36$	&$<33$	&$<43$ & & $^{\star}$C. Fassenacht, priv. comm. \\%
CLASS~B2108+213	&J	&4.57	&0.67	&$<32$	&$<32$	&$<36$	\\%
JVAS~B2114+022	&J	&1.31	&0.59	&137(7)	&108(9)	&36(9) & Star-forming lens galaxy & \\%
HE~2149$-$2745	&N	&1.7	&2.03	&84(7)	&77(9)	&37(9)	\\%
CY~2201$-$3201	&N	&0.83	&3.9	&$<37$	&$<30$	&$<44$	& \\%
Q~2237+030		&N	&1.78	&1.69	&$<30$	&$<28$	&$<40$	& & \\%
CLASS~B2319+052	&J	&1.36	&--		&56(6)	&57(7)	&$<33$	\\%
PSS~J2322+1944	&N	&1.49	&4.12	&83(4)	&105(3)	&81(5)	\\%
SDSS~J2343$-$0050	&N	&1.51	&0.787	&$<26$	&$<30$	&$<34$ & & \cite{Jackson:2008}\\%
\hline
\end{tabular*}
\end{table}

\end{landscape}

\begin{landscape}

\begin{table}
\caption{We give the FIR luminosities (40--120 $\umu$m) and star formation rates of the quasar sample, derived from both the least squares SED fit and the median, 16th and 84th percentiles from the MCMC analysis. The values are not corrected for lensing magnification ($\mu_{\rm FIR}$). Where objects are fit with two-temperature dust models, both temperatures are reported but the lower temperature is used for estimation of $L_{\rm FIR}$ (i.e. luminosity due to star formation) and SFR. Square brackets denote temperature fits where the source is synchrotron-dominated (noted in the comments), which are used only to derive upper limits on the dust emission.}
\bgroup
\def\arraystretch{1.3}
\setlength{\tabcolsep}{1em}
\label{table:luminosities}
\begin{tabular*}{1.35\textwidth}{p{2.5cm} | c c c c | c c c c | p{3cm}}
\cline{2-9}
& \multicolumn{4}{ c| }{Least squared} & \multicolumn{4}{ c| }{MCMC} & \\ \hline
Lens Name	& $T_{\rm dust}$ (K)	&	$\beta$		&	log $\mu\ L_{\rm FIR}$ (L$_{\odot}$)	&	log $\mu$ SFR	(M$_{\odot}$ yr$^{-1}$)	& $T_{\rm dust}$ (K)	&	$\beta$		&	log $\mu\ L_{\rm FIR}$ (L$_{\odot}$)	&	log $\mu$ SFR	(M$_{\odot}$ yr$^{-1}$)	&	Comments	 \\
\hline
HE~0047$-$1756  &  42.6  &  -  &  13.2  &  3.7  &  43.2$^{+3.4}_{-2.8}$  &  -  &  13.2$^{+0.1}_{-0.1}$  &  3.7$^{+0.1}_{-0.1}$  &  \\
CLASS~B0128+437  &  -  &  -  &  $<12.8$  &  $<3.3$ & - & - & - & - & \\
Q~J0158$-$4325  &  27.7  &  -  &  12.0  &  2.5  &  30.4$^{+7.9}_{-5.7}$  &  -  &  12.1$^{+0.3}_{-0.2}$  &  2.6$^{+0.3}_{-0.2}$  & \\
JVAS~B0218+357  &  [16.8]  &  -  &  $<11.5$  &  $<2.1$ &  - & - & - & - & synchrotron dominated \\
HE~0230$-$2130  &  39.4  &  1.3  &  13.1  &  3.6  &  41.8$^{+9.0}_{-6.3}$  &  1.2$^{+0.4}_{-0.4}$  & 13.1$^{+0.1}_{-0.1}$  &  3.65$^{+0.08}_{-0.06}$ & \\
SDSS~J0246$-$0825  &  36.0  &  -  &  12.7  &  3.3  &  36.8$^{+3.9}_{-3.1}$  &  -  &  12.8$^{+0.1}_{-0.1}$  &  3.3$^{+0.1}_{-0.1}$  & \\
CFRS03.1077  &  42.7  &  -  &  13.0  &  3.5  &  44.7$^{+8.0}_{-6.1}$  &  -  &  13.0$^{+0.1}_{-0.1}$  &  3.48$^{+0.07}_{-0.07}$  & \\
MG~J0414+0534 &  41.2  &  2.3  &  13.7  &  4.2  &  42.8$^{+7.4}_{-4.8}$  &  2.2$^{+0.4}_{-0.4}$  &  13.66$^{+0.04}_{-0.03}$  &  4.18$^{+0.04}_{-0.03}$  & \\
HE~0435$-$1223  &  37.0  &  -  &  12.9  &  3.4  &  37.3$^{+2.7}_{-2.3}$  &  -  &  12.9$^{+0.1}_{-0.1}$  &  3.5$^{+0.1}_{-0.1}$  &  \\ 
CLASS~B0445+123  &  -  &  -  &  $<12.1$  &  $<2.6$ &  - & - & - & - &  \\
HE~0512$-$3329  &  41.8  &  -  &  12.6  &  3.1  &  50.1$^{+18.4}_{-12.2}$  &  -  &  12.8$^{+0.3}_{-0.2}$  &  3.3$^{+0.3}_{-0.2}$  & \\
CLASS B0631+519  &  26.3  &  -  &  12.6  &  3.1  &  26.7$^{+1.4}_{-1.3}$  &  -  &  12.6$^{+0.1}_{-0.1}$  &  3.1$^{+0.1}_{-0.1}$  & \\
CLASS B0712+472  &  -  &  -  &  $<12.2$  &  $<2.8$ &  - & - & - & - &  \\
CLASS B0739+366  &  25.7  &  2.9  &  12.7  &  3.2  &  27.0$^{+6.2}_{-4.2}$  &  2.7$^{+0.7}_{-0.7}$  &  12.7$^{+0.1}_{-0.1}$  &  3.2$^{+0.1}_{-0.1}$  & \\
SDSS~J0746+4403  &  -  &  -  &  $<12.5$  &  $<3.0$ & - & - & - & - & \\
MG~J0751+2716  &  36.2  &  2.4  &  13.4  &  3.9  &  36.2$^{+1.9}_{-1.7}$  &  2.4$^{+0.2}_{-0.2}$  &  13.39$^{+0.01}_{-0.01}$  &  3.91$^{+0.01}_{-0.01}$  &  \\ 
SDSS~J0806+2006  &  -  &  -  &  12.4$^{+0.4}_{-0.2}$  &  2.9$^{+0.4}_{-0.2}$  & - & - & - & - \\
HS~0810+2554  &  89.1  &  1.0  &  13.5  &  4.0  &  89.0$^{+6.5}_{-6.0}$  &  1.0$^{+0.2}_{-0.2}$  &  13.50$^{+0.03}_{-0.03}$  &  4.01$^{+0.03}_{-0.03}$  & \\
SDSS~J0819+5356  &  36.2  &  -  &  12.6  &  3.2  &  38.1$^{+7.5}_{-6.3}$  &  -  &  12.7$^{+0.1}_{-0.1}$  &  3.2$^{+0.1}_{-0.1}$  & \\
SDSS~J0820+0812  &  35.8  &  -  &  12.7  &  3.2  &  38.1$^{+8.1}_{-5.2}$  &  -  &  12.7$^{+0.2}_{-0.1}$  &  3.2$^{+0.2}_{-0.1}$  & \\ 
HS~0818+1227  &  -  &  -  &  12.6$^{+0.4}_{-0.4}$  &  3.1$^{+0.4}_{-0.4}$  &  -  & - &  -  & -  & \\
APM~08279+5255  &  56.9, 160.3  &  1.4  &  14.1  &  4.6  &  57.1$^{+1.2}_{-1.2}$, 160.7$^{+2.5}_{-2.5}$ &  1.43$^{+0.03}_{-0.03}$  &  14.10$^{+0.01}_{-0.01}$  &   4.62$^{+0.01}_{-0.01}$  & \\
SDSS~J0832+0404  &  -  &  -  &  $<12.1$  &  $<2.6$ & - & - & - & - & \\ 
CLASS~B0850+054  &  42.4  &  -  &  12.8  &  3.3  & 48.8$^{+17.6}_{-10.0}$  &  -  &  12.8$^{+0.3}_{-0.1}$  &  3.4$^{+0.3}_{-0.1}$ & \\
SDSS~J0903+5028  &  60.2  &  -  &  13.8  &  4.3  &  60.5$^{+3.5}_{-3.1}$  &  -  &  13.83$^{+0.02}_{-0.02}$  &  4.35$^{+0.02}_{-0.02}$  & \\
SDSS~J0904+1512  &  -  &  -  &  12.4$^{+0.2}_{-0.2}$  &  3.0$^{+0.2}_{-0.2}$  & - & - & - & - & \\
SBS~0909+523  &  -  &  -  &  $<12.2$  &  $<2.7$ &  - & - & - & - &  \\
RX~J0911+0551  &  51.0  &  1.3  &  13.6  &  4.1  &  52.6$^{+7.8}_{-6.1}$  &  1.3$^{+0.2}_{-0.2}$  &  13.58$^{+0.04}_{-0.03}$  &  4.09$^{+0.04}_{-0.03}$   & \\ 
RX~J0921+4529  &  -  &  -  &  $<12.3$  &  $<2.8$ &  - & - & - & - &  \\
SDSS~J0924+0219  &  31.8  &  -  &  12.5  &  3.0  &  33.4$^{+5.9}_{-4.0}$  &  -  &  12.5$^{+0.2}_{-0.1}$  &  3.0$^{+0.2}_{-0.1}$  & \\
FBQS~0951+2635  &  -  &  -  &  $<12.1$  &  $<2.6$ &  - & - & - & - &  \\
Q~0957+561  &  27.9  &  2.3  &  12.6  &  3.1  &  31.1$^{+11.1}_{-6.8}$  &  2.0$^{+0.7}_{-0.6}$  &  12.7$^{+0.2}_{-0.1}$  &  3.2$^{+0.2}_{-0.1}$  &  \\
SDSS~J1001+5027  &  -  &  -  &  12.5$^{+0.3}_{-0.1}$  &  3.0$^{+0.3}_{-0.1}$  & - & - & - & - & \\
SDSS~J1004+1229  &  -  &  -  &  $<12.6$  &  $<3.1$ & - & - & - & - & \\
\hline
\end{tabular*}
\egroup
\end{table}
\end{landscape}
\clearpage

\begin{landscape}
\begin{table}
\ContinuedFloat
\contcaption{}
\bgroup
\def\arraystretch{1.3}
\setlength{\tabcolsep}{1em}
\begin{tabular*}{1.35\textwidth}{p{2.5cm} | c c c c | c c c c | p{3cm}}
\cline{2-9}
& \multicolumn{4}{ c| }{Least squared} & \multicolumn{4}{ c| }{MCMC} & \\ \hline
Lens Name	& $T_{\rm dust}$ (K)	&	$\beta$		&	log $\mu\ L_{\rm FIR}$ (L$_{\odot}$)	&	log $\mu$ SFR	(M$_{\odot}$ yr$^{-1}$)	& $T_{\rm dust}$ (K)	&	$\beta$		&	log $\mu\ L_{\rm FIR}$ (L$_{\odot}$)	&	log $\mu$ SFR	(M$_{\odot}$ yr$^{-1}$)	&	Comments	 \\
\hline
LBQS~1009$-$0252  &  -  &  -  &  $<12.6$  &  $<3.2$  &  -  &  -  &  -  &  -  & \\
SDSS~J1011+0143  &  -  &  -  &  $<12.7$  &  $<3.2$ & - & - & - & - & \\
Q~1017$-$207  &  -  &  -  &  $<12.6$  &  $<3.1$ &  - & - & - & - &  \\
SDSS~J1021+4913  &  38.8  &  -  &  12.4  &  3.0  &  48.4$^{+19.8}_{-13.4}$  &  -  &  12.6$^{+0.3}_{-0.2}$  &  3.1$^{+0.3}_{-0.2}$  & \\
IRAS~F10214+4724  &  40.6, 104.3  &  1.3  &  13.5  &  4.0  &  44.7$^{+5.9}_{-5.9}$, 90.2$^{+7.8}_{-6.0}$  &  1.2$^{+0.2}_{-0.2}$  &  13.5$^{+0.1}_{-0.1}$  &  4.1$^{+0.1}_{-0.1}$   &  \\
SDSS~J1029+2623  &  35.9  &  -  & 12.7 & 3.2 &  36.4$^{+3.1}_{-2.5}$  &  -  &  12.7$^{+0.1}_{-0.1}$  &  3.2$^{+0.1}_{-0.1}$   & \\
JVAS B1030+074  &  [20.2]  &  -  &  $<$11.9  &  $<$2.4  &  - & - & - & - & synchrotron dominated \\
SDSS~J1054+2733  &  27.6  &  -  &  12.4  &  2.9  & 28.0$^{+2.3}_{-2.0}$  &  -  &  12.4$^{+0.1}_{-0.1}$  &  3.0$^{+0.1}_{-0.1}$  & \\
SDSS~J1055+4628  &  -  &  -  &  $<12.1$  &  $<2.6$ & - & - & - & - & \\
HE~1104$-$1805  &  35.5  &  1.9  &  13.2  &  3.7  &  36.1$^{+4.6}_{-3.6}$  &  1.9$^{+0.3}_{-0.3}$  &  13.22$^{+0.04}_{-0.04}$  &  3.74$^{+0.04}_{-0.04}$  & \\
PG~1115+080  &  51.7  &  -  &  13.2  &  3.7   &  55.1$^{+17.5}_{-8.9}$  &  -  &  12.9$^{+0.2}_{-0.1}$  &  3.4$^{+0.2}_{-0.1}$  & \\
CLASS B1127+385  &  23.7  &  2.9  &  12.8  &  3.3 &  24.2$^{+4.3}_{-3.1}$  &  2.8$^{+0.6}_{-0.6}$  &  12.8$^{+0.1}_{-0.1}$  &  3.3$^{+0.1}_{-0.1}$  & \\
RX~J1131$-$1231  &  19.4  &  2.9  &  12.1  &  2.6  &  20.5$^{+5.7}_{-4.3}$  &  2.7$^{+1.0}_{-0.7}$  &  12.2$^{+0.3}_{-0.1}$  &  2.7$^{+0.3}_{-0.2}$  & \\
MG~J1131+0456  &  -  &  -  &  $<12.3$  &  $<2.9$  &  - & - & - & - &  \\
SDSS~J1131+1915  &  -  &  -  &  $<12.7$  &  $<3.2$  & - & - & - & - & \\
SDSS~J1138+0314  &  -  &  -  &  12.6$^{+0.2}_{-0.1}$  &  3.1$^{+0.2}_{-0.1}$  & - & - & - & - & \\
SDSS~J1155+6346  &  43.0  &  -  &  12.8  &  3.3  &  49.4$^{+20.5}_{-10.5}$  &  -  &  12.8$^{+0.1}_{-0.1}$  &  3.3$^{+0.1}_{-0.1}$ & \\
CLASS B1152+200  &  -  &  -  &  12.2$^{+0.8}_{-0.3}$  &  2.7$^{+0.8}_{-0.3}$  &  -  &  -  & - & - & \\
SDSS~J1206+4332  &  42.0  &  -  &  12.9  &  3.4  &  44.2$^{+8.5}_{-5.5}$  &  -  &  12.9$^{+0.2}_{-0.1}$  &  3.5$^{+0.2}_{-0.1}$  & \\
Q~1208+101  &  68.9  &  1.2  &  13.3  &  3.8  &  81.4$^{+33.1}_{-23.5}$  &  1.0$^{+0.6}_{-0.4}$  &  13.3$^{+0.1}_{-0.1}$  &  3.8$^{+0.1}_{-0.1}$  & \\
SDSS~J1216+3529  &  -  &  -  &  $<12.5$  &  $<3.0$  & - & - & - & - & \\
HST~12531$-$2914  &  -  &  -  &  $<12.2$  &  $<2.7$ &  -  &  -  &  -  &  -  & \\
SDSS~J1258+1657  &  33.3  &  -  &  12.9  &  3.4  &  34.2$^{+3.9}_{-3.2}$  &  -  &  12.9$^{+0.1}_{-0.1}$  &  3.4$^{+0.1}_{-0.1}$  & \\
SDSS~J1304+2001  &  35.5  &  -  &  12.4  &  2.9  &  31.0$^{+7.3}_{-4.6}$  &  -  &  12.4$^{+0.2}_{-0.1}$  &  2.9$^{+0.2}_{-0.1}$  & \\
SDSS~J1313+5151  &  -  &  -  &  12.7$^{+0.4}_{-0.2}$  &  3.2$^{+0.4}_{-0.2}$  &  -  &  -  &  -  &  -  & \\
SDSS~J1322+1052  &  30.4  &  -  &  12.6  &  3.2  &  30.8$^{+2.9}_{-2.4}$  &  -  &  12.6$^{+0.1}_{-0.1}$  &  3.2$^{+0.1}_{-0.1}$  & \\
SDSS~J1330+1810  &  35.6  &  -  &  12.7  &  3.3  &  36.2$^{+3.5}_{-2.7}$  &  -  &  12.8$^{+0.1}_{-0.1}$  &  3.3$^{+0.1}_{-0.1}$  & \\
SDSS~J1332+0347  &  -  &  -  &  $<12.2$  &  $<2.7$ & - & - & - & - & \\
LBQS~1333+0133  &  20.1  &  -  &  12.1  &  2.6  &  20.2$^{+1.4}_{-1.3}$  &  -  &  12.1$^{+0.1}_{-0.1}$  &  2.6$^{+0.1}_{-0.1}$  & \\
SDSS~J1339+1310  &  -  &  -  &  12.6$^{+0.3}_{-0.2}$  &  3.1$^{+0.3}_{-0.2}$   & - & - & - & - & \\
SDSS~J1349+1227  &  32.9  &  -  &  12.5  &  3.0  &  34.4$^{+5.9}_{-4.1}$  &  -  &  12.6$^{+0.2}_{-0.1}$  &  3.1$^{+0.2}_{-0.1}$  & \\
SDSS~J1353+1138  &  45.7  &  -  &  12.8  &  3.3  &  51.7$^{+15.2}_{-11.0}$  &  -  &  12.9$^{+0.2}_{-0.2}$  &  3.4$^{+0.2}_{-0.2}$  & \\
Q~1355$-$2257  &  25.7  &  -  &  11.8  &  2.3  &  31.2$^{+17.1}_{-7.6}$  &  -  &  12.0$^{+0.6}_{-0.2}$  &  2.5$^{+0.6}_{-0.2}$  & \\
CLASS B1359+154  &  58.8  &  1.7  &  13.5  &  4.1  &  60.3$^{+10.8}_{-10.2}$  &  1.7$^{+0.5}_{-0.4}$  &  13.55$^{+0.03}_{-0.03}$  &  4.06$^{+0.03}_{-0.03}$  & \\
SDSS~J1400+3134  &  69.6  &  -  &  13.4  &  3.9  &  75.0$^{+18.7}_{-17.2}$  &  -  &  13.1$^{+0.1}_{-0.1}$  &  3.6$^{+0.1}_{-0.1}$  & \\
\hline
\end{tabular*}
\egroup
\end{table}
\end{landscape}
\clearpage

\begin{landscape}
\begin{table}
\ContinuedFloat
\contcaption{}
\bgroup
\def\arraystretch{1.3}
\setlength{\tabcolsep}{1em}
\begin{tabular*}{1.35\textwidth}{p{2.5cm} | c c c c | c c c c | p{3cm}}
\cline{2-9}
& \multicolumn{4}{ c| }{Least squared} & \multicolumn{4}{ c| }{MCMC} & \\ \hline
Lens Name	& $T_{\rm dust}$ (K)	&	$\beta$		&	log $\mu\ L_{\rm FIR}$ (L$_{\odot}$)	&	log $\mu$ SFR	(M$_{\odot}$ yr$^{-1}$)	& $T_{\rm dust}$ (K)	&	$\beta$		&	log $\mu\ L_{\rm FIR}$ (L$_{\odot}$)	&	log $\mu$ SFR	(M$_{\odot}$ yr$^{-1}$)	&	Comments	 \\
\hline
SDSS~J1402+6321  &  -  &  -  &  $<11.5$  &  $<2.0$ & - & - & - & - & \\
SDSS~J1406+6126  &  -  &  -  &  $<12.5$  &  $<3.0$  & - & - &  -  &  -  & \\
HST~14113+52116  &  -  &  -  &  $<12.6$  &  $<3.1$  &  -  &  -  &  -  &  -  & \\
H~1413+117  &  35.5, 124.1  &  2.4  &  13.85  &  4.37  &  35.6$^{+0.6}_{-0.6}$, 125.6$^{+10.6}_{-8.9}$  &  2.4$^{+0.1}_{-0.1}$  &  13.85$^{+0.01}_{-0.01}$  &   4.37$^{+0.01}_{-0.01}$  &  \\
JVAS B1422+231  &  -  &  -  &  12.4  &  2.9  &  -  &  -  &  12.36$^{+0.40}_{-0.18}$  &  2.87$^{+0.40}_{-0.18}$  & \\
SDSS~J1455+1447  &  39.7  &  -  &  12.5  &  3.0  &  46.6$^{+14.9}_{-10.8}$  &  -  &  12.6$^{+0.3}_{-0.2}$  &  3.1$^{+0.3}_{-0.2}$ & \\
SBS~1520+530   &  42.1  &  -  &  12.7  &  3.2  &  46.2$^{+12.7}_{-7.2}$  &  -  &  12.8$^{+0.2}_{-0.1}$  &  3.3$^{+0.2}_{-0.1}$  & \\
SDSS~J1524+4409  &  32.0  &  -  &  12.4  &  2.9  &  34.6$^{+8.1}_{-7.4}$  &  -  &  12.2$^{+0.3}_{-0.2}$  &  2.7$^{+0.3}_{-0.2}$  & \\
HST~15433+5352  &  -  &  -  &  $<12.7$  &  $<3.2$  &  - & - & - & - &  \\
MG~J1549+3047  &  -  &  -  &  $<12.1$  &  $<2.6$ &  - &  -  &  -  &  -  & \\
CLASS B1555+375  &  -  &  -  &  $<12.3$  &  $<2.8$  & - & - & - & - & \\
CLASS B1600+434  &  35.8  &  -  &  12.3  &  2.8  &  38.9$^{+8.9}_{-7.5}$  &  -  &  12.4$^{+0.2}_{-0.2}$  &  2.9$^{+0.2}_{-0.2}$  &  \\
CLASS B1608+656  &  31.9  &  1.8  &  11.9  &  2.4  &  36.3$^{+13.7}_{-9.2}$  &  1.5$^{+0.9}_{-0.7}$  &  11.9$^{+0.2}_{-0.1}$  &  2.4$^{+0.2}_{-0.1}$  & \\
SDSS~J1620+1203  &  -  &  -  &  $<12.0$  &  $<2.6$  & - & - & - & - & \\  
PMN~J1632$-$0033  &  [26.8]  &  -  &  $<12.8$  &  $<3.3$  & - & - & - & - & synchrotron dominated \\
FBQS~1633+3134  & 65.1  &  -  &  12.9  &  3.5  &  66.2$^{+14.6}_{-15.1}$  &  -  &  12.9$^{+0.2}_{-0.2}$  &  3.5$^{+0.2}_{-0.2}$  & \\
SDSS~J1650+4251  &  28.1  &  -  &  12.5  &  3.0  &  28.3$^{+2.3}_{-1.9}$  &  -  &  12.5$^{+0.1}_{-0.1}$  &  3.0$^{+0.1}_{-0.1}$ & \\
MG~J1654+1346  &  -  &  -  &  $<12.4$  &  $<2.9$  &  - & - & - & - & \\
PKS J1830$-$211 & [29.7]  &  -  &  $<13.8$  &  $<4.3$ & - & - & - & -   & synchrotron dominated \\
PMN~J1838$-$3427  &  [32.8] &  -  &  $<13.1$  &  $<3.6$  &  - & - & - & - & synchrotron dominated \\
CLASS B1933+503  &  20.4  &  3.9  &  13.0  &  3.5  &  20.4$^{+2.3}_{-2.1}$  &  3.9$^{+0.7}_{-0.6}$  &  13.01$^{+0.04}_{-0.03}$  &  3.52$^{+0.04}_{-0.03}$  & \\ 
JVAS B1938+666  &  28.9  &  2.0  &  13.1  &  3.6  &  29.2$^{+2.7}_{-2.3}$  &  2.0$^{+0.3}_{-0.3}$  &  13.08$^{+0.04}_{-0.03}$  &  3.59$^{+0.04}_{-0.03}$  & \\
PMN~J2004$-$1349  &  30.7  &  -  &  12.5  &  3.0  &  31.8$^{+4.0}_{-3.1}$  &  -  &  12.5$^{+0.1}_{-0.1}$  &  3.0$^{+0.1}_{-0.1}$ &  \\
MG~J2016+112  &  88.6  &  -  &  13.4  &  3.9  &  99.5$^{+32.4}_{-18.3}$  &  -  &  13.38$^{+0.05}_{-0.05}$  &  3.90$^{+0.05}_{-0.05}$  & \\
WFI~J2026$-$4536  &  50.1  &  -  &  13.8  &  4.3  &  50.3$^{+1.3}_{-1.2}$  &  -  &  13.81$^{+0.02}_{-0.02}$  &  4.33$^{+0.02}_{-0.02}$  &  \\
WFI~J2033$-$4723  &  33.2  &  -  &  12.7  &  3.2  &  33.7$^{+3.0}_{-2.4}$  &  -  &  12.7$^{+0.1}_{-0.1}$  &  3.3$^{+0.1}_{-0.1}$  &  \\
CLASS B2045+265  &  -  &  -  &  $<12.7$  &  $<3.2$  &  -  &  -  &  -  &  - & \\ 
CLASS B2108+213  &  -  &  -  &  $<11.7$  &  $<2.2$  & - & - & - & - &  \\
JVAS B2114+022  &  30.4  &  -  &  12.1  &  2.6  &  31.6$^{+4.3}_{-3.1}$  &  -  &  12.1$^{+0.2}_{-0.1}$  &  2.6$^{+0.2}_{-0.1}$  & \\
HE~2149$-$2745  &  26.9  &  2.8  &  12.8  &  3.3  &  30.7$^{+12.0}_{-6.5}$  &  2.4$^{+0.9}_{-0.9}$  &  12.9$^{+0.1}_{-0.2}$  &  3.4$^{+0.1}_{-0.2}$  &  \\
CY~2201$-$3201  &  -  &  -  &  $<12.8$  &  $<3.3$  &- & - & - & - &  \\
Q~2237+030  &  -  &  -  &  $12.5^{+0.3}_{-0.4}$ & $3.1^{+0.3}_{-0.4}$  & - & - & - & - & \\
CLASS B2319+052  &  19.1  &  4.5  &  12.6  &  3.1  &  19.0$^{+4.1}_{-3.0}$  &  4.6$^{+1.2}_{-1.1}$  &  12.6$^{+0.1}_{-0.1}$  &  3.1$^{+0.1}_{-0.1}$  & \\  
PSS~J2322+1944  &  47.2  &  1.65  &  13.6  &  4.1  &  47.4$^{+2.5}_{-2.3}$  &  1.6$^{+0.1}_{-0.1}$  &  13.58$^{+0.01}_{-0.01}$  &  4.10$^{+0.01}_{-0.01}$  & \\
SDSS~J2343$-$0050  &  -  &  -  &  $<11.8$  &  $<2.4$  & - & - & - & - &  \\
\hline
\end{tabular*}
\egroup
\end{table}
\end{landscape}
\clearpage

\onecolumn
\section{SED\lowercase{s and ancillary data}}
\label{section:SEDs}
\begin{figure*}
\caption{Included here are 69 SEDs for the quasars in our sample with fitted SEDs, excluding synchrotron-dominated sources (see Fig.~\ref{fig:syncseds}). The legend details the free parameters of the model and their least squared values, excluding the normalization (FIR luminosities are given in Table~\ref{table:luminosities}). Where $\beta$ is not given, it is fixed to 1.5. For objects with T fixed, the dust temperature is set to 38~K, the median of the sample. For composite spectra, modified black-body fits are in red, synchrotron in cyan, and the total spectrum in green. The {\it Herschel}/SPIRE bandwidth is in grey. As we discuss in Section~\ref{section:sedfitting}, we do not attempt to fit complex synchrotron components in cases where there is a suggestion that the synchrotron emission is falling off towards the sub-mm or if there is only one detection.}
\label{figure:SEDs}
\includegraphics[width=\textwidth]{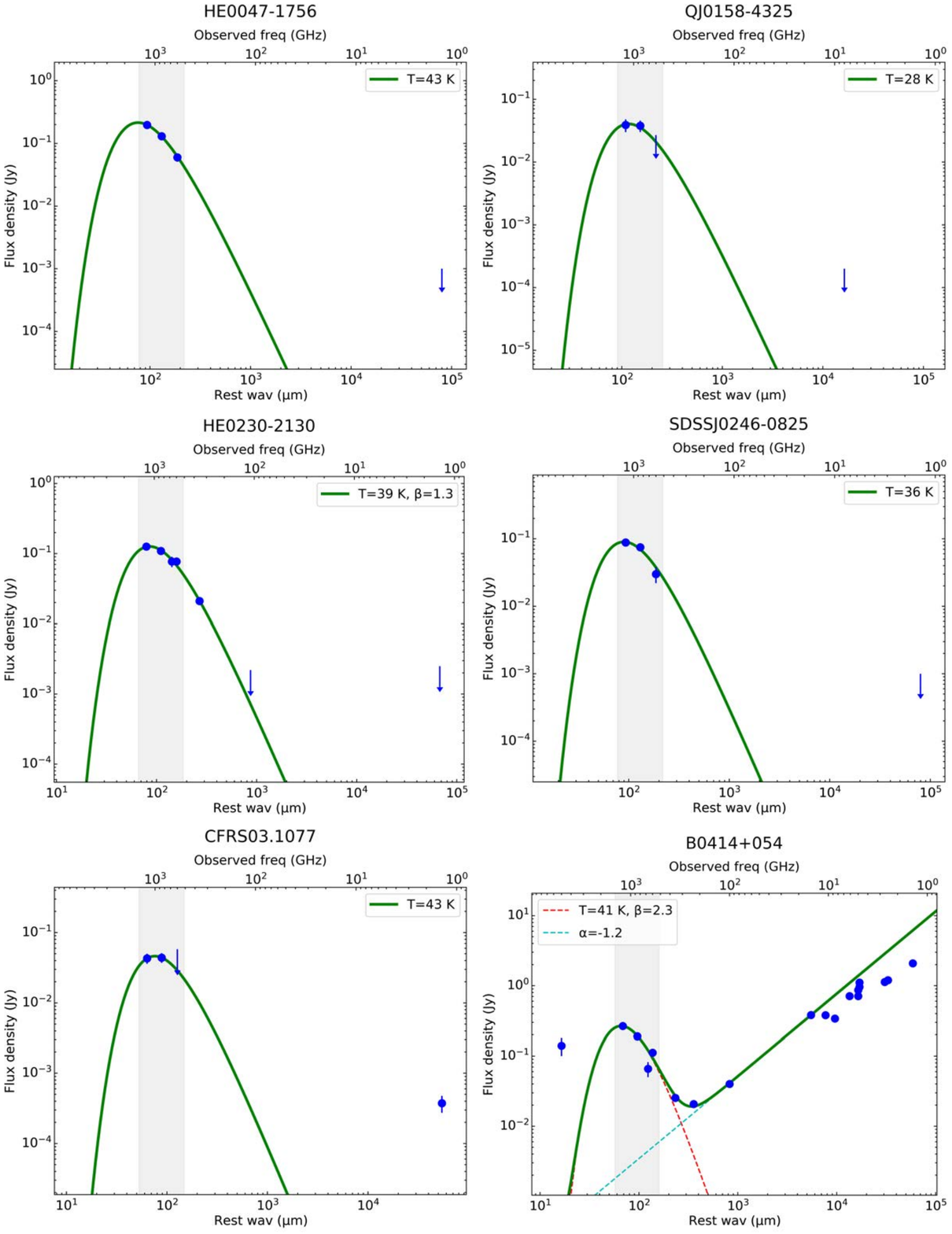}
\end{figure*}
\begin{figure*}
\ContinuedFloat
\contcaption{}
\includegraphics[width=\textwidth]{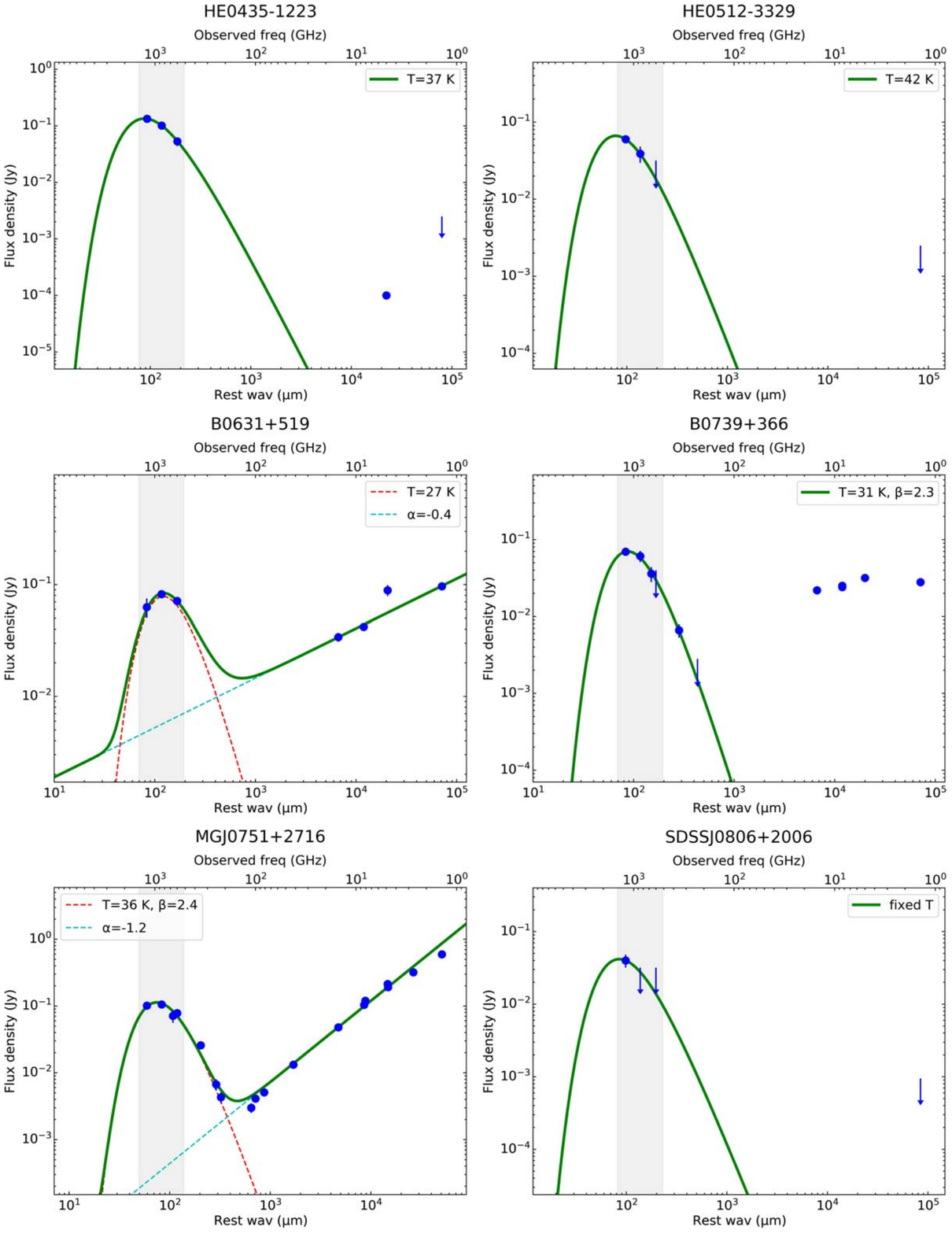}
\end{figure*}
\begin{figure*}
\ContinuedFloat
\contcaption{}
\includegraphics[width=\textwidth]{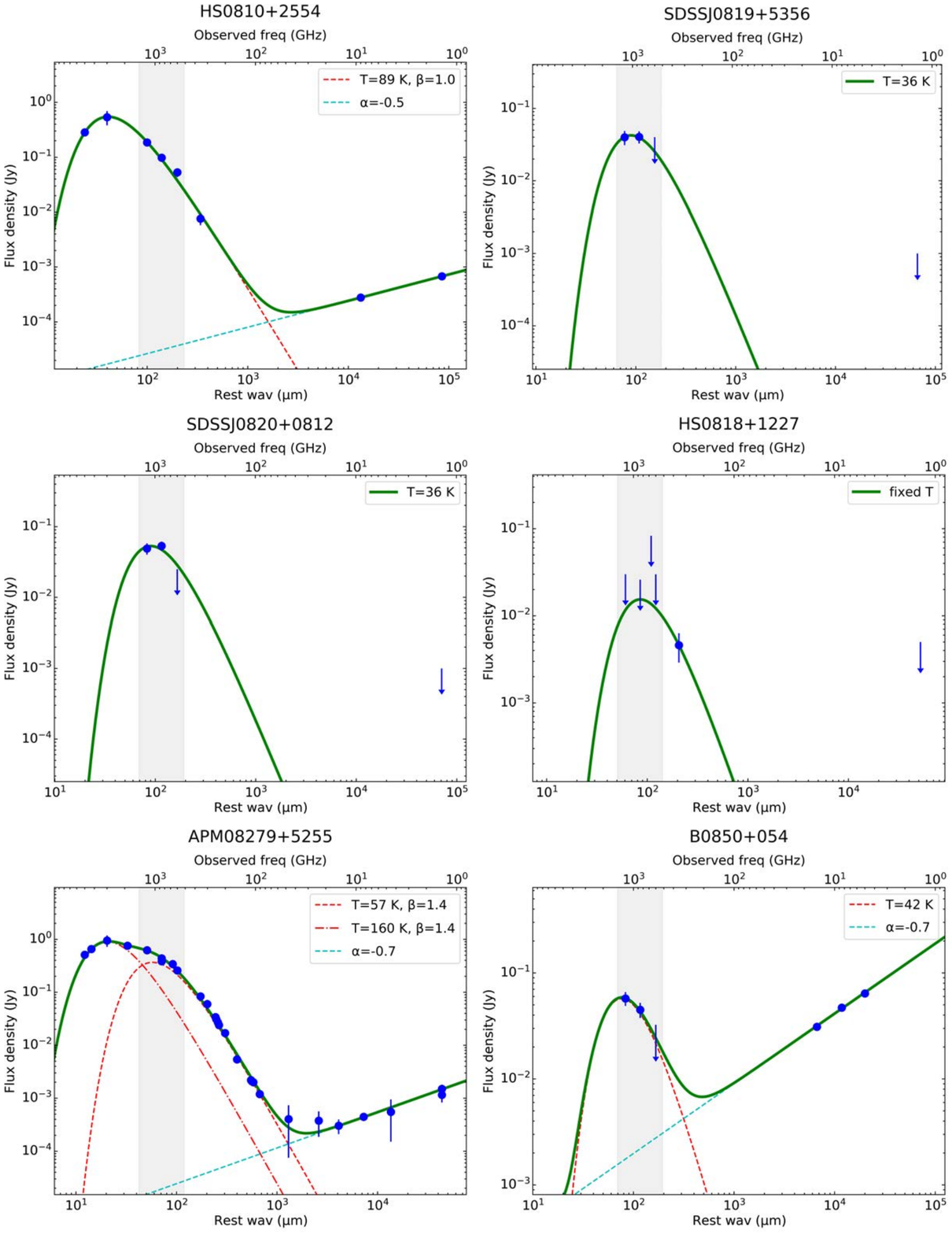}
\end{figure*}
\begin{figure*}
\ContinuedFloat
\contcaption{}
\includegraphics[width=\textwidth]{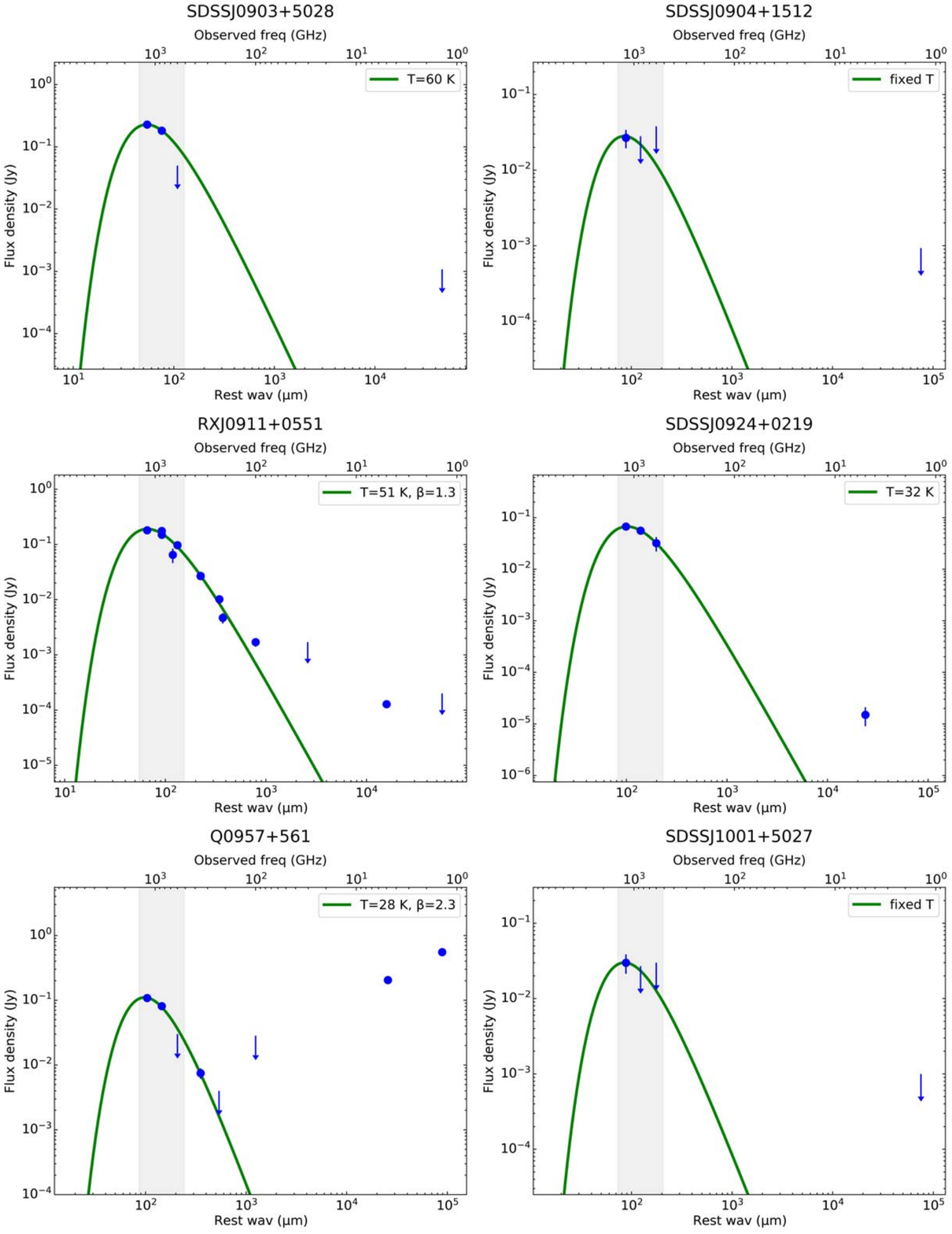}
\end{figure*}
\begin{figure*}
\ContinuedFloat
\contcaption{}
\includegraphics[width=\textwidth]{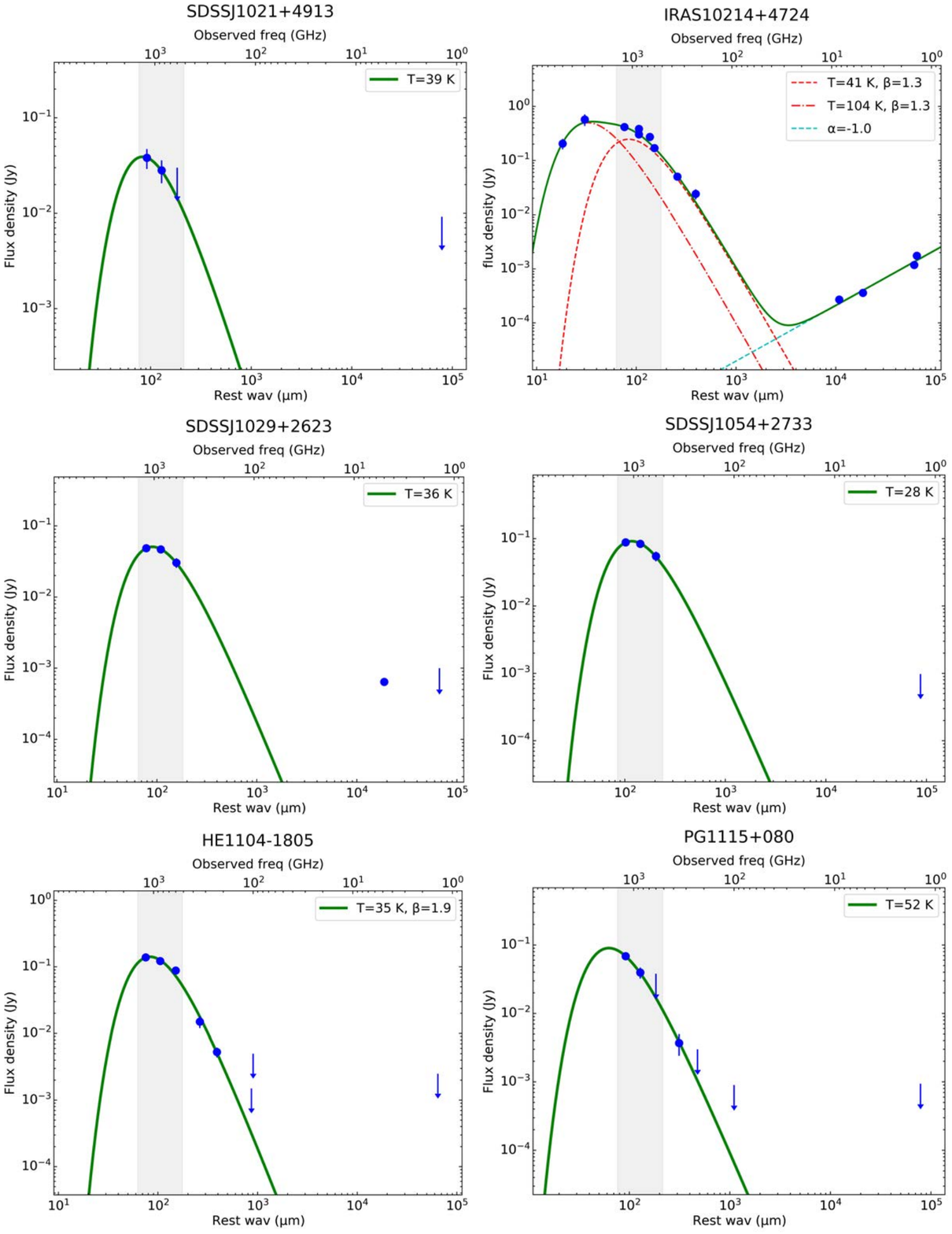}
\end{figure*}
\begin{figure*}
\ContinuedFloat
\contcaption{}
\includegraphics[width=\textwidth]{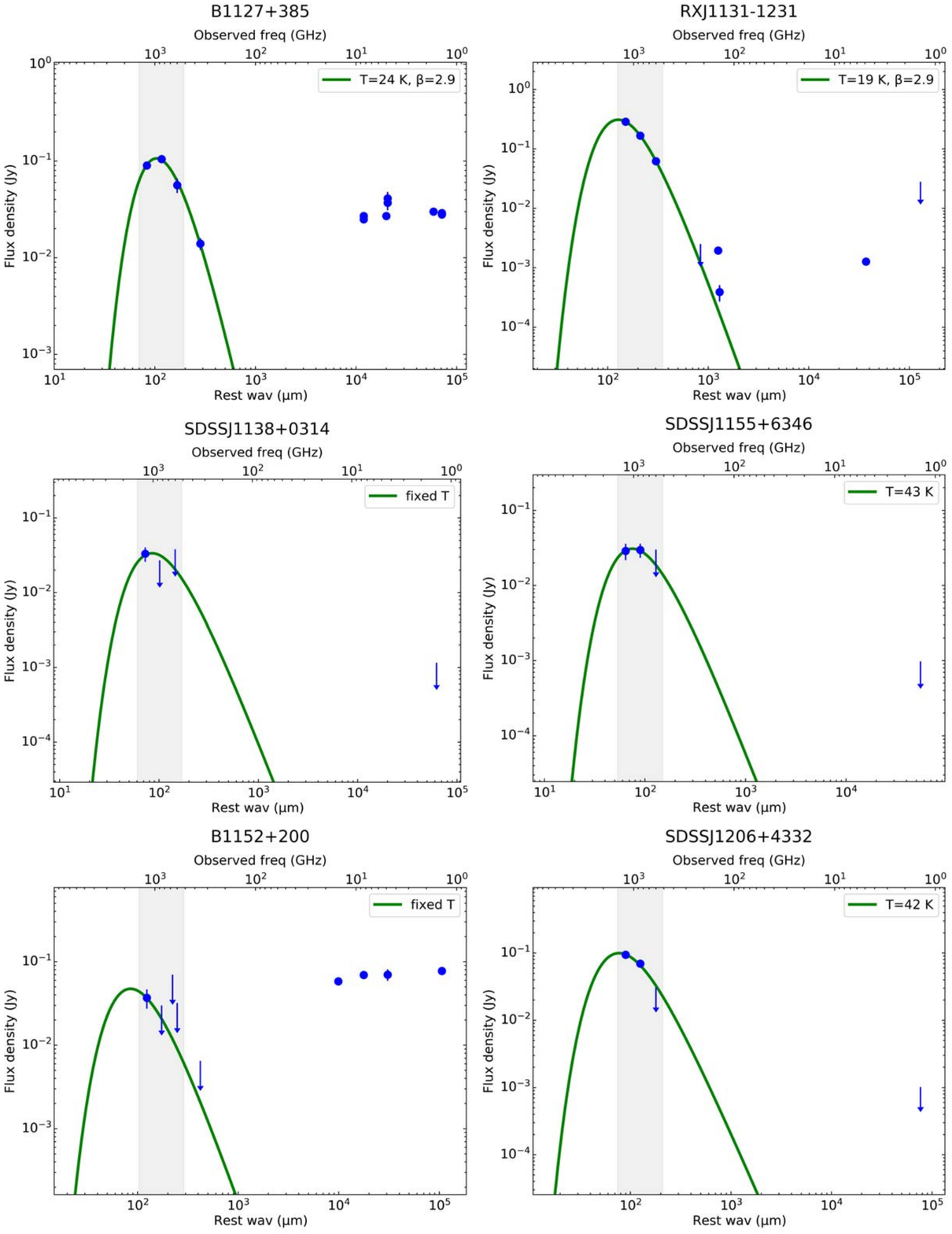}
\end{figure*}
\begin{figure*}
\ContinuedFloat
\contcaption{}
\includegraphics[width=\textwidth]{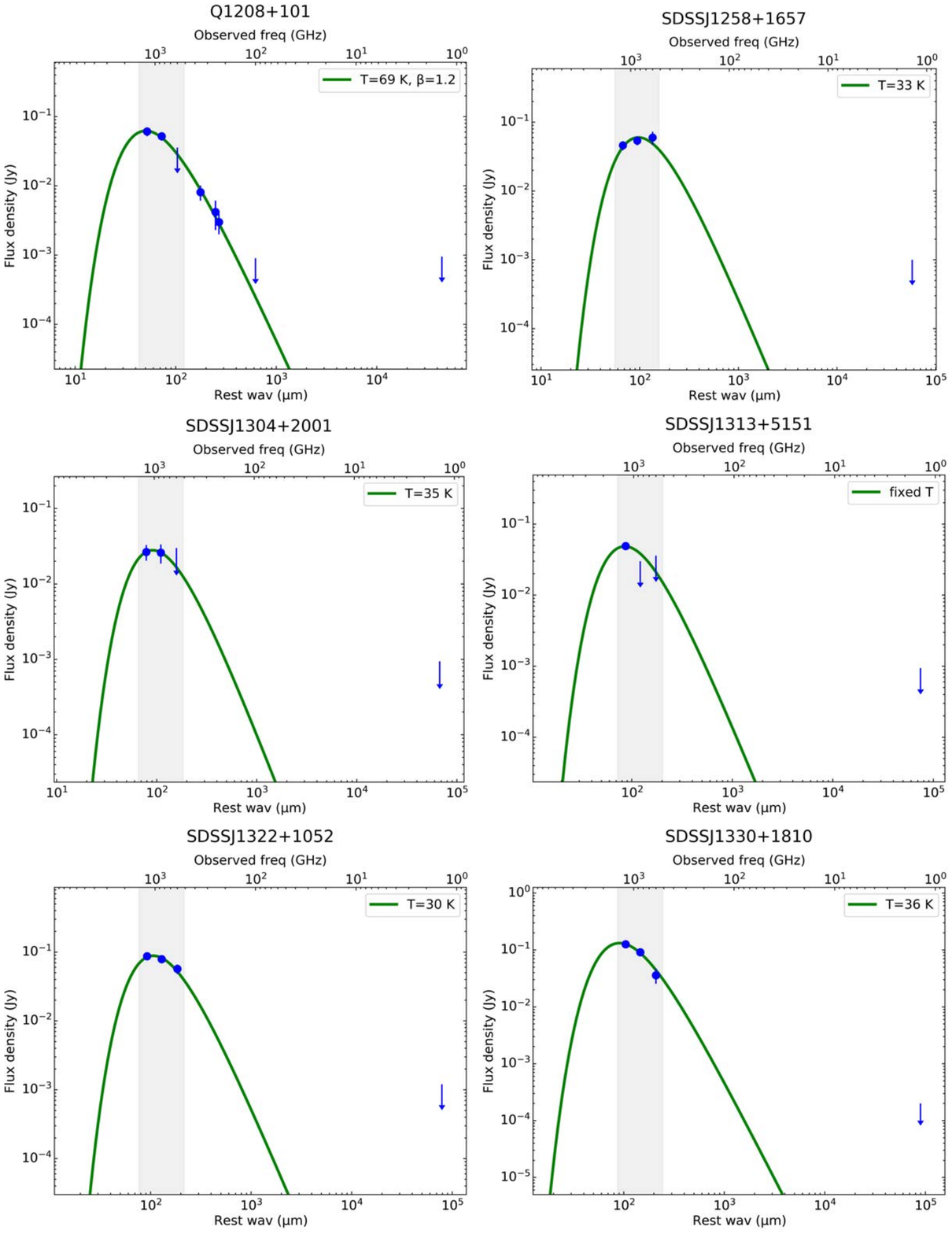}
\end{figure*}
\begin{figure*}
\ContinuedFloat
\contcaption{}
\includegraphics[width=\textwidth]{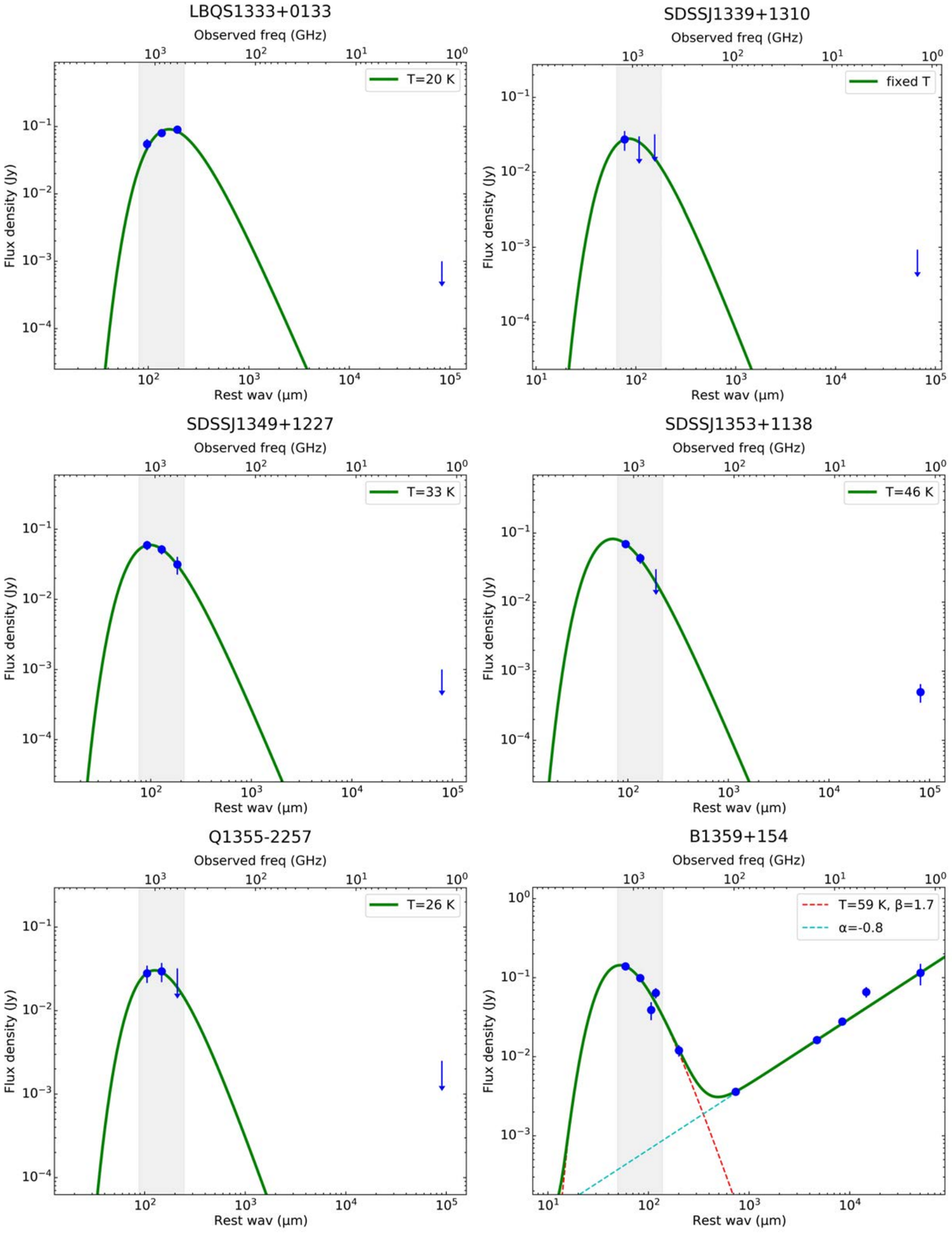}
\end{figure*}
\begin{figure*}
\ContinuedFloat
\contcaption{}
\includegraphics[width=\textwidth]{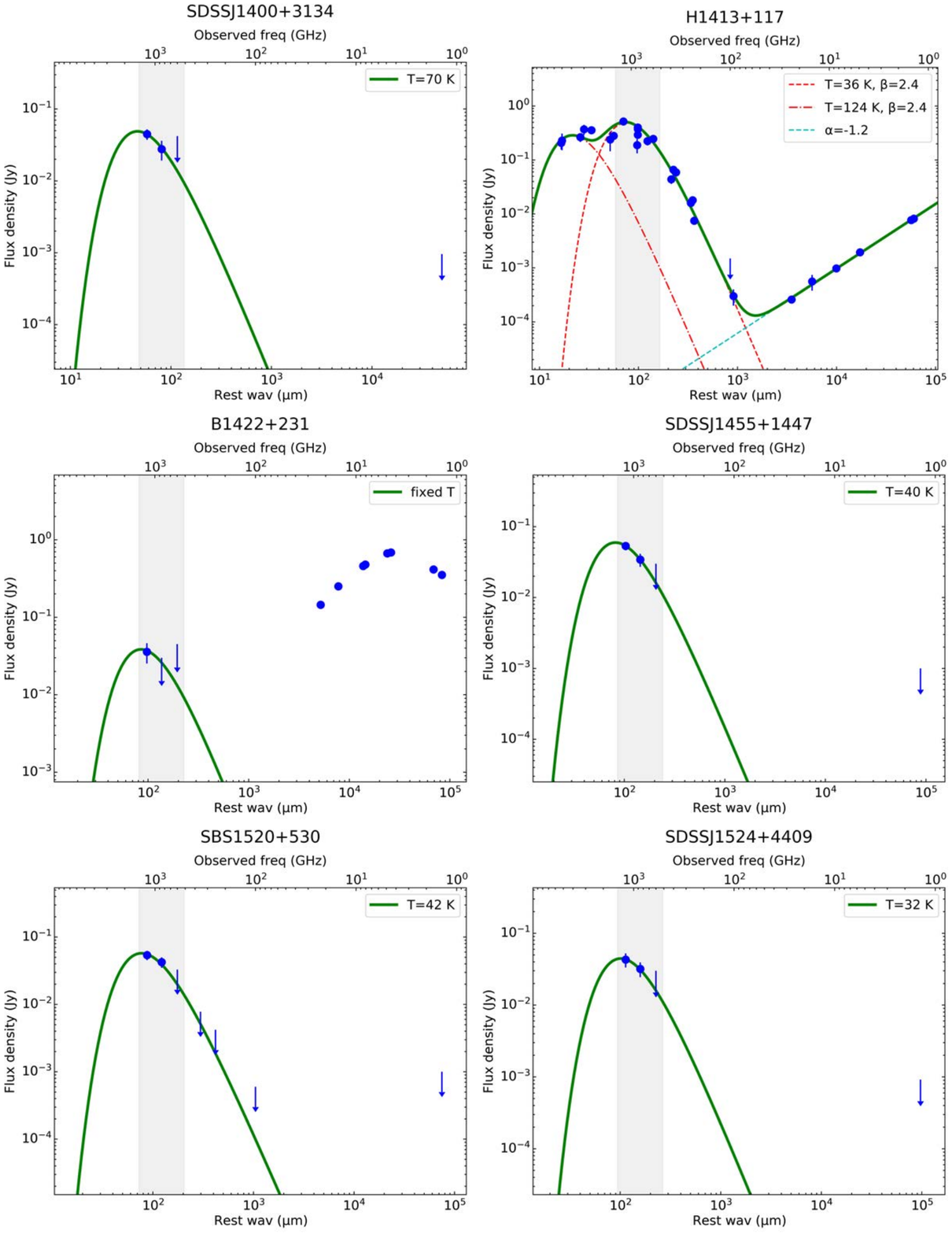}
\end{figure*}
\begin{figure*}
\ContinuedFloat
\contcaption{}
\includegraphics[width=\textwidth]{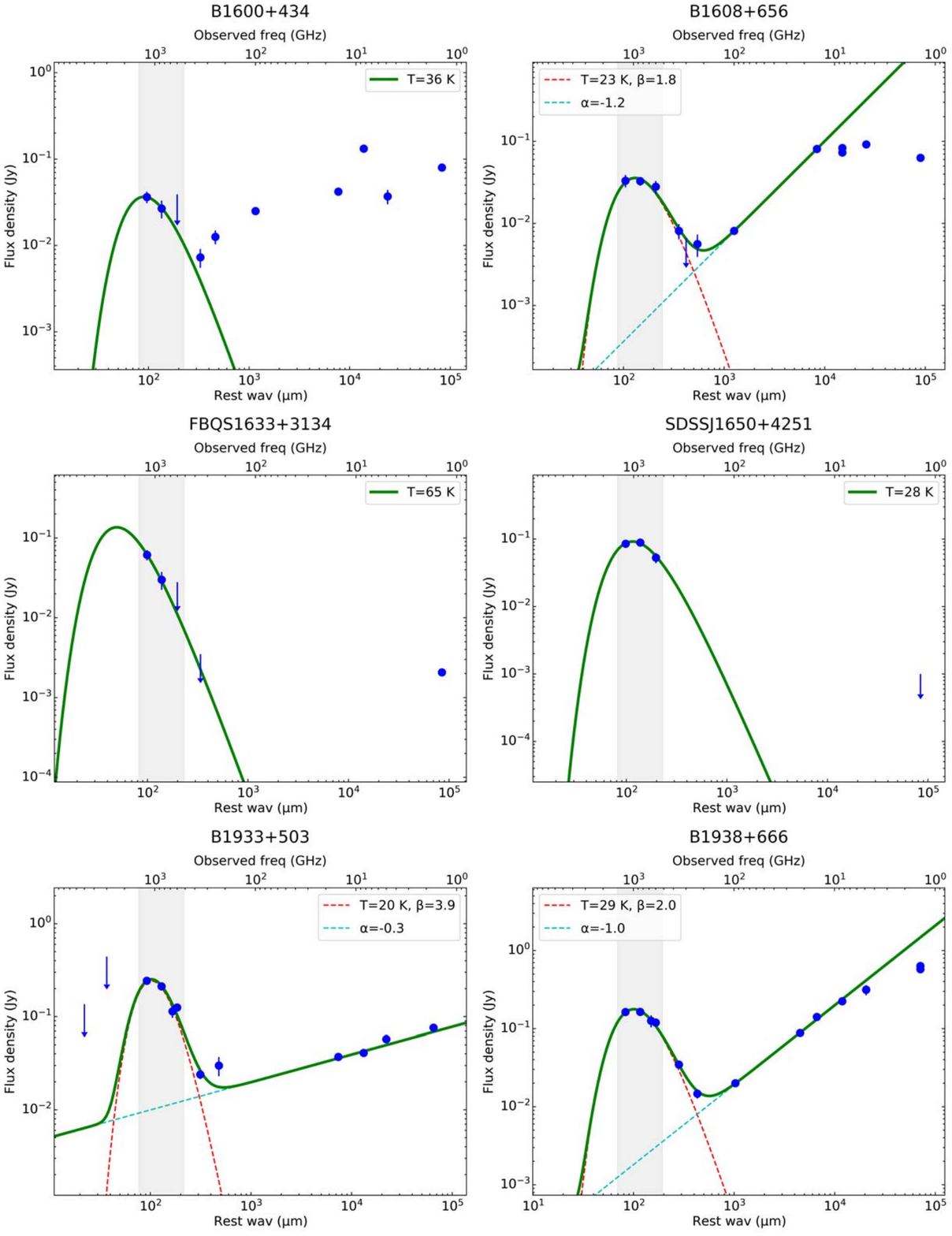}
\end{figure*}
\begin{figure*}
\ContinuedFloat
\contcaption{}
\includegraphics[width=\textwidth]{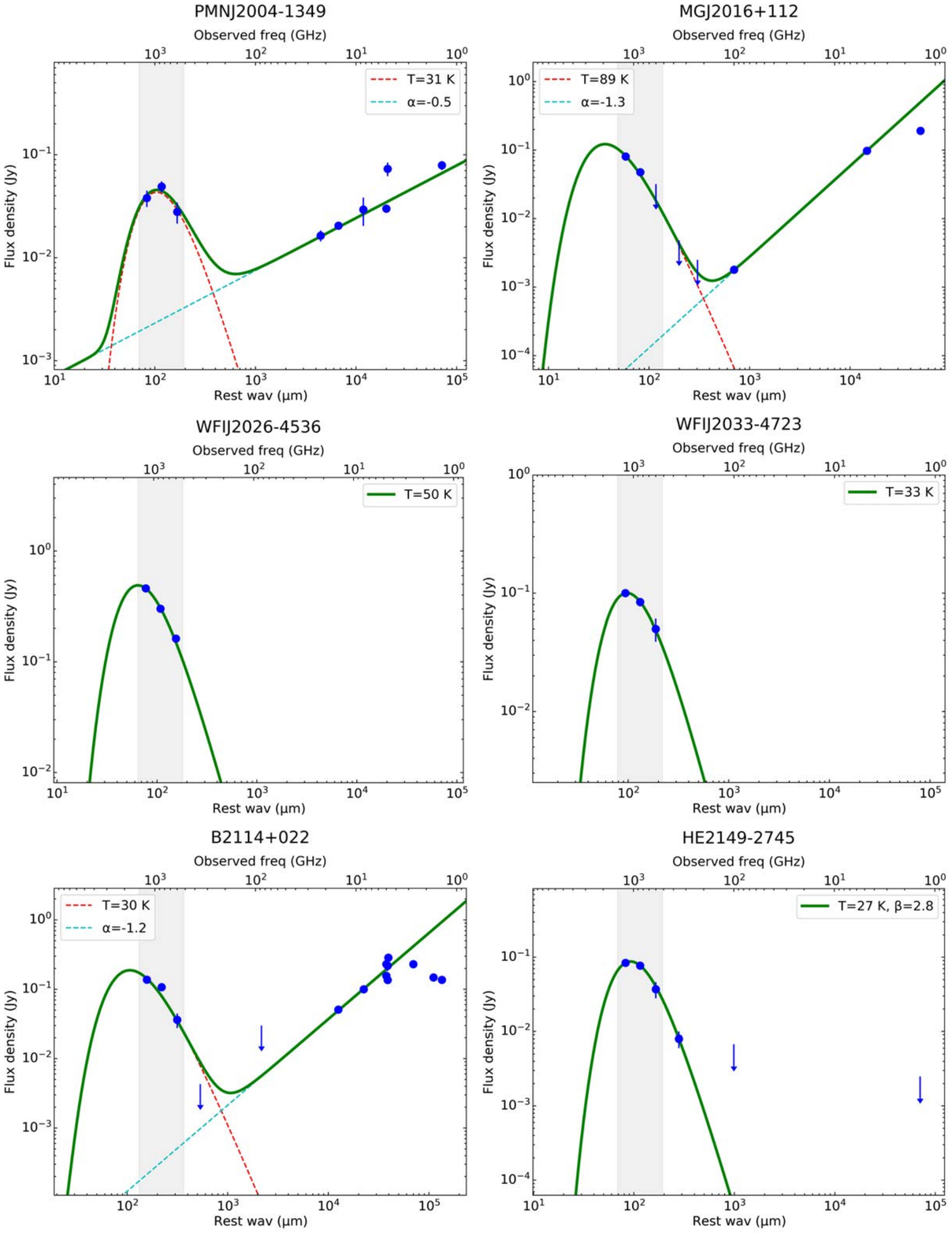}
\end{figure*}
\begin{figure*}
\ContinuedFloat
\contcaption{}
\includegraphics[width=\textwidth]{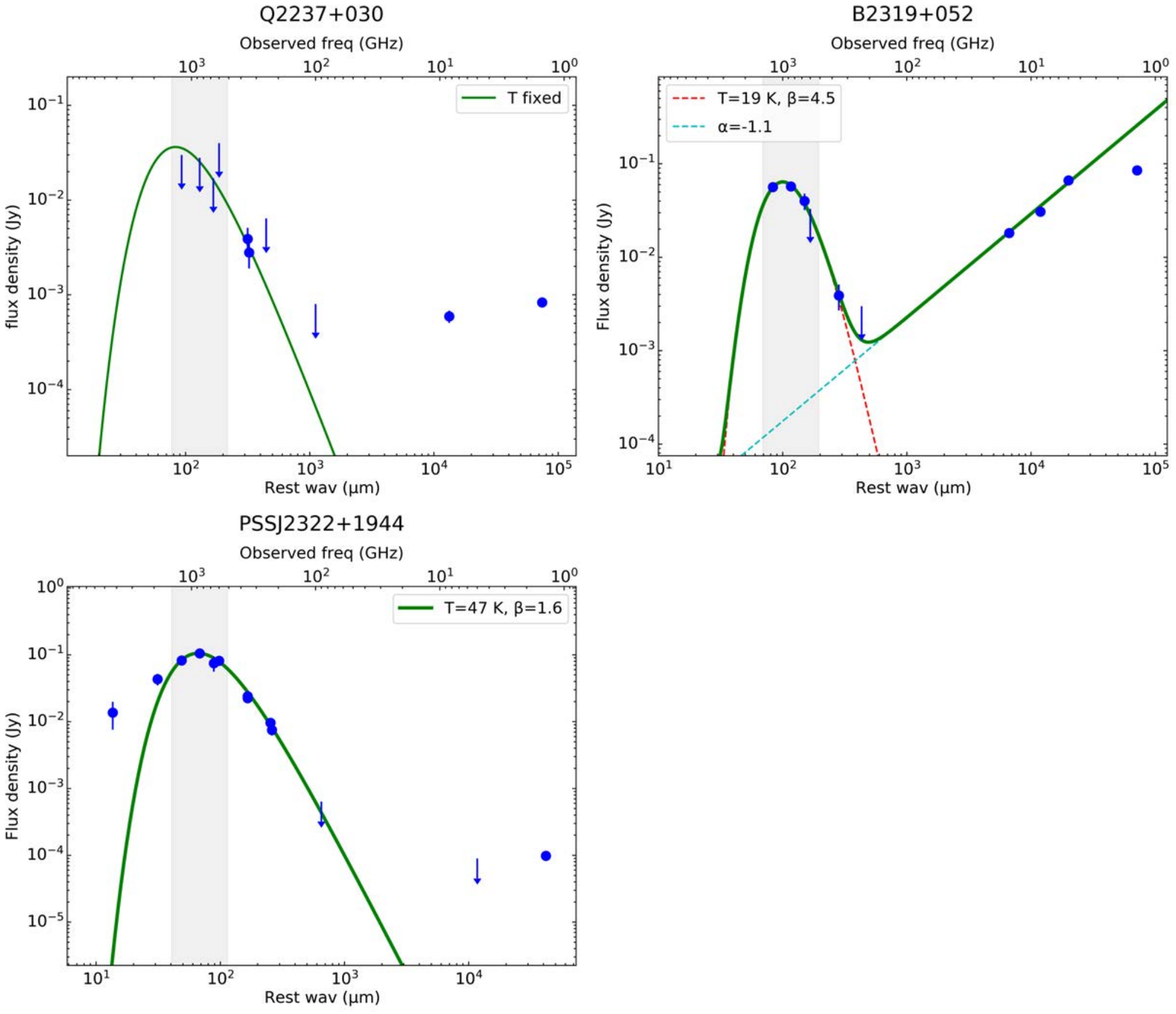}
\end{figure*}
\clearpage

\begin{figure*}
\caption{Included here are SEDs for 5 quasars which appear to have synchrotron-dominated emission in the FIR.}
\label{fig:syncseds}
\begin{subfigure}[t]{\textwidth}
\includegraphics[width=0.5\textwidth]{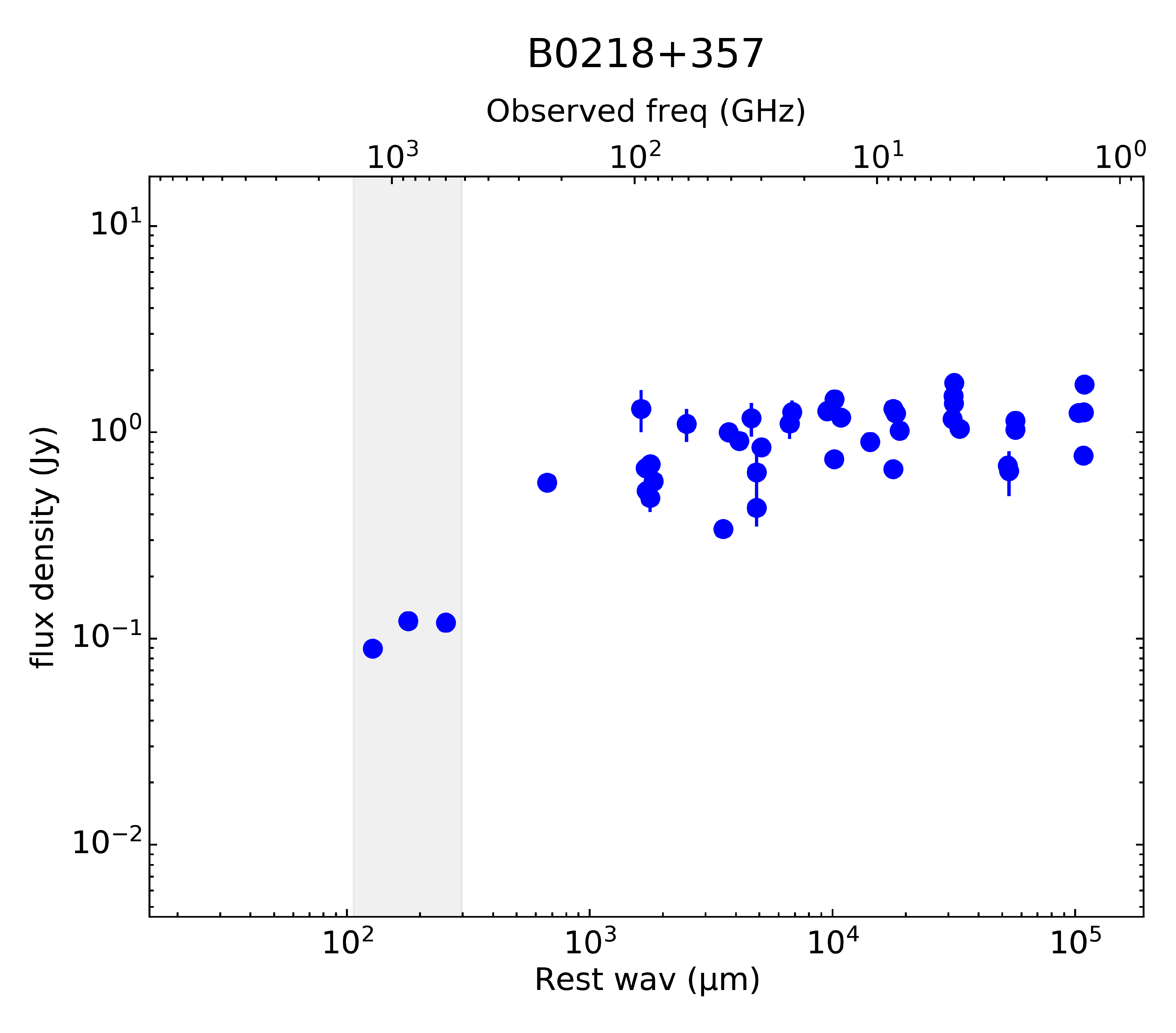}
\includegraphics[width=0.5\textwidth]{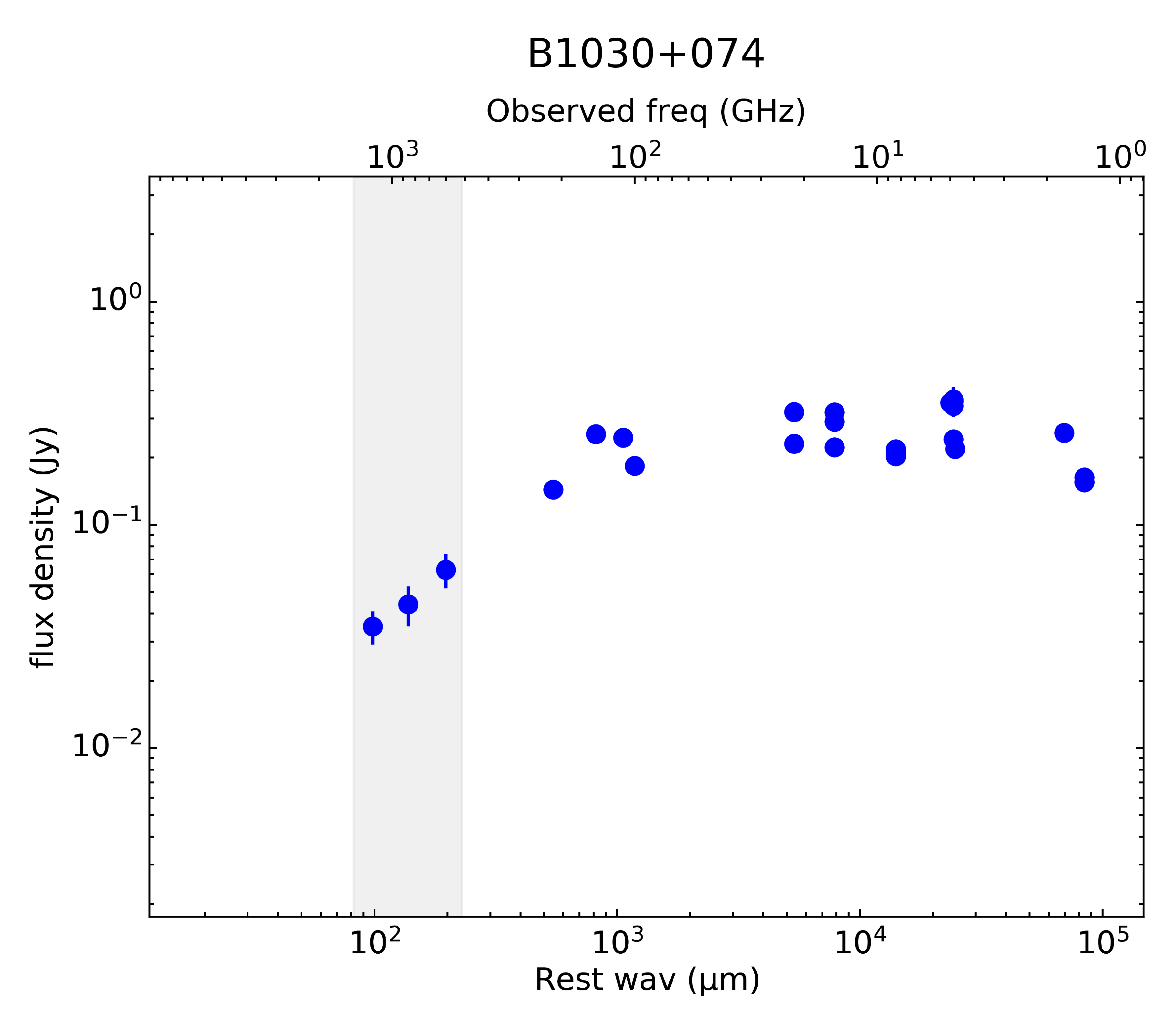}
\end{subfigure}\hfill
\begin{subfigure}[t]{\textwidth}
\includegraphics[width=0.5\textwidth]{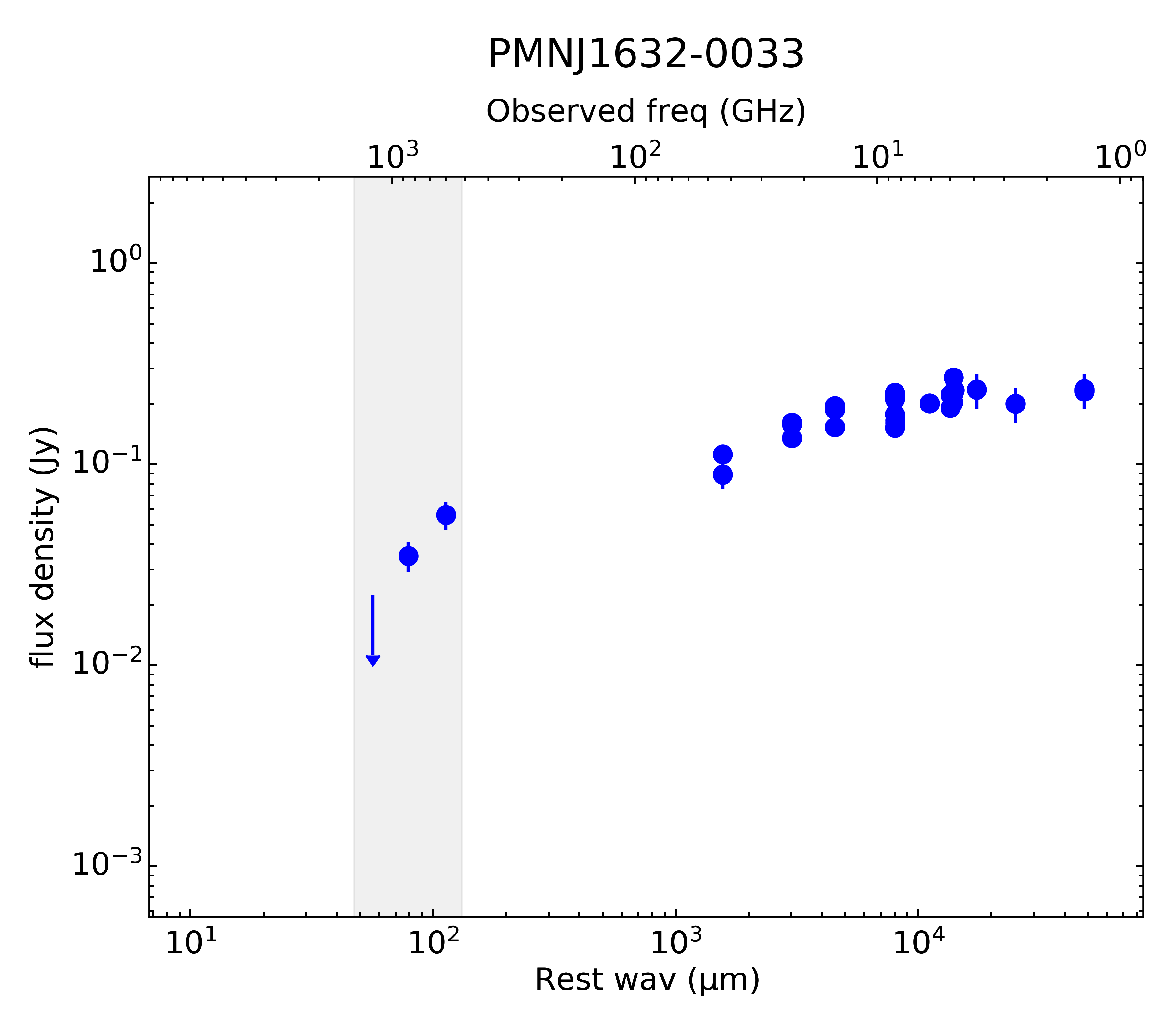}
\includegraphics[width=0.5\textwidth]{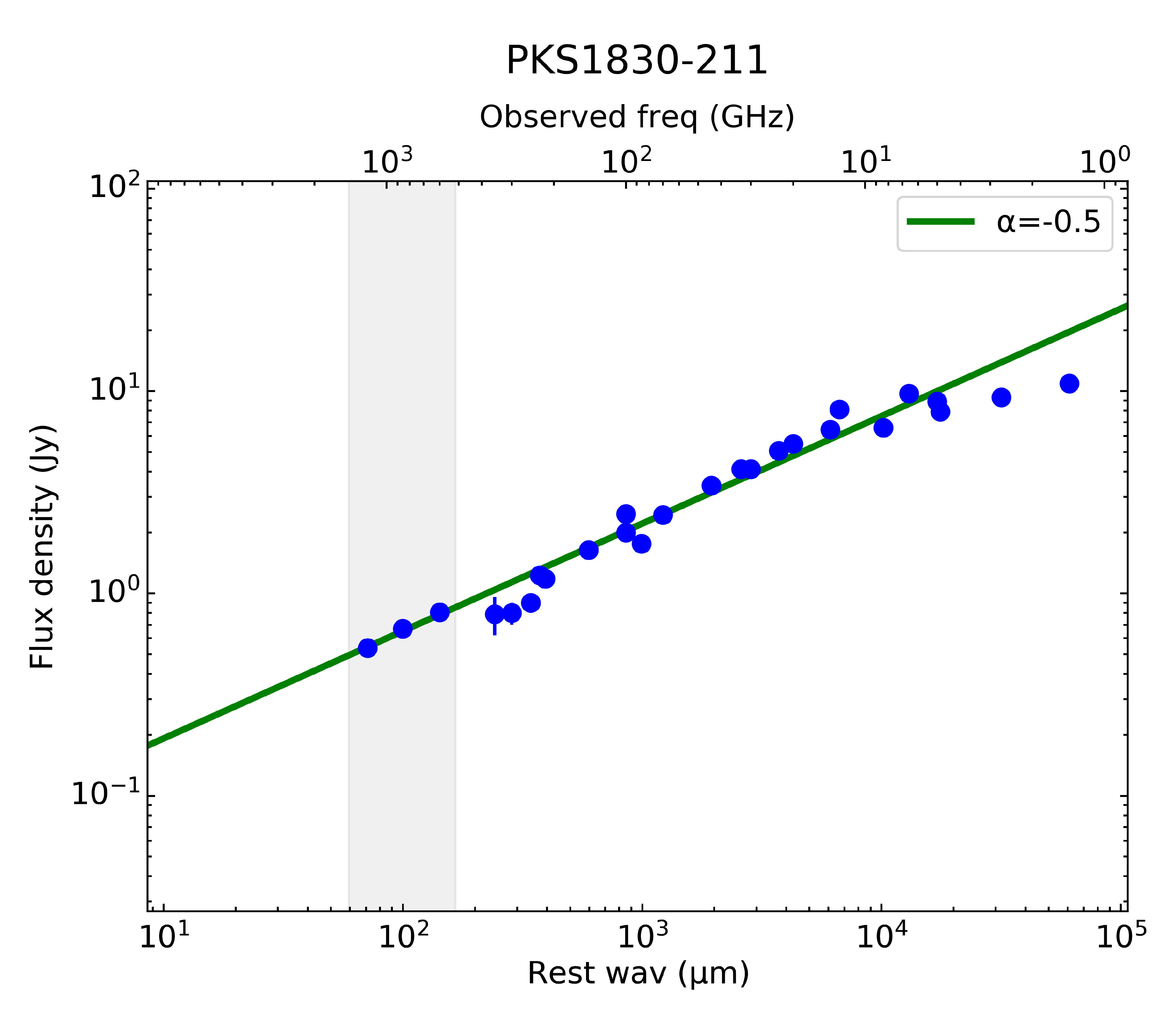}
\end{subfigure}\hfill
\begin{subfigure}[t]{\textwidth}
\includegraphics[width=0.5\textwidth]{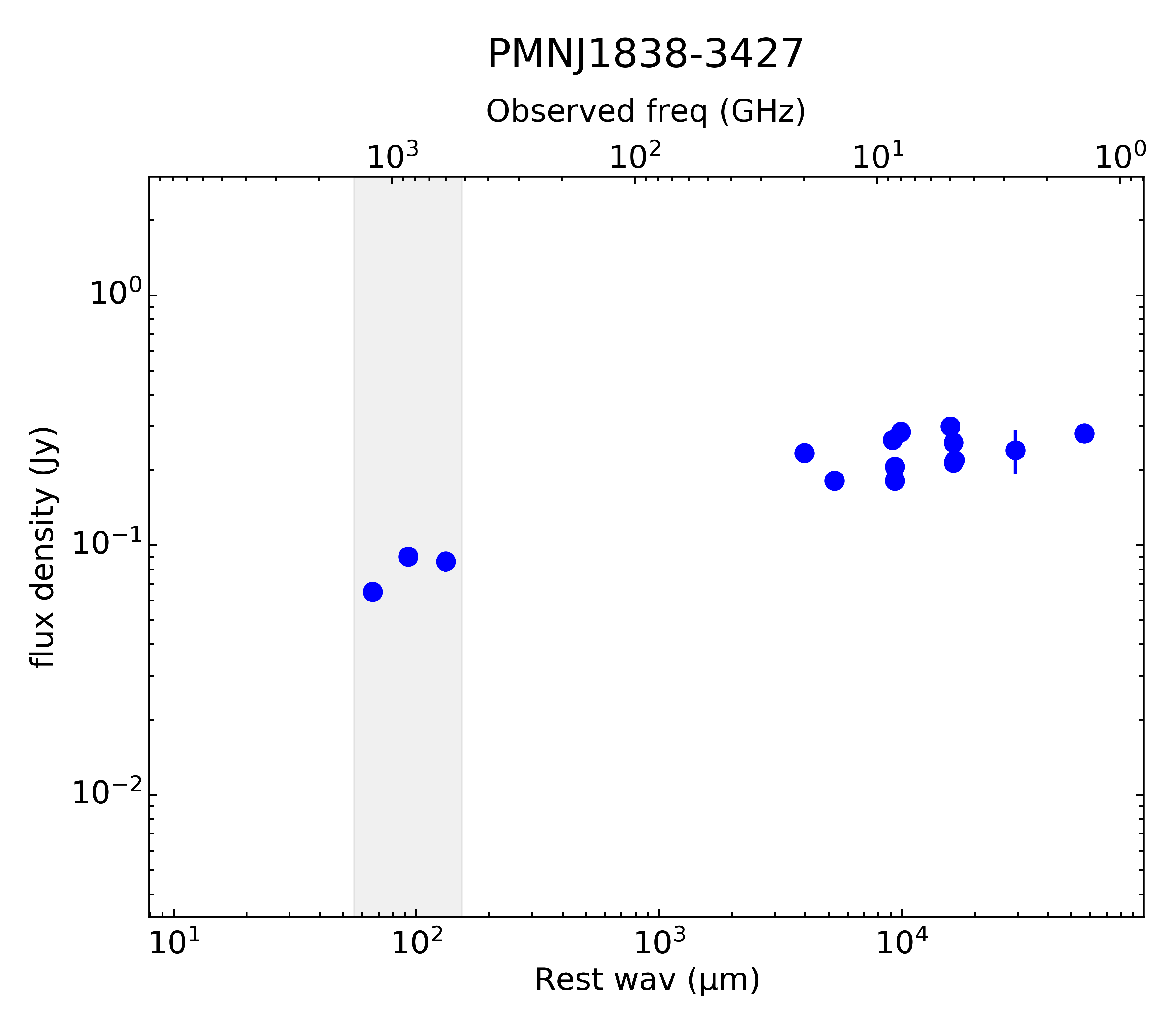}
\end{subfigure}\hfill
\end{figure*}

\clearpage
\twocolumn
\begin{table}
\bgroup
\def\arraystretch{1.}
\setlength{\tabcolsep}{1.2em}
\caption{Data from the literature shown in our SEDs in Fig.~\ref{figure:SEDs}. Data points in the FIR--sub-mm frequency range that have been excluded from our SED fitting are denoted by $\dagger$. Any radio data not given here are upper limits from FIRST \protect\cite[$<1$~mJy]{Becker:1995} or NVSS \protect\cite[$<2.5$~mJy]{Condon:1998} at 1.4~GHz. BI02 refers to \protect\cite{Barvainis:2002}. If errors are not given in the literature, we assume a flux calibration error of 10~percent.}
\label{table:auxdata}
\begin{tabular*}{0.48\textwidth}{l l l}
\hline \noalign {\smallskip}
$\nu$ (GHz) & $S_{\nu}$ (Jy) & Reference \\ 
\hline \noalign {\smallskip}
    \multicolumn{3}{c}{\it QJ0158-4325} \\ 
    \hline \noalign {\smallskip} 
8.46 & $<0.0002$ & \cite{Morgan:1999} \\ 
\hline \noalign {\smallskip}
    \multicolumn{3}{c}{\it B0218+357} \\ 
    \hline \noalign {\smallskip} 
229 & $0.57\pm0.03$ & \cite{Agudo:2014} \\ 
94 & $1.3\pm0.3$ & \cite{Wright:2009} \\
90 & $0.67\pm0.07$ & \cite{Kuhr:1981}\\
89.3 & $0.52\pm0.08$ & \cite{Jethava:2007} \\
86.4 & $0.48\pm0.07$ & \cite{Jethava:2007} \\
86.2 & $0.70\pm0.03$ & \cite{Agudo:2014} \\ 
83.6 & $0.58\pm0.09$ & \cite{Jethava:2007} \\
61.1 & $1.0\pm0.2$ & \cite{Wright:2009} \\
43.2 & $0.34\pm0.03$ & \cite{Jethava:2007} \\
41 & $1.0\pm0.1$ & \cite{Wright:2009} \\
37 & $0.91\pm0.09$ & \cite{Nieppola:2011} \\
33 & $1.170\pm0.216$ & \cite{Wright:2009} \\
31.4 & $0.64\pm0.18$ & \cite{Kuhr:1981} \\
31.4 & $0.43\pm0.08$ & \cite{Kuhr:1981} \\
30 & $0.846\pm0.043$ & \cite{Lowe:2007} \\
23 & $1.104\pm0.175$ & \cite{Wright:2009} \\
22.4 & $1.253\pm0.172$ & \cite{Patnaik:1992} \\ 
16.1 & $1.27\pm0.13$ & \cite{Davies:2009} \\ 
15.064 & $0.74\pm07$ & \cite{Kuhr:1981} \\
15.0 & $1.445\pm0.145$ & \cite{Richards:2011} \\
14.1 & $1.18\pm0.12$ & \cite{Jethava:2007} \\
10.695 & $0.90\pm0.04$ & \cite{Kuhr:1981} \\
8.6 & $1.30\pm0.13$ & \cite{Jethava:2007} \\
8.6 & $0.663\pm0.022$ & \cite{Zeiger:2010} \\
8.4 & $1.236\pm0.046$ & \cite{Patnaik:1992} \\
8.085 & $1.02\pm0.050$ & \cite{Kuhr:1981} \\
4.9 & $1.16\pm0.01$ & \cite{Kuhr:1981} \\
4.85 & $1.498\pm0.188$ & \cite{Gregory:1991} \\
4.83 & $1.74\pm0.17$ & \cite{Griffith:1990} \\
4.585 & $1.04\pm0.05$ & \cite{Kuhr:1981} \\
4.84 & $1.383\pm0.083$ & \cite{Patnaik:1992} \\
2.9 & $0.69\pm0.06$ & \cite{Jethava:2007} \\
2.867 & $0.650\pm0.16$ & \cite{Zeiger:2010} \\
2.695 & $1.03\pm0.02$ & \cite{Kuhr:1981} \\
2.695 & $1.14\pm0.06$ & \cite{Kuhr:1981} \\
1.415 & $0.770\pm0.077$ & \cite{Patnaik:1992} \\
1.41 & $1.25\pm0.01$ & \cite{Kuhr:1981} \\
1.41 & $1.24\pm0.06$ & \cite{Kuhr:1981} \\
\hline \noalign {\smallskip} 
\multicolumn{3}{c}{\it HE0230-2130} \\ 
\hline \noalign {\smallskip} 
667.0 & $0.077\pm0.013$ & BI02 \\ 
353.0 & $0.0210\pm0.0017$ & BI02 \\ 
109.0 & $<0.0022$ & \cite{Riechers:2011b} \\ 
1.4 & $<0.0025$ & NVSS \\ 
\hline \noalign {\smallskip}
\multicolumn{3}{c}{\it CFRS03.1077} \\ 
\hline \noalign {\smallskip} 
1.4 & $0.00038\pm0.00012$ & FIRST \\ 
\hline
\end{tabular*}
\egroup
\end{table}
\begin{table}
\bgroup
\ContinuedFloat
\def\arraystretch{1.}
\setlength{\tabcolsep}{1.2em}
\contcaption{}
\begin{tabular*}{0.5\textwidth}{l l l}
\hline \noalign {\smallskip}
$\nu$ (GHz) & $S_{\nu}$ (Jy) & Reference \\
\hline \noalign {\smallskip} 
    \multicolumn{3}{c}{\it B0414+054} \\ 
    \hline \noalign {\smallskip} 
5000 & $0.18\pm0.04$ & \cite{Lawrence:1995} \\ 
3000 & $<0.786$ & \cite{Lawrence:1995} \\ 
667 & $0.066\pm0.016^{\dagger}$ & BI02 \\
353 & $0.0253\pm0.0018$ & BI02 \\ 
231 & $0.0207\pm0.0013$ & BI02 \\ 
100 & $0.040\pm0.002$ & BI02 \\ 
15.0 & $0.381\pm0.006$ & \cite{Hewitt:1992} \\ 
10.7 & $0.38\pm0.02$ & \cite{Kellerman:1973} \\ 
8.6 & $0.340\pm0.034$ & \cite{Tingay:2003} \\ 
6.1 & $0.71\pm0.02$ & \cite{Castangia:2011} \\ 
5.0 & $0.864\pm0.015$ & \cite{Hewitt:1992} \\ 
5.0 & $0.710\pm0.071$ & \cite{Wright:1990} \\ 
4.8 & $1.110\pm0.154$ & \cite{Gregory:1991} \\ 
4.8 & $0.959\pm0.051$ & \cite{Griffith:1995} \\ 
2.7 & $1.130\pm0.113$ & \cite{Wright:1990} \\ 
2.5 & $1.200\pm0.012$ & \cite{Tingay:2003} \\ 
1.4 & $2.087\pm0.074$ & NVSS \\ 
\hline \noalign {\smallskip} 
    \multicolumn{3}{c}{\it HE0435-1223} \\ 
    \hline \noalign {\smallskip} 
5.0 & $1.13\pm0.04\times10^{-4}$ & \cite{Jackson:2015} \\ 
\hline \noalign {\smallskip}
    \multicolumn{3}{c}{\it B0631+519} \\ 
    \hline \noalign {\smallskip} 
15.0 & $0.034\pm0.003$ & \cite{York:2005} \\ 
8.4 & $0.042\pm0.002$ & \cite{York:2005} \\ 
4.8 & $0.089\pm0.010$ & \cite{Gregory:1991} \\ 
1.4 & $0.0966\pm0.0029$ & NVSS \\ 
\hline \noalign {\smallskip} 
    \multicolumn{3}{c}{\it B0739+366} \\ 
    \hline \noalign {\smallskip} 
667 & $0.036\pm0.008$ & BI02 \\ 
353 & $0.0066\pm0.0013$ & BI02 \\ 
231 & $<0.0028$ & BI02 \\ 
15 & $0.0219\pm0.0019$ & \cite{Marlow:2001} \\ 
8.4 & $0.0251\pm0.0022$ & \hspace{5mm} " \\ 
8.4 & $0.0241\pm0.0021$ & \hspace{5mm} " \\ 
5.0 & $0.0317\pm0.0027$ & \hspace{5mm} " \\ 
1.4 & $0.0279\pm0.0010$ & NVSS \\ 
\hline \noalign {\smallskip} 
    \multicolumn{3}{c}{\it MGJ0751+2716} \\ 
    \hline \noalign {\smallskip} 
667 & $0.071\pm0.015$ & BI02 \\ 
353 & $0.0258\pm0.0030$ & BI02 \\ 
246 & $0.0067\pm0.0013$ & \cite{Barvainis:2002b} \\ 
220 & $0.0043\pm0.0008$ & \cite{Alloin:2007} \\ 
110 & $0.0030\pm0.0005$ & \cite{Alloin:2007} \\ 
100 & $0.0041\pm0.0005$ & BI02 \\ 
82 & $0.0051\pm0.0004$ & \cite{Alloin:2007} \\ 
42 & $0.0132\pm0.0010$ & \cite{Carilli:2005} \\ 
15 & $0.048\pm0.004$ & \cite{Lehar:1997} \\ 
8.1 & $0.12\pm0.01$ & \cite{Condon:1983} \\ 
8.3 & $0.104\pm0.001$ & \cite{Lehar:1997} \\ 
4.8 & $0.191\pm0.001$ & \cite{Lehar:1997} \\ 
4.8 & $0.214\pm0.028$ & \cite{Gregory:1991} \\ 
2.7 & $0.32\pm0.02$ & \cite{Condon:1983} \\
1.4 & $0.595\pm0.018$ & NVSS \\
\hline \noalign {\smallskip}
    \multicolumn{3}{c}{\it HS0810+2554} \\ 
    \hline \noalign {\smallskip} 
5000.0 & $0.284\pm0.040$ & \cite{Moshir:1990} \\ 
3000.0 & $0.538\pm0.156$ & \cite{Moshir:1990} \\ 
353.0 & $0.0076\pm0.0018$ & \cite{Priddey:2003} \\ 
9.0 & $0.000278\pm0.000007$ & \cite{Jackson:2015} \\ 
1.4 & $0.000679\pm0.000058$ & \cite{Jackson:2015} \\ 
\hline
\end{tabular*}
\egroup
\end{table}
\begin{table}
\bgroup
\ContinuedFloat
\def\arraystretch{1.}
\setlength{\tabcolsep}{1.2em}
\contcaption{}
\begin{tabular*}{0.5\textwidth}{l l l}
\hline \noalign {\smallskip}
$\nu$ (GHz) & $S_{\nu}$ (Jy) & Reference \\
\hline \noalign {\smallskip} 
    \multicolumn{3}{c}{\it HS0818+1227} \\ 
    \hline \noalign {\smallskip} 
667.0 & $<0.083$ & BI02 \\ 
353.0 & $0.0046\pm0.0017$ & BI02 \\ 
1.4 & $<0.001$ & FIRST \\ 
\hline \noalign {\smallskip} 
    \multicolumn{3}{c}{\it APM08279+5255} \\ 
    \hline \noalign {\smallskip} 
5000 & $0.511\pm0.051$ & \cite{Irwin:98} \\ 
4300 & $0.654\pm0.009$ & PACS catalog \\ 
3000 & $0.951\pm0.228$ & \cite{Irwin:98} \\ 
1875 & $0.759\pm0.010$ & PACS catalog \\ 
857 & $0.386\pm0.032$ & \cite{Beelen:2006} \\ 
667 & $0.342\pm0.026$ & \cite{Beelen:2006} \\ 
353 & $0.084\pm0.003$ & \cite{Lewis:1998} \\ 
302 & $0.060\pm0.012$ & \cite{Krips:2007} \\ 
250 & $0.034\pm0.001$ & \cite{Lis:2011} \\ 
246 & $0.031\pm0.002$ & \cite{vanderWerf:2011} \\ 
237 & $0.027\pm0.001$ & \cite{vanderWerf:2011} \\ 
231 & $0.024\pm0.002$ & \cite{Lewis:1998} \\ 
201 & $0.017\pm0.001$ & \cite{vanderWerf:2011} \\ 
153 & $0.0054\pm0.0003$ & \cite{vanderWerf:2011} \\ 
111 & $0.0022\pm0.0002$ & \cite{Riechers:2010} \\ 
109 & $0.0021\pm0.0002$ & \hspace{5mm} " \\ 
105 & $0.0020\pm0.0001$ & \hspace{5mm} " \\ 
90.8 & $0.00120\pm0.00013$ & \cite{Garcia-Burillo:2006} \\ 
46.9 & $0.000405\pm0.000330$ & \cite{Riechers:2009} \\ 
23.5 & $0.000376\pm0.000190$ & \hspace{5mm} " \\ 
14.9 & $0.000303\pm0.000093$ & \hspace{5mm} " \\ 
8.4 & $0.000446\pm0.000020$ & \hspace{5mm} " \\ 
4.5 & $0.000551\pm0.000400$ & \hspace{5mm} " \\ 
1.4 & $0.00116\pm0.00033$ & \hspace{5mm} " \\ 
1.4 & $0.0015\pm0.0003$ & NVSS \\ 
\hline \noalign {\smallskip} 
    \multicolumn{3}{c}{\it B0850+054} \\ 
    \hline \noalign {\smallskip} 
15.0 & $0.031\pm0.001$ & \cite{Biggs:2003} \\ 
8.5 & $0.047\pm0.001$ & \hspace{5mm} " \\ 
5.0 & $0.064\pm0.002$ & \hspace{5mm} " \\ 
    \hline \noalign {\smallskip} 
    \multicolumn{3}{c}{\it RXJ0911+0551} \\ 
    \hline \noalign {\smallskip} 
857 & $0.150\pm0.021$ & \cite{Wu:2009} \\ 
667 & $0.065\pm0.019$ & BI02 \\ 
353 & $0.027\pm0.004$ & \cite{Hainline:2004} \\ 
353 & $0.0267\pm0.0014$ & BI02 \\ 
230 & $0.0102\pm0.0018$ & BI02 \\ 
212.5 & $0.0047\pm0.0010$ & \cite{Tuan-Anh:2013} \\ 
100.0 & $0.0017\pm0.0003$ & BI02\\ 
30.4 & $<0.0017$ & \cite{Riechers:2011} \\ 
5.0 & $1.28(\pm0.05)\times10^{-4}$ & \cite{Jackson:2015} \\ 
1.4 & $<0.0002$ & \cite{Jackson:2015} \\ 
\hline \noalign {\smallskip} 
\multicolumn{3}{c}{\it SDSSJ0924+0219} \\ 
\hline \noalign {\smallskip} 
5.0 & $1.5(\pm0.4)\times10^{-5}$ & \cite{Jackson:2015} \\
\hline \noalign {\smallskip} 
\multicolumn{3}{c}{\it Q0957+561} \\ 
\hline \noalign {\smallskip} 
353 & $0.0075\pm0.0014$ & BI02 \\ 
231 & $<0.004$ & BI02 \\ 
100 & $<0.0284$ & \cite{Planesas:1999} \\ 
4.8 & $0.205\pm0.022$ & \cite{Gregory:1991} \\ 
1.4 & $0.552\pm0.017$ & NVSS \\ 
\hline
\end{tabular*}
\egroup
\end{table}
\begin{table}
\bgroup
\ContinuedFloat
\def\arraystretch{1.}
\setlength{\tabcolsep}{1.2em}
\caption{cont.}
\begin{tabular*}{0.5\textwidth}{l l l}
\hline \noalign {\smallskip}
$\nu$ (GHz) & $S_{\nu}$ (Jy) & Reference \\
\hline \noalign {\smallskip}
    \multicolumn{3}{c}{\it IRAS~F10214+4724} \\ 
    \hline \noalign {\smallskip} 
5000 & $0.2\pm0.045$ & \cite{Moshir:1990} \\ 
3000 & $0.57\pm0.14$ & \cite{Moshir:1990} \\ 
857 & $0.38\pm0.05$ & \cite{Benford:1999} \\ 
667 & $0.27\pm0.05$ & \cite{Rowan-Robinson:1993} \\ 
375 & $0.050\pm0.005$ & \hspace{5mm} " \\ 
272 & $0.024\pm0.005$ & \hspace{5mm} " \\
8.4 & $0.00027\pm0.00005$ & \cite{Lawrence:1993} \\
4.86 & $0.00036\pm0.00006$ & \hspace{5mm} " \\
1.49 & $0.00118\pm0.0001$ & \hspace{5mm} " \\
1.4 & $0.0018\pm0.0001$ & FIRST \\ 
\hline \noalign {\smallskip} 
\multicolumn{3}{c}{\it SDSSJ1029+2623} \\ 
\hline \noalign {\smallskip} 
5.0 & $6.43(\pm0.23)\times10^{-4}$ & \cite{Kratzer:2011} \\ 
\hline \noalign {\smallskip} 
\multicolumn{3}{c}{\it CLASS~B1030+074} \\ 
\hline \noalign {\smallskip} 
216 & $0.114\pm0.010$ & \cite{Xanthopoulos:2001} \\ 
144 & $0.225\pm0.012$ & \hspace{5mm} " \\ 
112 & $0.246\pm0.012$ & \hspace{5mm} " \\ 
100 & $0.184\pm0.002$ & \cite{Barvainis:2002b} \\ 
22 & $0.320\pm0.030$ & \cite{Lee:2017} \\ 
22 & $0.231\pm0.012$ & \cite{Xanthopoulos:1998} \\ 
15 & $0.319\pm0.016$ & \hspace{5mm} " \\   
15 & $0.223\pm0.015$ & \hspace{5mm} " \\
15 & $0.290\pm0.004$ & \cite{Richards:2011} \\
8.4 & $0.210\pm0.011$ & \cite{Xanthopoulos:1998} \\ 
8.4 & $0.218\pm0.011$ & \cite{Xanthopoulos:1998} \\
8.4 & $0.203\pm0.0003$ & \cite{Xanthopoulos:2001} \\
5.0 & $0.353\pm0.018$ &  \cite{Xanthopoulos:1998} \\  
4.85 & $0.364\pm0.051$ & \cite{Gregory:1991} \\ 
4.85 & $0.356\pm0.052$ & \cite{Becker:1991} \\ 
4.85 & $0.242\pm0.016$ & \cite{Griffith:1995} \\
4.85 & $0.341\pm0.005$ & \cite{Xanthopoulos:2001} \\
4.775 & $0.219\pm0.20$ & \cite{Bennett:1986} \\
1.7 & $0.258\pm0.013$ & \cite{Xanthopoulos:1998} \\ 
1.4 & $0.163\pm0.004$ & \cite{White:1992} \\ 
1.4 & $0.155\pm0.004$ & NVSS \\ 
1.4 & $0.156\pm0.005$ & \cite{Xanthopoulos:2001}\\
\hline \noalign {\smallskip}
    \multicolumn{3}{c}{\it HE1104-1805} \\ 
    \hline \noalign {\smallskip} 
353 & $0.015\pm0.003$ & BI02 \\ 
231 & $0.0053\pm0.0009$ & BI02 \\ 
100 & $<0.0022$ & BI02 \\ 
1.4 & $<0.0025$ & NVSS \\ 
\hline \noalign {\smallskip} 
\multicolumn{3}{c}{\it PG1115+080} \\ 
\hline \noalign {\smallskip} 
353 & $0.0037\pm0.0013$ & BI02 \\ 
231 & $<0.003$ & BI02 \\ 
104 & $<0.0015$ & \cite{Riechers:2011b}\\
100 & $<0.005$ & BI02 \\ 
1.4 & $<0.00094$ & FIRST \\ 
\hline
\end{tabular*}
\egroup
\end{table}
\begin{table}
\bgroup
\ContinuedFloat
\def\arraystretch{1.}
\setlength{\tabcolsep}{1.2em}
\contcaption{}
\begin{tabular*}{0.5\textwidth}{l l l}
\hline \noalign {\smallskip}
$\nu$ (GHz) & $S_{\nu}$ (Jy) & Reference \\
\hline \noalign {\smallskip} 
\multicolumn{3}{c}{\it B1127+385} \\ 
\hline \noalign {\smallskip}
667 & $<0.065$ & BI02 \\
353.0 & $0.014\pm0.002$ & BI02 \\ 
231 & $<0.0028$ & BI02 \\
8.4 & $0.027\pm0.002$ & \cite{Koopmans:1999} \\ 
5.0 & $0.027\pm0.002$ & \cite{Koopmans:1999} \\ 
4.85 & $0.041\pm0.007$ & \cite{Gregory:1991} \\ 
4.85 & $0.037\pm0.006$ & \cite{Becker:1991} \\ 
1.7 & $0.030\pm0.002$ & \cite{Koopmans:1999} \\ 
1.4 & $0.029\pm0.001$ & \cite{Becker:1995} \\ 
\hline \noalign {\smallskip} 
\multicolumn{3}{c}{\it RX~J1131-1231} \\ 
\hline \noalign {\smallskip} 
216.0 & $<0.0025$ & \cite{Leung:2017} \\ 
144.1 & $0.00195\pm0.00020\dagger$ & \cite{Paraficz:2017} \\ 
139.3 & $0.00039\pm0.00012\dagger$ & \cite{Leung:2017} \\ 
4.9 & $0.00127\pm0.00004$ & \cite{Leung:2017} \\ 
1.4 & $<0.028$ & FIRST \\ 
\hline \noalign {\smallskip}
    \multicolumn{3}{c}{\it B1152+200} \\ 
    \hline \noalign {\smallskip} 
667 & $<0.07$ & BI02 \\ 
353 & $<0.0065$ & BI02 \\ 
14.94 & $0.0581\pm0.0004$ & \cite{Myers:1999} \\ 
8.46 & $0.0695\pm0.0005$ &  \cite{Myers:1999} \\ 
4.85 & $0.070\pm0.0011$ & \cite{Gregory:1991} \\
1.4 & $0.0774\pm0.008$ & NVSS \\
\hline \noalign {\smallskip}
\multicolumn{3}{c}{\it Q1208+101} \\ 
\hline \noalign {\smallskip} 
353 & $0.0081\pm0.0020$ & BI02 \\ 
250 & $0.0042\pm0.0019$ & \cite{Andreani:1999} \\ 
231 & $0.003\pm0.001$ & BI02 \\ 
100 & $<0.0009$ & BI02 \\ 
1.4 & $<0.00095$ & FIRST \\ 
\hline \noalign {\smallskip}
\multicolumn{3}{c}{\it SDSSJ1330+1810} \\ 
\hline \noalign {\smallskip} 
1.4 & $<0.0002$ & \cite{Stacey:2015} \\
\hline \noalign {\smallskip}
\multicolumn{3}{c}{\it SDSSJ1353+1138} \\ 
\hline \noalign {\smallskip} 
1.4 & $0.0005\pm0.00015$ & FIRST \\ 
\hline \noalign {\smallskip} 
\multicolumn{3}{c}{\it B1359+154} \\ 
\hline \noalign {\smallskip} 
667 & $0.039\pm0.010^{\dagger}$ & BI02 \\ 
353 & $0.012\pm0.002$ & BI02 \\ 
96.4 & $0.00360\pm0.00018$ & \cite{Riechers:2011b} \\ 
14.94 & $0.01625\pm0.0009$ & \cite{Myers:1999} \\
8.46 & $0.0279\pm0.0014$ & \cite{Myers:1999} \\
4.8 & $0.066\pm0.010$ & \cite{Becker:1991} \\ 
1.4 & $0.115\pm0.035$ & NVSS \\ 
\hline
\end{tabular*}
\egroup
\end{table}
\begin{table}
\bgroup
\ContinuedFloat
\def\arraystretch{1.}
\setlength{\tabcolsep}{1.2em}
\contcaption{}
\begin{tabular*}{0.5\textwidth}{l l l}
\hline \noalign {\smallskip}
$\nu$ (GHz) & $S_{\nu}$ (Jy) & Reference \\
\hline \noalign {\smallskip} 
\multicolumn{3}{c}{\it H1413+117} \\ 
\hline \noalign {\smallskip} 
5080 & $0.207\pm0.051$ & \cite{Weiss:2003} \\ 
5000 & $0.230\pm0.078$ & \cite{Barvainis:1995} \\ 
3261 & $0.266\pm0.050$ & \cite{Weiss:2003} \\ 
3000 & $0.370\pm0.078$ & \cite{Barvainis:1995} \\ 
2520 & $0.356\pm0.048$ & \cite{Weiss:2003} \\ 
1630 & $0.240\pm0.095$ & \cite{Weiss:2003} \\ 
1500 & $0.280\pm0.014$ & \cite{Rowan-Robinson:2000} \\ 
870 & $0.189\pm0.056$ & \cite{Barvainis:1992} \\ 
857 & $0.376\pm0.015$ & \cite{Weiss:2003} \\ 
857 & $0.293\pm0.014$ & \cite{Benford:1999} \\ 
685 & $0.224\pm0.038$ & \cite{Barvainis:1992} \\ 
394 & $0.044\pm0.008$ & \cite{Barvainis:1992} \\ 
375 & $0.066\pm0.007$ & \cite{Hughes:1997} \\ 
353 & $0.059\pm0.008$ & BI02 \\ 
250 & $0.016\pm0.002$ & \cite{Weiss:2003} \\ 
240 & $0.018\pm0.002$ & \cite{Barvainis:1995} \\ 
230 & $0.0075\pm0.0006$ & \cite{Weiss:2003} \\ 
100 & $<0.0015$ & \cite{Weiss:2003} \\ 
92.9 & $0.0003\pm0.0001$ & Stacey et al. in prep \\ 
24.0 & $0.00026\pm0.00003$ & \cite{Solomon:2003} \\ 
14.9 & $0.00056\pm0.00018$ & \cite{Barvainis:1997} \\ 
8.5 & $0.00098\pm0.00008$ & \hspace{5mm} " \\ 
4.9 & $0.00195\pm0.00013$ & \hspace{5mm} " \\ 
1.5 & $0.00768\pm0.00050$ & \hspace{5mm} " \\ 
1.4 & $0.0082\pm0.0006$ & NVSS \\ 
\hline \noalign {\smallskip} 
\multicolumn{3}{c}{\it B1422+231} \\ 
\hline \noalign {\smallskip} 
22.5 & $0.145\pm0.015$ & \cite{Tinti:2005} \\ 
15.0 & $0.251\pm0.013$ & \hspace{5mm} " \\ 
8.5 & $0.460\pm0.016$ & \hspace{5mm} " \\
8.1 & $0.479\pm0.016$ & \hspace{5mm} " \\
4.9 & $0.669\pm0.020$ & \hspace{5mm} " \\
4.5 & $0.686\pm0.021$ & \hspace{5mm} " \\
1.7 & $0.414\pm0.012$ & \hspace{5mm} " \\
1.4 & $0.352\pm0.011$ & \hspace{5mm} " \\
\hline \noalign {\smallskip}
    \multicolumn{3}{c}{\it SBS1520+530} \\ 
    \hline \noalign {\smallskip} 
353 & $<0.0078$ & BI02 \\ 
250 & $<0.0042$ & BI02 \\ 
100 & $<0.0006$ & BI02 \\ 
1.4 & $<0.001$ & FIRST \\ 
\hline \noalign {\smallskip} 
    \multicolumn{3}{c}{\it B1600+434} \\ 
    \hline \noalign {\smallskip} 
353 & $0.0073\pm0.0018$ & BI02 \\ 
300 & $0.0126\pm0.0023$ & BI02 \\ 
100 & $0.0250\pm0.0003\dagger$ & BI02 \\ 
15 & $0.042\pm0.002$ & \cite{Waldram:2003} \\
8.4 & $0.132\pm0.013$ & \cite{Jackson:1995} \\
4.8 & $0.037\pm0.007$ & \cite{Gregory:1991} \\ 
1.4 & $0.079920\pm0.000145$ & FIRST \\ 
\hline
\end{tabular*}
\egroup
\end{table}
\begin{table}
\bgroup
\ContinuedFloat
\def\arraystretch{1.}
\setlength{\tabcolsep}{1.2em}
\contcaption{}
\begin{tabular*}{0.5\textwidth}{l l l}
\hline \noalign {\smallskip}
$\nu$ (GHz) & $S_{\nu}$ (Jy) & Reference \\
\hline \noalign {\smallskip} 
\multicolumn{3}{c}{\it B1608+656} \\ 
\hline \noalign {\smallskip} 
353.0 & $0.0081\pm0.0017$ & BI02 \\ 
300.0 & $<0.0066$ & \cite{Xanthopoulos:2001} \\ 
231.0 & $0.0056\pm0.0017$ & BI02 \\ 
100.0 & $0.0081\pm0.0004$ & BI02 \\ 
15 & $0.081\pm0.004$ & \cite{Snellen:1995} \\
8.4 & $0.083\pm0.004$ & \cite{Snellen:1995} \\ 
8.4 & $0.0732\pm0.0020$ & \cite{Myers:1995} \\ 
4.8 & $0.0920\pm0.0090$ & \cite{Gregory:1991} \\ 
1.4 & $0.063\pm0.003$ & \cite{Snellen:1995} \\
\hline \noalign {\smallskip} 
    \multicolumn{3}{c}{\it PMN~J1632-0033} \\ 
    \hline \noalign {\smallskip} 
43.34 & $0.089\pm0.014$ &  \cite{Winn:2002}\\
43.34 & $0.112\pm0.006$ &  \cite{Winn:2003b}\\
22.46 & $0.135\pm0.014$ &  \cite{Winn:2002} \\
22.46 & $0.161\pm0.008$ &  \cite{Winn:2003b} \\
22.46 & $0.158\pm0.008$ &  \cite{Winn:2003b} \\
14.94 & $0.153\pm0.007$ &  \cite{Winn:2002} \\
14.94 & $0.187\pm0.009$ & \cite{Winn:2003b} \\
14.94 & $0.195\pm0.010$ & \cite{Winn:2003b} \\
8.64 & $0.177\pm0.005$ &  \cite{Winn:2002} \\
8.46 & $0.220\pm0.011$ & \cite{Winn:2003b} \\
8.46 & $0.227\pm0.011$ & \hspace{5mm} " \\
8.46 & $0.211\pm0.011$ & \hspace{5mm} " \\
8.46 & $0.152\pm0.008$ & \cite{McKean:2007} \\
8.45 & $0.165\pm0.030$ &  \cite{Winn:2002} \\
8.44 & $0.160\pm0.040$ &  \hspace{5mm} " \\
6.1 & $0.201\pm0.006$ & \hspace{5mm} " \\
5.0 & $0.191\pm0.020$ &  \hspace{5mm} " \\
5.0 & $0.222\pm0.011$ & \cite{Winn:2003} \\
4.8 & $0.233\pm0.007$ &  \cite{Winn:2002} \\
4.86 & $0.223\pm0.011$ & \cite{Winn:2002} \\
4.86 & $0.204\pm0.010$ & \cite{McKean:2007} \\
4.85 & $0.270\pm0.040$ & \cite{Becker:1991} \\
3.9 & $0.235\pm0.047$ & \cite{Larionov:1994} \\
2.7 & $0.200\pm0.040$ & \cite{Wright:1990} \\
1.4 & $0.230\pm0.012$ &  \cite{Winn:2002} \\
1.4 & $0.236\pm0.047$ & \cite{White:1992} \\
\hline \noalign {\smallskip} 
    \multicolumn{3}{c}{\it FBQS1633+3134} \\ 
    \hline \noalign {\smallskip} 
353.0 & $<0.0035$ & BI02 \\ 
1.4 & $0.00177\pm0.00014$ & FIRST \\
\hline
\end{tabular*}
\egroup
\end{table}
\begin{table}
\bgroup
\ContinuedFloat
\def\arraystretch{1.}
\setlength{\tabcolsep}{1.2em}
\contcaption{}
\begin{tabular*}{0.5\textwidth}{l l l}
\hline \noalign {\smallskip}
$\nu$ (GHz) & $S_{\nu}$ (Jy) & Reference \\
\hline \noalign {\smallskip}
\multicolumn{3}{c}{\it PKS~1830$-$211} \\
\hline \noalign {\smallskip}
353 & $0.79\pm0.17$ & \cite{Giommi:2012} \\
300 & $0.8\pm0.1$ &  \cite{Marti-Vidal:2013}\\
250 & $0.9\pm0.1$ &  \cite{Marti-Vidal:2013}\\
229 & $1.23\pm0.06$ & \cite{Agudo:2010} \\
217 & $1.18\pm0.08$ & \cite{Giommi:2012} \\ 
143 & $1.64\pm0.07$ & \cite{Giommi:2012} \\
100 & $2.0\pm0.02$ &  \cite{Muller:2006} \\
100 & $2.47\pm0.11$ & \cite{Giommi:2012} \\
86 & $1.76\pm0.09$ & \cite{Agudo:2010} \\
70 & $2.44\pm0.22$ & \cite{Giommi:2012} \\
44 & $3.42\pm0.36$ & \cite{Giommi:2012} \\
33 & $4.12\pm0.24$ & \cite{Massardi:2009} \\
30 & $4.11\pm0.33$ & \cite{Giommi:2012} \\
23 & $5.08\pm0.22$ & \cite{Massardi:2009} \\
20 & $5.50\pm0.36$ & \cite{Massardi:2008} \\
14 & $6.45\pm0.65$ & \cite{Henkel:2008} \\
12.8 & $8.12\pm0.81$ & \cite{Henkel:2008} \\
8.4 & $6.59\pm0.66$ & \cite{Wright:1990} \\
6.55 & $9.74\pm0.97$ & \cite{Ellingsen:2012} \\
5 & $8.9\pm0.9$ & \cite{Wright:1990} \\
4.9 & $7.92\pm0.10$ & \cite{Griffith:1994} \\
2.7 & $9.3\pm0.93$ & \cite{Wright:1990} \\
1.4 & $10.90\pm0.33$ & NVSS \\
\hline \noalign {\smallskip}
\multicolumn{3}{c}{\it PMN~J1838$-$3427} \\
\hline \noalign {\smallskip}
20 & $0.234\pm0.012$ & \cite{Murphy:2010} \\
14.94 & $0.181\pm0.009$ & \cite{Winn:2000}  \\
8.64 & $0.264\pm0.009$ & \hspace{5mm} "  \\
8.46 & $0.206\pm0.006$ & \hspace{5mm} "  \\
8.46 & $0.181\pm0.005$ & \hspace{5mm} " \\
8 & $0.284\pm0.014$ & \cite{Murphy:2010} \\
5 & $0.299\pm0.015$ & \cite{Murphy:2010} \\
4.86 & $0.214\pm0.006$ & \cite{Winn:2000} \\
4.85 & $0.258\pm0.021$ & \cite{Wright:1990} \\
4.80 & $0.219\pm0.007$ & \cite{Winn:2000} \\
2.7 & $0.240\pm0.048$ & \cite{Wright:1990} \\
1.4 & $0.280\pm0.008$ & NVSS \\
\hline \noalign {\smallskip} 
    \multicolumn{3}{c}{\it B1933+503} \\ 
    \hline \noalign {\smallskip} 
3000 & $<0.443$ & \cite{Chapman:1999} \\ 
667 & $0.114\pm0.017$ & \hspace{5mm} " \\ 
353 & $0.0240\pm0.0026$ & \hspace{5mm} " \\ 
231 & $0.030\pm0.007$ & \hspace{5mm} " \\ 
15 & $0.0371\pm0.0011$ & \cite{Sykes:1998} \\ 
8.4 & $0.0410\pm0.0011$ & \hspace{5mm} " \\ 
5.0 & $0.0575\pm0.0057$ & \hspace{5mm} " \\ 
1.7 & $0.0759\pm0.0075$ & \hspace{5mm} " \\ 
\hline \noalign {\smallskip} 
    \multicolumn{3}{c}{\it B1938+666} \\ 
    \hline \noalign {\smallskip} 
667 & $0.126\pm0.022$ & BI02 \\ 
353 & $0.0346\pm0.0020$ & BI02 \\ 
231 & $0.0147\pm0.0020$ & BI02 \\ 
97 & $0.0200\pm0.0014$ & \cite{Riechers:2011b} \\ 
22 & $0.088\pm0.009$ & \cite{King:1997} \\ 
15 & $0.141\pm0.14$ & \cite{King:1997} \\
8.4 & $0.224\pm0.022$ & \cite{Patnaik:1992} \\
4.8 & $0.314\pm0.047$ & \cite{Becker:1991} \\ 
4.8 & $0.316\pm0.028$ & \cite{Gregory:1991} \\ 
1.4 & $0.5768\pm0.0173$ & NVSS \\
1.4 & $0.634\pm0.063$ & FIRST \\
\hline
\end{tabular*}
\egroup
\end{table}
\begin{table}
\bgroup
\ContinuedFloat
\def\arraystretch{1.}
\setlength{\tabcolsep}{1.2em}
\contcaption{}
\begin{tabular*}{0.5\textwidth}{l l l}
\hline \noalign {\smallskip}
$\nu$ (GHz) & $S_{\nu}$ (Jy) & Reference \\
\hline \noalign {\smallskip} 
    \multicolumn{3}{c}{\it PMNJ2004-1349} \\
    \hline \noalign {\smallskip} 
22.46 & $0.0164\pm0.002$ & \cite{Winn:2001} \\
14.96 & $0.0205\pm0.0010$ & \hspace{5mm} " \\
8.46 & $0.0294\pm0.0090$  & \hspace{5mm} " \\
5.0 & $0.030\pm0.002$ & \hspace{5mm} " \\
4.8 & $0.073\pm0.011$ & \cite{Griffith:1994} \\ 
1.4 & $0.079\pm0.08$ & NVSS \\
\hline \noalign {\smallskip} 
    \multicolumn{3}{c}{\it MGJ2016+112} \\ 
    \hline \noalign {\smallskip} 
353.0 & $<0.0048$ & \cite{Barvainis:2002b} \\ 
231.0 & $<0.0025$ & \hspace{5mm} " \\ 
100.0 & $0.0018\pm0.0002$ & \hspace{5mm} " \\ 
4.8 & $0.098\pm0.010$ & \cite{Bennett:1986} \\  
1.4 & $0.1911\pm0.0058$ & NVSS \\ 
\hline \noalign {\smallskip} 
    \multicolumn{3}{c}{\it B2114+022} \\ 
    \hline \noalign {\smallskip} 
353.0 & $<0.0043$ & BI02 \\
87 & $<0.030$ & \cite{Xanthopoulos:2001} \\
15.0 & $0.051\pm0.002$ & \cite{Augusto:2001} \\ 
8.4 & $0.100\pm0.002$ & \hspace{5mm} " \\ 
5.0 & $0.156\pm0.016$ & \hspace{5mm} " \\
5.0 & $0.230\pm0.023$ & \cite{Wright:1990} \\ 
4.85 & & \cite{Griffith:1990} \\
4.85 & $0.136\pm0.003$ & \cite{Vollmer:2008} \\
4.775 & & \cite{Bennett:1986} \\
2.6 & & \cite{Wright:1990} \\
1.7 & $0.148\pm0.015$ & \cite{Augusto:2001} \\
1.4 & $0.137\pm0.004$ & NVSS \\
\hline \noalign {\smallskip} 
    \multicolumn{3}{c}{\it HE2149-2745} \\ 
    \hline \noalign {\smallskip} 
353.0 & $0.008\pm0.002$ & BI02 \\ 
100.0 & $<0.0064$ & BI02 \\ 
1.4 & $<0.0025$ & NVSS \\ 
\hline \noalign {\smallskip} 
    \multicolumn{3}{c}{\it Q2237+030} \\ 
    \hline \noalign {\smallskip} 
667 & $<0.017$ & BI02 \\
353 & $0.0039\pm0.0012$ & BI02 \\
250 & $<0.0064$ & BI02 \\
100 & $<0.0008$ & BI02 \\
8.4 & $0.000593\pm0.000088$ & \cite{Falco:1996} \\
1.5 & $0.000832\pm0.000087$ & \cite{Falco:1996} \\
\hline \noalign {\smallskip} 
    \multicolumn{3}{c}{\it B2319+052} \\ 
    \hline \noalign {\smallskip} 
667.0 & $0.040\pm0.008$ & BI02 \\ 
353.0 & $0.0039\pm0.0012$ & BI02 \\ 
231.0 & $<0.003$ & BI02 \\ 
15.0 & $0.0182\pm0.0006$ & \cite{Rusin:2001} \\ 
8.4 & $0.0308\pm0.0001$ & \hspace{5mm} " \\ 
5.0 & $0.0666\pm0.0001$ & \hspace{5mm} " \\ 
1.4 & $0.0853\pm0.0004$ & \hspace{5mm} " \\
\hline
\end{tabular*}
\egroup
\end{table}
\begin{table}
\bgroup
\ContinuedFloat
\def\arraystretch{1.}
\setlength{\tabcolsep}{1.2em}
\contcaption{}
\begin{tabular*}{0.5\textwidth}{l l l}
\hline \noalign {\smallskip}
$\nu$ (GHz) & $S_{\nu}$ (Jy) & Reference \\
\hline \noalign {\smallskip} 
    \multicolumn{3}{c}{\it PSSJ2322+1944} \\ 
    \hline \noalign {\smallskip} 
4300 & $0.0137\pm0.0061^{\dagger}$ & PACS catalog \\ 
1875 & $0.0434\pm0.0084^{\dagger}$ & PACS catalog \\ 
660 & $0.075\pm0.019$ & \cite{Cox:2002} \\ 
353 & $0.0225\pm0.0025$ & \cite{Isaak:2002} \\ 
353 & $0.024\pm0.002$ & \cite{Cox:2002} \\ 
231 & $0.0096\pm0.0005$ & \cite{Omont:2001} \\ 
225 & $0.0075\pm0.0013$ & \cite{Cox:2002} \\ 
90 & $<0.00064$ & \cite{Cox:2002} \\ 
5.0 & $<9\times10^{-5}$ & \cite{Carilli:2001} \\ 
1.4 & $9.8(\pm1.5)\times10^{-5}$ & \cite{Carilli:2001} \\ 
\hline \noalign  {\smallskip} 
\end{tabular*}
\label{table:SEDrefs}
\egroup
\end{table}
\clearpage

\section{N\lowercase{otes on individual sources}}
\label{section:individual}
We discuss the results for a few individual sources of note.

\subsection{HS~0810+2554}
\label{section:0810}

HS~0810+2554 is an outlier of our sample in several respects. It has a high effective dust temperature ($T_{\rm d} = 89.0^{+6.5}_{-6.0}$~K) and the lowest dust emissivity index ($\beta = 1.0\pm0.2$) of our sample. Such a high dust temperature is more consistent with dust heated by the AGN than star formation (e.g. APM~08279+5255, \citealt{Weiss:2007}).

Despite its weak radio flux-density ($\sim\umu$Jy), this source has a mas-scale radio jet that is the major contributor to its radio emission (Hartley et al. in prep). Radio observations with the VLA and the Multi-Element Radio Linked Interferometer Network (e-MERLIN) also imply a compact radio-emitting region with a scale of 70~pc \citep{Jackson:2015}. However, we find this source falls below the radio--infrared correlation, rather than above as would be expected for a source with a radio excess.

The emissivity index of 1.0 is lower than the typically observed values of $\beta=$ 1.5--2 for star-forming galaxies. These properties could be a result of composite dust emission from both AGN and star formation heating, similar to that observed in APM~08279+5255, IRAS~F10214+4724 and the Cloverleaf \citep{Beelen:2006}. Differential magnification could be responsible for a boosting of the more compact AGN-heated component.

The measured $L_{\rm FIR}$ and star formation rate given in Table~\ref{table:luminosities} of this Appendix are from a single-temperature model, and likely an overestimate the actual properties of this quasar. Additional data, taken at mm and sub-mm wavelengths will be needed to properly separate the two components of the true SED.

We try fitting a two-component dust model, one of fixed temperature 38~K (the median fitted dust temperature of the sample) and leave the second temperature as a free parameter, both with fixed $\beta=1.5$. We find a fit for the warmer component of $84.4^{+6.5}_{-5.7}$~K and derive a FIR luminosity of $3.7^{+1.7}_{-2.3}\times 10^{12}$~L$_{\odot}$ for the cold component: almost an order of magnitude lower than that from the single-temperature model. The result is consistent within 2$\sigma$ with the radio--infrared correlation and falls amid non-jetted sources.

\subsection{RX~J1131-1231}

RX~J1131$-$1231 is one of the lower redshift quasars in our sample, at $z=0.67$. The quasar is known to have a radio jet based on JVLA observations \citep{Wucknitz:2008}. \cite{Leung:2017} observed RX~J1131$-$1231 with the Plateau de Bure Interferometer (PdBI) and Combined Array for Research in Millimeter-wave Astronomy (CARMA) at 2.2 and 3~mm, respectively. They derive star-forming properties of this source by fitting an SED, assuming both the PdBI measurement and CARMA upper-limit describe the Rayleigh-Jeans slope of the modified black-body. We consider these data and also include a recent ALMA observation at 2.1~mm \citep{Paraficz:2017}, however, we find significant differences between the ALMA 2.1~mm and PdBI 2.2~mm measurements. \citeauthor{Paraficz:2017} propose the difference is due to a contribution from synchrotron emission at the base of the jet associated with the AGN, which could be either highly variable, or so compact ($\sim$10$^{-4}$~pc) that micro-lensing may be changing the flux-density over time-scales of months (the observations were performed 5--7 months apart: PdBI between December 2014 and February 2015; ALMA in July 2015). We include only the {\it Herschel}/SPIRE measurements and the CARMA upper limit to constrain the thermal dust emission, finding a relatively low dust temperature of $T_{\rm d} = 21^{+6}_{-4}$~K and a high emissivity index of $\beta=2.7^{+1.0}_{-0.7}$, however, this is not robust as the peak is poorly constrained. We do not attempt to fit a synchrotron component due to the uncertainties discussed here and by \cite{Paraficz:2017}. Further high and low wavelength data are needed to better constrain the dust temperature and $L_{\rm FIR}$, and characterise the millimetre emission for this object.

\subsection{H~1413+117}

The {\it Cloverleaf} quasar (H~1413+117) has been studied extensively over the past $\sim$20 years as it is one of the most FIR-luminous gravitationally-lensed quasars known, and so there are many measurements in the literature that cover the full infrared SED. The SED can be resolved into two dust peaks that are presumably due to heating by both star formation ($T_d = 35.6\pm0.6$~K) and AGN activity ($T_d = 125.6^{+10.6}_{-8.9}$~K). We find this quasar falls above the radio--infrared correlation with a $q_{\rm IR}$ value of $1.42\pm0.01$, consistent with jet emission known to exist in this source based on radio observations with e-MEJIN (Stacey et al. in prep).

With the addition of the {\it Herschel}/SPIRE data, we see clear differences in the measurements around the lower-temperature peak over a period of $\sim$20 years. This is most obvious with the four measurements around 350~$\umu$m, which have increased intermittently to a factor of 2 relative to the first measurement by \cite{Barvainis:1992}. This is likely the effect of calibration errors in previous measurements rather than intrinsic variability.

The spread of the data causes uncertainty in the SED fitting as the AGN contribution and Rayleigh-Jeans slope are not well constrained. However, the new {\it Herschel}/SPIRE data constrain the peak of the SF-heated dust component accurately, thus, assuming the contribution from the AGN component is small at those wavelengths, the effect of the uncertainty on the derived FIR luminosity due to star formation is not significant.

\subsection{PKS~1830-211}
\label{section:1830}

PKS~1830$-$211 is a radio-powerful gravitationally-lensed blazar; a radio source that is being viewed directly down the line-of-sight of the relativistic jet \citep{Marti-Vidal:2013}. The SED appears to be dominated by synchrotron emission from the radio through to the FIR measured with {\it Herschel}/SPIRE. While there appears to be a tentative suggestion that the synchrotron component begins to fall off towards the FIR, this is far from clear because this source is highly variable. We fit the data $>10$~GHz with both a simple power-law, typical of optically-thin synchrotron emission, and a function that includes a modified black-body (leaving the temperature as a free parameter) to account for the possibility of a contribution from thermal dust. We assume this grey-body represents an upper limit on the star-forming properties, meaning there could be underlying dust-obscured star formation in the host galaxy at a rate as high as $\sim1000~{\rm M_{\odot}~yr^{-1}}$. The data are also consistent with a simple power law of $\alpha=-0.5$ that begins to flatten $\sim10$~GHz.

\end{appendix}

\bsp	
\label{lastpage}
\end{document}